\documentclass[pra,onecolumn,floatfix,a4paper,superscriptaddress]{revtex4}
\usepackage{bm,color,graphicx,amsmath,txfonts}
\usepackage{here}
\usepackage{array}
\usepackage{graphicx}
\usepackage[colorlinks, citecolor=blue,linkcolor=blue]{hyperref}
\usepackage{braket}
\usepackage{dsfont}
\usepackage[left=	2cm,top=2cm,right=1.2cm,bottom=2cm]{geometry}
\usepackage{tikz}
\everymath{\displaystyle}

\newcommand{\ke}[1]{|#1\rangle}

\begin{document}

\makeatletter
\renewcommand{\@biblabel}[1]{\makebox[2em][l]{\textsuperscript{\textcolor{black}{\fontsize{10}{12}\selectfont[#1]}}}}
\makeatother

\let\oldbibliography\thebibliography
\renewcommand{\thebibliography}[1]{%
  \addcontentsline{toc}{section}{\refname}%
  \oldbibliography{#1}%
  \setlength\itemsep{0pt}%
}

\title{Optimal multiparameter quantum estimation  in accelerating Unruh-DeWitt detectors}

\author{Omar Bachain}
\address{LPHE-Modeling and Simulation, Faculty of Sciences, Mohammed V University in Rabat, Rabat, Morocco}

\author{Elhabib \surname{Jaloum}}
\address{LPTHE-Department of Physics, Faculty of Sciences, Ibnou Zohr University, Agadir 80000, Morocco}

\author{Mohamed \surname{Amazioug} }
\email{m.amazioug@uiz.ac.ma}
\address{LPTHE-Department of Physics, Faculty of Sciences, Ibnou Zohr University, Agadir 80000, Morocco}

\author{Reem Altuijri}
\address{Department of Physics, College of Science, Princess Nourah bint Abdulrahman University, P. O. Box 84428, Riyadh 11671, Saudi Arabia}

\author{Rachid Ahl Laamara}
\address{LPHE-Modeling and Simulation, Faculty of Sciences, Mohammed V University in Rabat, Rabat, Morocco}
\address{Centre of Physics and Mathematics, CPM, Faculty of Sciences, Mohammed V University in Rabat, Rabat, Morocco}

\author{Abdel-Haleem Abdel-Aty}
\address{Department of Physics, College of Sciences, University of Bisha, Bisha 61922, Saudi Arabia}

\date{\today}

\begin{abstract}

The quantum Fisher information matrix (QFIM) is central to multiparameter quantum metrology, dictating the attainable sensitivity via the quantum Cram\'er-Rao bound. In this work, we investigate the ultimate precision limits for relativistic quantum thermometry in a bipartite system of uniformly accelerated Unruh-DeWitt detectors. Utilizing the symmetric logarithmic derivative (SLD) formalism within the QFIM framework, we analyze the individual and simultaneous estimation of the Unruh temperature $T$ and the initial-state parameter $\Delta_0$. In the noiseless case, we demonstrate that these two parameters are quantum compatible, allowing the multiparameter quantum Cram\'er-Rao bound to be saturated without a loss of precision. We then examine the impact of environmental effects by comparing Markovian and non-Markovian dynamics. In the Markovian regime, dissipation leads to a monotonic degradation of estimation precision; conversely, non-Markovian memory effects induce temporal oscillations and transient precision enhancements due to information backflow. The robustness of the estimation protocols is further analyzed under correlated noisy channels, including amplitude damping, phase flip, and phase damping. We show that dissipative noise results in more significant precision loss than purely dephasing mechanisms, whereas classical correlations in the noise mitigate this degradation. Finally, we discuss the estimation of the detector energy-level spacing $\omega$ as a natural extension, highlighting its sensitivity to the environmental structure. Our results provide a unified framework for relativistic multiparameter quantum metrology within the context of open quantum systems.

\end{abstract}
\maketitle

\section{INTRODUCTION }    \label{sec:1}

Quantum parameter estimation (QPE) has emerged as a rapidly advancing interdisciplinary field that unifies classical estimation theory with the principles of quantum mechanics~\cite{Pirandola2018,Giovannetti2011,Mukhopadhyay2025}. 
Its central objective is to surpass classical precision limits in the estimation of unknown parameters by exploiting uniquely quantum resources. 
Such enhancement is typically achieved through the careful design of estimation protocols that harness quantum entanglement~\cite{Rodriguez2024,Demkowicz2014,Zhang2023}, nonclassical states~\cite{Pezze2018,Rahman2025}, and various forms of quantum correlations~\cite{Amazioug2024,Elghaayda2025,Sahota2015,Ciampini2016}.  
To quantitatively assess the performance of QPE protocols, the quantum Cram\'er--Rao bound (CRB) serves as a cornerstone of the theoretical framework, providing asymptotically attainable lower bounds on the estimation variance~\cite{Mondal2025,Safranek2018,Genoni2013,Asjad2023,Cavazzoni2025,He2025,Sidhu2020,Demkowicz2020,Paris2009}. 
The inverse of the quantum CRB defines the quantum Fisher information (QFI), which quantifies the sensitivity of a quantum state to infinitesimal variations in the parameter of interest~\cite{Gessner2023,Gebhart2024,Montenegro2025}. 
Beyond its foundational role in parameter estimation, the QFI has become a versatile analytical tool with wide-ranging applications, including quantum lidar~\cite{Reichert2024,Qian2023,AbdRabbou2022}, quantum telescopy~\cite{Khabiboulline2019,Gottesman2012}, and quantum thermometry~\cite{Jahromi2023,Rubio2021,Chang2024}.  

In recent years, considerable attention has been devoted to the study of QPE in relativistic settings~\cite{Ahmadi2014,Du2021,Patterson2023,Zhao2020,Chen2025}, where parameters such as acceleration~\cite{Zhao2020}, temperature~\cite{Feng2022,Tian2015}, and phenomena related to the Unruh--Hawking effect~\cite{Aspachs2010} are of particular interest. 
Among these directions, the characterization of uniformly accelerating observers occupies a prominent position, as it provides valuable insights into quantum information processing within relativistic frameworks~\cite{Zhao2020,Liu2021,Wang2025,Zhao2021}. 
For example, Zhao \textit{et al.} investigated the quantum estimation of both acceleration and temperature using a uniformly accelerated Unruh-DeWitt detector interacting with a massless scalar field in the Minkowski vacuum~\cite{Zhao2020,Wu2025,Louko2008,MartinMartinez2013,Lin2017,Foo2021,Sachs2017,Foo2020,Du2026,Li2025,Bachain2025}. 
Their results revealed that optimal acceleration estimation is achieved at specific acceleration regimes and demonstrated that the presence of a reflecting boundary enhances the precision of both acceleration and temperature estimation. 
Subsequently, Liu \textit{et al.} examined the estimation of the initial weight parameter, phase parameter, and the inverse of acceleration in a similar detector--field configuration~\cite{Liu2021}. 
However, these studies predominantly focus on single-parameter estimation strategies based on the quantum CRB, while the more intricate problem of multiparameter estimation and its evolution over time for uniformly accelerated Unruh--DeWitt detectors remains largely unexplored.  

Extending quantum estimation theory from single-parameter to multiparameter scenarios is fundamentally nontrivial~\cite{Albarelli2020,Demkowicz2020,Chen2024npj}. 
In contrast to the single-parameter case, optimal measurements associated with different parameters are generally incompatible, which may render the quantum CRB non-saturable in multiparameter settings~\cite{Albarelli2020,Demkowicz2020,Chen2024npj,Chen2022PRA}. 
Formally, the standard quantum CRB is derived by quantizing the classical CRB through the symmetric logarithmic derivative (SLD), leading to the SLD-CRB~\cite{Helstrom1976}. 
Nevertheless, this quantization procedure is not unique~\cite{Hayashi2017}. 
An alternative formulation based on the right logarithmic derivative (RLD) gives rise to the RLD-CRB~\cite{Yuen1973}, yet this bound may also fail to be tight, as the corresponding optimal estimators do not necessarily correspond to physically realizable positive-operator-valued measures (POVMs)~\cite{Suzuki2019,Ragy2016}. 
To overcome these limitations, the Holevo Cram\'er--Rao bound (HCRB) is often employed, as it provides the ultimate attainable precision limit in multiparameter quantum estimation and is generally tighter than both the SLD-CRB and RLD-CRB~\cite{Albarelli2020,Demkowicz2020,Yang2019}.

In this paper, we investigate multiparameter quantum estimation in a bipartite system composed of two uniformly accelerated Unruh--DeWitt detectors interacting with a scalar field. 
We focus on the simultaneous and individual estimation of the Unruh temperature $T$ and the initial-state parameter $\Delta_0$, employing the quantum Fisher information matrix (QFIM) to derive the corresponding quantum Cram\'er--Rao bounds \cite{Bakmou2019,Alanazi2024}. 
We first analyze the noiseless scenario and demonstrate the quantum compatibility of the parameters, identifying the conditions under which the multiparameter bound is saturable.
We then examine the dynamical behavior of the estimation precision in both Markovian and non-Markovian regimes, revealing qualitatively distinct temporal features associated with dissipative and memory-induced effects. 
In particular, we show that non-Markovian dynamics can induce transient enhancements of precision due to information backflow. 
To assess the robustness of the estimation protocols, we further incorporate correlated noisy channels, including amplitude damping, phase flip, and phase damping \cite{Pirandola2008,Damodarakurup2009}. 
We characterize how different noise mechanisms affect the attainable precision and compare their impact on individual and simultaneous strategies  \cite{jaloum2025quantum,jaloum2026controlling}. 
Finally, as a natural extension, we explore the estimation of the detector energy-level spacing $\omega$, highlighting its sensitivity to environmental structure and dynamical regimes. 
Our results provide a comprehensive framework for relativistic multiparameter quantum metrology in accelerated open quantum systems.
\begin{figure}[H]
	\centering
	\includegraphics[
	width=0.5\textwidth,
	trim=0 180 0 150,
	clip
	]{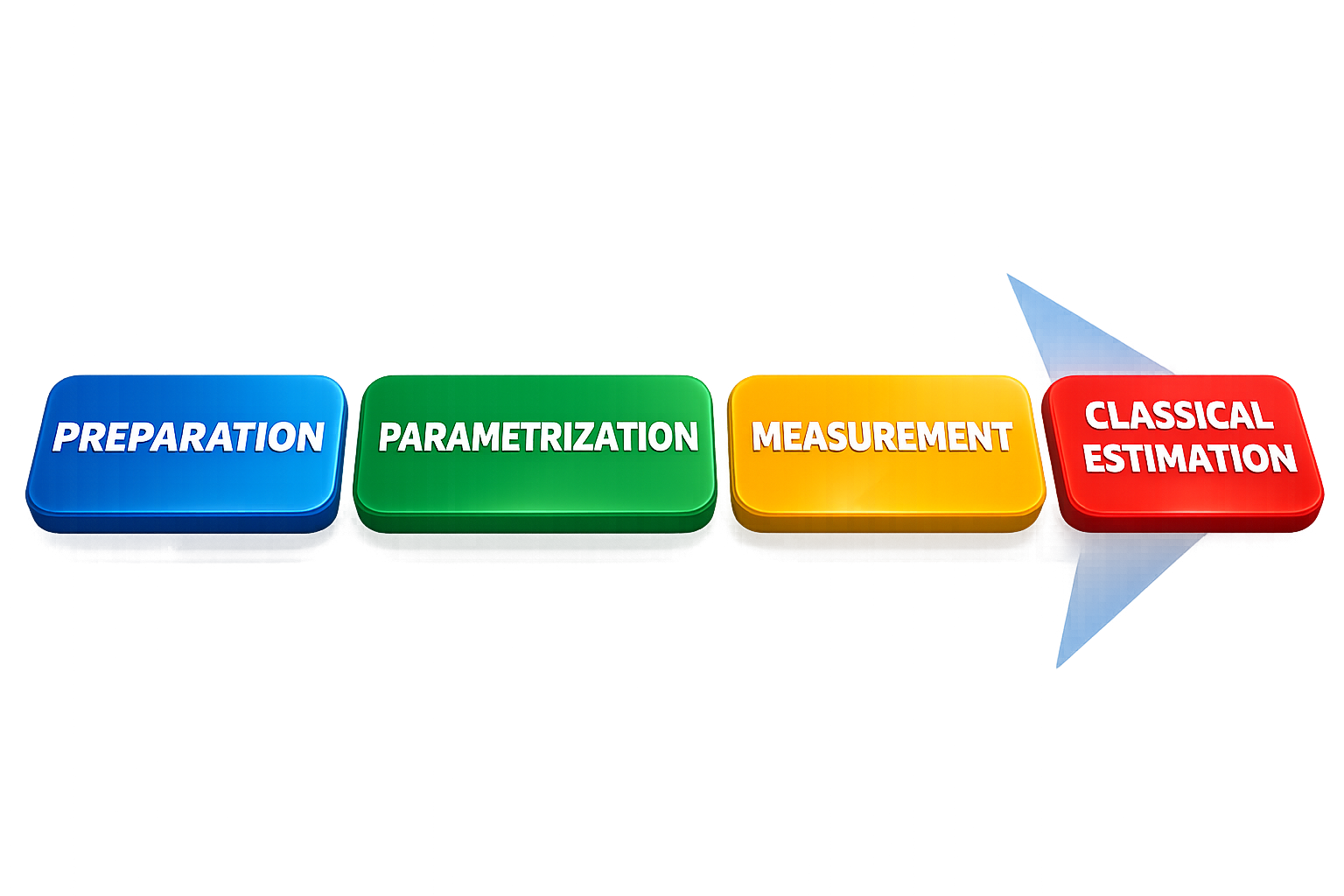}
	\caption{
		Schematic diagram of the complete quantum metrological process, consisting of four sequential steps: probe state preparation, parameter encoding, measurement, and classical estimation.
	}
\end{figure}
The remainder of this paper is organized as follows. 
In Sec.~\ref{sec:2}, we present the theoretical model describing two uniformly accelerated Unruh--DeWitt detectors and outline the dynamical framework governing their interaction with the scalar field. 
Section~\ref{sec:3} introduces the quantum Fisher information matrix (QFIM) formalism and establishes the multiparameter estimation framework adopted throughout this work. 
In Sec.~\ref{sec:4}, we analyze the QFIM in the absence of noisy channels and discuss the fundamental precision limits in the noiseless scenario. 
Section~\ref{sec:5} is devoted to the investigation of individual and simultaneous parameter estimation in both Markovian and non-Markovian regimes. 
In Sec.~\ref{sec:6}, we examine the robustness of the estimation strategies under correlated quantum channels, including amplitude damping, phase flip, and phase damping. 
Finally, Sec.~\ref{sec:8} summarizes the main results and presents concluding remarks.

\section{THEORETICAL MODEL}\label{sec:2}

To investigate the dynamics of two uniformly accelerated Unruh-DeWitt (UdW) detectors in a $(3+1)$-dimensional Minkowski spacetime \cite{Benatti2004,Bhuvaneswari2022,Elghaayda2025}, we employ the open quantum systems formalism. Within this framework, the detectors constitute the reduced subsystem whose evolution becomes non-unitary due to decoherence and dissipative effects induced by their interaction with a quantum scalar field environment.
Each detector is modeled as an effective two-level quantum system. The total Hamiltonian governing the composite system is written as \cite{Koga2019,Zhou2022,Li2024,Li2025_Tetra}
\begin{equation}
	\mathcal{H}_{\text{tot}} = \mathcal{H}_{S} + \mathcal{H}_{F} + \mathcal{H}_{\text{int}},
\end{equation}
where $\mathcal{H}_{S}$ denotes the free Hamiltonian of the two non-interacting detectors in their co-moving frame, $\mathcal{H}_{F}$ corresponds to the free massless scalar field, and $\mathcal{H}_{\text{int}}$ represents the interaction term.
The free detector Hamiltonian takes the form
\begin{equation}
	\mathcal{H}_{S}=\frac{\omega}{2}\left(\tau_3^{(1)} \otimes \mathbb{I}^{(2)}+\mathbb{I}^{(1)}\otimes\tau_3^{(2)}   \right),
\end{equation}
where $\tau_3^{(k)}$ are Pauli operators acting on detector $k=1,2$, and $\omega$ is the energy-level spacing.
The scalar field Hamiltonian $\mathcal{H}_{F}$ describes a free massless Klein–Gordon field $\varphi(x)$ satisfying
\begin{equation}
	g^{\mu\nu}\nabla_\mu \nabla_\nu \varphi(x)=0.
\end{equation}

The interaction Hamiltonian is assumed in monopole (dipole-type) coupling form \cite{HuYu2013}
\begin{equation}
	\mathcal{H}_{\text{int}}=\lambda\left[  \left( \tau_2^{(1)} \otimes \mathbb{I}^{(2)}\right)\varphi(x_1) +\left( \mathbb{I}^{(1)}\otimes\tau_2^{(2)}\right)\varphi(x_2) \right],
\end{equation}
where $\lambda$ is a small dimensionless coupling constant.
We consider the reduced density matrix of the detectors,
\[
\varrho_{12}(\hat{t})=\mathrm{Tr}_{F}\!\left[\varrho_{\text{tot}}(\hat{t})\right],
\]
with $\hat{t}$ denoting proper time along the detectors’ trajectories. Initially, the total state is taken to be separable,
\[
\varrho_{\text{tot}}(0)=\varrho_{12}(0)\otimes |0\rangle\langle 0|,
\]
where $|0\rangle$ is the Minkowski vacuum.
In the weak-coupling and Markovian limit, the reduced dynamics obey the Gorini–Kossakowski–Sudarshan–Lindblad master equation \cite{Gorini1976,Lindblad1976}
\begin{equation}
	\frac{d\varrho_{12}}{d\hat{t}}=-i\left[\mathcal{H}_{\mathrm{LS}},\varrho_{12}\right]
	+\mathcal{D}[\varrho_{12}],
\end{equation}
where $\mathcal{H}_{\mathrm{LS}}$ includes the Lamb-shift contribution and $\mathcal{D}$ is the dissipator.
The dissipator reads
\begin{equation}
	\mathcal{D}[\varrho]=
	\sum_{i,j=1}^{3}
	\sum_{\alpha,\beta=1}^{2}
	\frac{\Gamma_{ij}}{2}
	\left(
	2\tau_j^{(\beta)}\varrho\tau_i^{(\alpha)}
	-
	\{\tau_i^{(\alpha)}\tau_j^{(\beta)},\varrho\}
	\right),
\end{equation}
where $\Gamma_{ij}$ denotes the Kossakowski matrix.
To determine $\Gamma_{ij}$, we introduce the positive-frequency Wightman function
\begin{equation}
	W^{+}(x,x')=\langle 0|\varphi(x)\varphi(x')|0\rangle.
\end{equation}

Its Fourier transform along the detector trajectory is defined as
\begin{equation}
	\mathcal{G}(\nu)=\int_{-\infty}^{+\infty} d\hat{t} \, e^{i\nu \hat{t}} W^{+}(\hat{t}).
\end{equation}
	The asymptotic stationary state obtained from the master equation results from the interplay between dissipative effects induced by the quantum field and correlations generated through the Markovian evolution \cite{BenattiFloreanini2004PRA,BenattiFloreaniniPiani2003PRL}. 
	
	For two spatially separated detectors, the interatomic distance 
	$L = |x_{1} - x_{2}|$ 
	acts as a control parameter for correlation generation, since the cross Wightman functions — and consequently the Kossakowski coefficients — depend explicitly on $L$ through a modulation function 
	$\mathcal{G}(\omega,L)=\mathcal{G}_0(\omega) \, h(\omega,L)$ 
	\cite{Yu2011PRL,HuYu2013}. 
	Correlation generation is enhanced at small separations and disappears in the limit $L\to\infty$ \cite{BenattiFloreanini2005}. 
	Moreover, for sufficiently small but finite $L$, nonclassical correlations may persist asymptotically despite environmental dissipation \cite{BenattiFloreaniniMarzolino2010PRA}. 
	Accordingly, by restricting to small inter-detector separations, one may neglect explicit distance dependence and treat all Kossakowski matrices as identical,
	$\Gamma_{ij}^{(11)}=\Gamma_{ij}^{(22)}=\Gamma_{ij}^{(12)}=\Gamma_{ij}^{(21)}=\Gamma_{ij}$,
	where indices $(1,2)$ label the two detectors \cite{HuYu2011}.

The matrix $\Gamma_{ij}$ admits the decomposition
\begin{equation}
	\Gamma_{ij}=
	\frac{\alpha_+}{2}\delta_{ij}
	-i\frac{\alpha_-}{2}\epsilon_{ijk}\delta_{3k}
	+\alpha_0\delta_{3i}\delta_{3j},
\end{equation}
where
\begin{equation}
	\alpha_{\pm}=\mathcal{G}(\omega)\pm \mathcal{G}(-\omega),
	\qquad
	\alpha_0=\mathcal{G}(0)-\frac{\alpha_+}{2}.
\end{equation}

The Lamb-shift Hamiltonian becomes
\begin{equation}
	\mathcal{H}_{\mathrm{LS}}=\frac{1}{2}\tilde{\omega}\tau_3,
\end{equation}
with renormalized frequency
\[
\tilde{\omega}=\omega+i\big[\mathcal{K}(-\omega)-\mathcal{K}(\omega)\big],
\]
where $\mathcal{K}(\nu)$ denotes the Hilbert transform of $\mathcal{G}(\nu)$.
	Along uniformly accelerated trajectories, the Wightman function satisfies the Kubo–Martin–Schwinger (KMS) condition
	\begin{equation}
		W^{+}(\hat{t})=W^{+}(\hat{t}+i\beta),
	\end{equation}
	with $\beta=1/T$.
	This periodicity in imaginary time implies the detailed-balance relation
	\begin{equation}
		\mathcal{G}(-\omega)=e^{-\beta\omega}\mathcal{G}(\omega).
	\end{equation}
	Consequently,
	\begin{equation}
		\frac{\alpha_-}{\alpha_+}
		=\tanh\!\left(\frac{\beta\omega}{2}\right).
	\end{equation}
	This expression coincides formally with the thermal detailed-balance condition for a two-level system at temperature $T$. 
	Here, $T$ corresponds to the Unruh temperature perceived by the accelerated detectors. 
	Therefore, although the field is in the Minkowski vacuum, the detectors relax toward a thermal stationary state determined by the ratio $\tanh(\beta\omega/2)$. 
	In this operational sense, the UdW detector acts as a probe of the effective thermal character of the vacuum in non-inertial frames.
In the Bloch representation, the stationary two-detector density matrix assumes an X-structure \cite{Bhuvaneswari2022},
\begin{equation}
	\varrho_{12}^{(\infty)}=
	\begin{pmatrix}
		x& 0 & 0 & 0 \\
		0 &z & \delta & 0 \\
		0 & \delta  & z & 0 \\
		0 & 0 & 0 & y
	\end{pmatrix}.\label{densit M}
\end{equation}

Its matrix elements are given by
\begin{align*}
	x=
	\frac{(3+\Delta_0)(\eta-1)^2}{4(3+\eta^2)},&
	\qquad
	z=
	\frac{(3+\Delta_0)(\eta+1)^2}{4(3+\eta^2)},
\\
	y=
	\frac{3-\Delta_0-(\Delta_0+1)\eta^2}{4(3+\eta^2)},&
	\qquad
	\delta=
	\frac{\Delta_0-\eta^2}{2(3+\eta^2)},
\end{align*}
where we defined
$
\eta=\tanh\!\left(\frac{\beta\omega}{2}\right).
$
The stationary state is therefore fully determined by the thermal parameter $\eta$ — encoding the Unruh effect — and by the initial-state parameter
\[
\Delta_0=\sum_i
\mathrm{Tr}\!\left[\varrho_{12}(0)\,
\tau_i^{(1)}\otimes\tau_i^{(2)}\right].
\]

Positivity of the initial state requires
$-3\leq \Delta_0 \leq 1.
$
\section{QUANTUM FISHER INFORMATION MATRIX}    \label{sec:3}

We present here the theoretical framework of multiparameter quantum estimation and derive a compact analytical expression of the quantum Fisher information matrix using a density-operator vectorization technique. This formalism allows one to avoid diagonalization of the quantum state and provides a computationally efficient route for arbitrary finite-dimensional systems.

Let $\mathcal{H}$ be an $n$-dimensional Hilbert space and let $\mathcal{B}(\mathcal{H})$ denote the space of linear operators acting on it. For any operator $X \in \mathcal{B}(\mathcal{H})$, we define the vectorization map $\mathcal{V}$ (denoted vec[X]) as \cite{Gilchrist2009}

\begin{equation}
	X
	=\begin{pmatrix}
		x_{11}  & x_{12}  & \cdots & x_{1n} \\
		x_{21}  & x_{22}  & \cdots & x_{2n}  \\
		\vdots            & \vdots            & \ddots & \vdots            \\
		x_{n1}            & x_{n2}            & \cdots & x_{nn}
	\end{pmatrix},\quad
\mathcal{V}(X)=	(x_{11},x_{21},\ldots,x_{n1},x_{12},\ldots,x_{n2},\ldots,x_{1n},\ldots,x_{nn})^{T}.
\end{equation}

If $X=\sum_{k,l} x_{kl} |k\rangle\langle l|$, the vectorized form can equivalently be written as

\begin{equation}
	\mathcal{V}(X)
	=
	(I\otimes X)
	\sum_{i=1}^{n} |i\rangle\otimes |i\rangle .\label{vec}
\end{equation}

Using standard properties of the Kronecker product, one obtains

\begin{align}
	\mathcal{V}(XY) &= (I\otimes X)\mathcal{V}(Y)
	= (Y^{T}\otimes I)\mathcal{V}(X), \\
	\mathcal{V}(AXB) &= (B^{T}\otimes A)\mathcal{V}(X), \\
	\mathrm{Tr}(X^{\dagger}Y)
	&=
	\mathcal{V}(X)^{\dagger}\mathcal{V}(Y).
\end{align}

Consider now a quantum state $\varrho(\boldsymbol{\lambda})$ depending smoothly on a set of real parameters
\[
\boldsymbol{\lambda}=(\lambda_1,\lambda_2,\ldots,\lambda_m).
\]
The precision of any unbiased estimator $\hat{\boldsymbol{\lambda}}$ is constrained by the multiparameter quantum Cram\'er–Rao bound \cite{Paris2009}

\begin{equation}
	\mathrm{Cov}(\hat{\boldsymbol{\lambda}})
	\ge
	\mathcal{F}^{-1},
\end{equation}

where $\mathcal{F}$ denotes the quantum Fisher information matrix (QFIM) with entries

\begin{equation}
	\mathcal{F}_{\mu\nu}
	=
	\frac{1}{2}
	\mathrm{Tr}
	\left[
	\varrho
	\left(
	\mathcal{L}_{\mu}\mathcal{L}_{\nu}
	+
	\mathcal{L}_{\nu}\mathcal{L}_{\mu}
	\right)
	\right].
\end{equation}

The operators $\mathcal{L}_{\mu}$ are the symmetric logarithmic derivatives (SLDs), implicitly defined through

\begin{equation}
	\partial_{\lambda_\mu}\varrho
	=
	\frac{1}{2}
	\left(
	\mathcal{L}_{\mu}\varrho
	+
	\varrho \mathcal{L}_{\mu}
	\right).
\end{equation}

In the single-parameter scenario, the inequality reduces to

\begin{equation}
	\mathrm{Var}(\lambda)
	\ge
	\frac{1}{\mathcal{F}(\lambda)},
\end{equation}

where $\mathcal{F}(\lambda)=\mathrm{Tr}[\varrho \mathcal{L}_{\lambda}^{2}]$ \cite{Helstrom1969,Holevo2001}.
If the density operator admits the spectral resolution
\[
\varrho=\sum_k r_k |k\rangle\langle k|,
\]
the QFIM may be expressed as \cite{Banchi2014,Sommers2003}

\begin{equation}
	\mathcal{F}_{\mu\nu}
	=
	2
	\sum_{r_k+r_l>0}
	\frac{
		\langle k|\partial_{\lambda_\mu}\varrho|l\rangle
		\langle l|\partial_{\lambda_\nu}\varrho|k\rangle
	}
	{r_k+r_l}.
\end{equation}

The corresponding SLD operators read

\begin{equation}
	\mathcal{L}_{\mu}
	=
	2
	\sum_{r_k+r_l>0}
	\frac{
		\langle k|\partial_{\lambda_\mu}\varrho|l\rangle
	}
	{r_k+r_l}
	|k\rangle\langle l|.
\end{equation}

An equivalent integral representation is given by 

\begin{equation}
	\mathcal{F}_{\mu\nu}
	=
	2
	\int_{0}^{\infty}
	\mathrm{Tr}
	\left(
	e^{-\varrho t}
	\partial_{\lambda_\mu}\varrho
	e^{-\varrho t}
	\partial_{\lambda_\nu}\varrho
	\right)
	dt.
\end{equation}

Although exact, these representations generally require either diagonalization or matrix exponentiation. To circumvent this difficulty, we adopt the vectorization approach developed in \cite{Safranek2018}.
Define the operator

\begin{equation}
	\eta
	=
	\varrho^{T}\otimes I
	+
	I\otimes\varrho .\label{nabla}
\end{equation}

By exploiting the properties of the vectorization map, the QFIM can be compactly written as

\begin{equation}
	\mathcal{F}_{\mu\nu}
	=
	2
	\mathcal{V}(\partial_{\lambda_\mu}\varrho)^{T}
	\,
	\eta^{-1}
	\,
	\mathcal{V}(\partial_{\lambda_\nu}\varrho).\label{F}
\end{equation}

The SLDs are obtained directly in vectorized form as

\begin{equation}
	\mathcal{V}(\mathcal{L}_{\mu})
	=
	2
	\eta^{-1}
	\mathcal{V}(\partial_{\lambda_\mu}\varrho).\label{SLD}
\end{equation}

This formulation provides a fully analytical expression once the inverse of $\eta$ is computed, and proves particularly advantageous for finite-dimensional composite systems.
In multiparameter estimation, the quantum Cram\'er–Rao bound is not always attainable due to the possible incompatibility of optimal measurements \cite{Rehacek2018,Ragy2016}. A sufficient condition for saturability is the commutativity of the SLD operators,

\begin{equation}
	[\mathcal{L}_{\mu},\mathcal{L}_{\nu}]=0,
\end{equation}

which guarantees the existence of a common eigenbasis. More generally, the weaker condition

\begin{equation}
	\mathrm{Tr}\big(\varrho [\mathcal{L}_{\mu},\mathcal{L}_{\nu}]\big)=0
\end{equation}

is sufficient to ensure the attainability of the multiparameter bound \cite{Ragy2016,Matsumoto2002,Crowley2014}.
In the following, we focus on the simultaneous estimation of the anisotropy parameter $\gamma$ and the temperature $T$. The evaluation of the quantum Fisher information matrix requires, as a preliminary step, the explicit determination of the operator $\eta$ defined in Eq.~(\ref{nabla}), which takes the following block structure 
\begin{equation}
	\eta_{16\times 16} =
	\begin{pmatrix}
		\eta_{11} & 0_{4} & 0_{4} &  0_{4} \\
		0_{4} & \eta_{22} & \eta_{23} & 0_{4} \\
		0_{4} & \eta_{32} & \eta_{33} & 0_{4} \\
		 0_{4} & 0_{4} & 0_{4} & \eta_{44}
	\end{pmatrix},\label{eta}
\end{equation}
where $\eta_{ij}$ $(i,j=1,2,3,4)$ denote the $4\times4$ block matrices given by
\begin{align}
	\eta_{11}=	\begin{pmatrix}
		2x & 0 & 0 & 0 \\
		0 & x+z & \delta & 0 \\
		0 & \delta & x+z & 0 \\
		0 & 0 & 0 & x+y
	\end{pmatrix},& \quad 
\eta_{22} =\eta_{33}= 
\begin{pmatrix}
	x+z & 0 & 0 & 0 \\
	0 & 2z & \delta & 0 \\
	0 & \delta & 2z & 0 \\
	0 & 0 & 0 & y+z
\end{pmatrix},\\
\eta_{44} =
\begin{pmatrix}
	x+y & 0 & 0 & 0 \\
	0 & y+z & \delta & 0 \\
	0 & \delta & y+z & 0 \\
	0 & 0 & 0 & 2y
\end{pmatrix}&,\quad
\eta_{23} =\eta_{32} =
\begin{pmatrix}
\delta & 0 & 0 & 0 \\
0 & \delta & 0 & 0 \\
0 & 0 & \delta & 0 \\
0 & 0 & 0 & \delta
\end{pmatrix}.
\end{align}
The inverse of the matrix 
$\eta$ introduced in Eq.~(\ref{eta}) reads
\begin{equation}
	\eta^{-1}_{16\times 16} =
	\begin{pmatrix}
		(\eta^{-1})_{11} & 0_{4\times 4} & 0_{4\times 4}& 0_{4\times 4} \\
	0_{4\times 4} & (\eta^{-1})_{22} & (\eta^{-1})_{23} & 0_{4\times 4} \\
	0_{4\times 4} & (\eta^{-1})_{32} & (\eta^{-1})_{33} & 0_{4\times 4} \\
	0_{4\times 4} & 0_{4\times 4} & 0_{4\times 4} & (\eta^{-1})_{44}
	\end{pmatrix},
\end{equation}
with
\begin{align}
	(\eta^{-1})_{11}=	\begin{pmatrix}
	\alpha_1 & 0 & 0 & 0 \\
		0 & \alpha_2	& \alpha_3	& 0 \\
		0 & 	\alpha_3 	&	\alpha_2 & 0 \\
		0 & 0 & 0 & \alpha_4
	\end{pmatrix}&, \qquad
	 (\eta^{-1})_{22}=(\eta^{-1})_{33}=\begin{pmatrix}
	\alpha_2 & 0 & 0 & 0 \\
	0 & \beta_2	& \beta_3	& 0 \\
	0 & 	\beta_3 	&	\beta_2 & 0 \\
	0 & 0 & 0 & \beta_4
	\end{pmatrix},\\
		(\eta^{-1})_{44}=	\begin{pmatrix}
		\alpha_4 & 0 & 0 & 0 \\
		0 & \beta_4	& \gamma_3	& 0 \\
		0 & 	\gamma_3 	&	\beta_4 & 0 \\
		0 & 0 & 0 & \gamma_4
	\end{pmatrix}&,\qquad (\eta^{-1})_{23}=(\eta^{-1})_{32}=\begin{pmatrix}
	\alpha_3 & 0 & 0 & 0 \\
	0 & \xi_2	& \xi_3	& 0 \\
	0 & 	\xi_3 	&	\xi_2 & 0 \\
	0 & 0 & 0 & \gamma_3
	\end{pmatrix},
\end{align}
where the coefficients $\alpha_1$, $\alpha_2$, $\alpha_3$, $\alpha_4$,  $\beta_2$, $\beta_3$, $\beta_4$,  $\gamma_3$, $\gamma_4$, $\xi_1$,  $\xi_2$, and  $\xi_3$   are respectively given by
\begin{align}
	\alpha_1 &= \dfrac{1}{2x}, 
	& \alpha_2 &= \dfrac{x+z}{\left(x+z-\delta\right)\left(x+z+\delta\right)}, \\
	\alpha_3 &= -\dfrac{\delta}{\left(x+z-\delta\right)\left(x+z+\delta\right)}, 
	& \alpha_4 &= \dfrac{1}{x+y}, \\
	\beta_2  &= \dfrac{2z^2 - \delta^2}{4z\left(z^2 - \delta^2\right)}, 
	& \beta_3 &= \dfrac{\delta}{4\left(\delta^{2}-z^{2}\right)}, \\
	\beta_4  &= \dfrac{y + z}{\left(y + z - \delta\right)\left(y + z + \delta\right)}, 
	& \gamma_3 &= -\dfrac{\delta}{\left(y+z-\delta\right)\left(y+z+\delta\right)}, \\
	\gamma_4 &= \dfrac{1}{2y}, 
	& \xi_2 &= \dfrac{\delta}{4\left(\delta^{2}-z^{2}\right)}, \\
	\xi_3 &= \dfrac{\delta^{2}}{4z\left(z^{2}-\delta^{2}\right)}.
\end{align}
Using definition~(\ref{vec}), the vectorized form of the partial derivative of the density matrix with respect to the parameter $\Delta_0$ can be written as
\begin{equation}
	\mathcal{V}\!\left[\partial_\gamma \varrho \right]
	=
	\left(
	\partial_{\Delta_0} x,\; 0,\; 0,\; 0,\;
	0,\; \partial_{\Delta_0} z,\; \partial_{\Delta_0} \delta,\; 0,\;
	0,\; \partial_{\Delta_0} \delta,\; \partial_{\Delta_0} z,\; 0,\;
	0,\; 0,\; 0,\; \partial_{\Delta_0} y
	\right)^{T}.
\end{equation}

Similarly, the derivative with respect to the temperature $T$ reads
\begin{equation}
	\mathcal{V}\!\left[\partial_T \varrho \right]
	=
\left(
\partial_T x,\; 0,\; 0,\; 0,\;
0,\; \partial_T z,\; \partial_T \delta,\; 0,\;
0,\; \partial_T \delta,\; \partial_T z,\; 0,\;
0,\; 0,\; 0,\; \partial_T y
\right)^{T}.
\end{equation}

Substituting these expressions into Eq.~(\ref{F}), the quantum Fisher information (QFI) matrix can be expressed in compact form as
\begin{equation}
	\mathcal{F}
	=
	\begin{pmatrix}
		\mathcal{F}_{\Delta_0\Delta_0} & \mathcal{F}_{\Delta_0 T} \\
		\mathcal{F}_{T\Delta_0} & \mathcal{F}_{TT}
	\end{pmatrix}
	=
	\begin{pmatrix}
		2\,\mathcal{V}[\partial_{\Delta_0} \varrho]^T 
		\eta^{-1} 
	\mathcal{V}[\partial_{\Delta_0} \varrho] 
		&
		2\,\mathcal{V}[\partial_{\Delta_0} \varrho]^T 
		\eta^{-1} 
		\mathcal{V}[\partial_T \varrho]
		\\[6pt]
		2\,\mathcal{V}[\partial_T \varrho]^T 
		\eta^{-1} 
		\mathcal{V}[\partial_{\Delta_0} \varrho]
		&
		2\,\mathcal{V}[\partial_T \varrho]^T 
		\eta^{-1} 
		\mathcal{V}[\partial_T \varrho]
	\end{pmatrix}.
\end{equation}

It can be readily verified that the matrix elements follow directly from the symmetry properties of the vectorized derivatives and the structure of $\eta^{-1}$.
In quantum metrology, an estimator is said to be optimal if it saturates the quantum Cram\'er--Rao bound (QCRB). 
This bound establishes the ultimate lower limit on the covariance matrix of any unbiased estimator 
$\hat{\boldsymbol{\theta}} = (\Delta_0, T)$ and can be written as
\begin{equation}
	\mathrm{Cov}\!\left(\hat{\boldsymbol{\theta}}\right) \ge \mathcal{F}^{-1}.\label{cov}
\end{equation}

The explicit expression of the inverse QFI matrix is given by
\begin{equation}
	\mathcal{F}^{-1}
	=
	\frac{1}{\det(\mathcal{F})}
	\begin{pmatrix}
		\mathcal{F}_{TT} & -\mathcal{F}_{\Delta_0 T} \\
		- \mathcal{F}_{\Delta_0 T} & \mathcal{F}_{\Delta_0\Delta_0}
	\end{pmatrix}.
\end{equation}

Consequently, inequality~(\ref{cov}) leads to the following lower bounds on the variances~\cite{Prussing1986}
\begin{equation}
	\mathrm{Var}(\Delta_0) \ge 
	\frac{\mathcal{F}_{TT}}{\det(\mathcal{F})},\label{ineq1}
\end{equation}
\begin{equation}
	\mathrm{Var}(T) \ge 
	\frac{\mathcal{F}_{\Delta_0\Delta_0}}{\det(\mathcal{F})},\label{ineq2}
\end{equation}
together with the correlation constraint
\begin{equation}
	\left(
	\mathrm{Var}(\Delta_0) - \frac{\mathcal{F}_{TT}}{\det(\mathcal{F})}
	\right)
	\left(
	\mathrm{Var}(T) - \frac{\mathcal{F}_{\Delta_0\Delta_0}}{\det(\mathcal{F})}
	\right)
	\ge
	\left(
	\mathrm{Cov}(\Delta_0, T) + \frac{\mathcal{F}_{\Delta_0 T}}{\det(\mathcal{F})}
	\right)^{2}.\label{ineq3}
\end{equation}

Moreover, by exploiting Eq.~(\ref{SLD}), the matrix representations of the symmetric logarithmic derivatives (SLDs) associated with the parameters $\Delta_0$ and $T$ can be explicitly constructed
\begin{equation}
\mathcal{L}_{\Delta_0}=	\begin{pmatrix}
		\dfrac{\partial_{\Delta_0} x}{x} & 0 & 0 & 0 \\[6pt]
		0 & \dfrac{z\,\partial_{\Delta_0}z  - \delta\,\partial_{\Delta_0}\delta }{z^2 - \delta^2} & \dfrac{ z\,\partial_{\Delta_0}\delta -  \delta\, \partial_{\Delta_0}z}{z^2 - \delta^2} & 0 \\[12pt]
		0 & \dfrac{ z\,\partial_{\Delta_0} \delta -  \delta\,\partial_{\Delta_0}z}{z^2 - \delta^2} & \dfrac{ z\, \partial_{\Delta_0}z-  \delta\,\partial_{\Delta_0}\delta}{z^2 - \delta^2} & 0 \\[6pt]
		0 & 0 & 0 & \dfrac{\partial_{\Delta_0}y}{y}
	\end{pmatrix}
	=\begin{pmatrix}
	\dfrac{1}{3+\Delta_0} & 0 & 0 & 0 \\[1em]
	0 &
	\dfrac{1+\Delta_0}{(\Delta_0+3)(\Delta_0-1)} &
	-\dfrac{2}{(\Delta_0+3)(\Delta_0-1)} &
	0 \\[1em]
	0 &
	-\dfrac{2}{(\Delta_0+3)(\Delta_0-1)} &
	\dfrac{1+\Delta_0}{(\Delta_0+3)(\Delta_0-1)} &
	0 \\[1em]
	0 & 0 & 0 &
	\dfrac{1}{3+\Delta_0}
	\end{pmatrix},
\end{equation}
and
\begin{align}
		\mathcal{L}_{T}&=	\begin{pmatrix}
		\dfrac{\partial_{T} x}{x} & 0 & 0 & 0 \\[6pt]
		0 & \dfrac{z\,\partial_{T}z  - \delta\,\partial_{T}\delta }{z^2 - \delta^2} & \dfrac{ z\,\partial_{T}\delta -  \delta\, \partial_{T}z}{z^2 - \delta^2} & 0 \\[12pt]
		0 & \dfrac{ z\,\partial_{T} \delta -  \delta\,\partial_{T}z}{z^2 - \delta^2} & \dfrac{ z\, \partial_{T}z-  \delta\,\partial_{T}\delta}{z^2 - \delta^2} & 0 \\[6pt]
		0 & 0 & 0 & \dfrac{\partial_{T}y}{y}
	\end{pmatrix}\\
&=	\begin{pmatrix}
	\dfrac{(1 + 2 e^{\omega/T})\, \omega}
	{T^{2}\left(1 + 2 \cosh\left(\frac{\omega}{T}\right)\right)}
	& 0 & 0 & 0 \\[1.2em]
	0 &
	\dfrac{\omega \sinh\left(\frac{\omega}{T}\right)}
	{T^{2}\left(1 + 2 \cosh\left(\frac{\omega}{T}\right)\right)}
	&
	\dfrac{\omega \sinh\left(\frac{\omega}{T}\right)}
	{T^{2}\left(1 + 2 \cosh\left(\frac{\omega}{T}\right)\right)}
	& 0 \\[1.2em]
	0 &
	\dfrac{\omega \sinh\left(\frac{\omega}{T}\right)}
	{T^{2}\left(1 + 2 \cosh\left(\frac{\omega}{T}\right)\right)}
	&
	\dfrac{\omega \sinh\left(\frac{\omega}{T}\right)}
	{T^{2}\left(1 + 2 \cosh\left(\frac{\omega}{T}\right)\right)}
& 0 \\[1.2em]
	0 & 0 & 0 &
	-\dfrac{(2 + e^{\omega/T})\, \omega}
	{\left(1 + e^{\omega/T} + e^{2\omega/T}\right) T^{2}}
	\end{pmatrix}.
\end{align}
The symmetric logarithmic derivatives $\mathcal{L}_{\Delta_0}$ and $\mathcal{L}_T$ share the same set of eigenvectors, which in the computational basis are given by
\begin{align}
	\mathcal{B}_{\Delta_0}=	\mathcal{B}_{T}&=\left\{
	(0,0,0,1),\;
	(0,1,1,0),\;
		(0,-1,1,0),\;
	(1,0,0,0)
	\right\}\\
	&=\left\{\ke{11}, \, \frac{1}{\sqrt{2}}\left(\ket{10}+\ket{01} \right),\,  \frac{1}{\sqrt{2}}\left(-\ket{10}+\ket{01} \right),\, \ke{00} \right\}
\end{align}
These vectors form a common eigenbasis of the two operators,
which directly implies that the corresponding symmetric logarithmic derivatives commute,
\begin{equation}
	[\mathcal{L}_{\Delta_0}, \mathcal{L}_T] = 0,
\end{equation}
which ensures the quantum compatibility of the parameters $\Delta_0$ and $T$. Consequently, a single measurement basis—given by the common eigenbasis above—simultaneously optimizes the estimation of both parameters.
Therefore, the bounds imposed by inequalities (\ref{ineq1}), (\ref{ineq2}), and (\ref{ineq3}) can be jointly saturated. In particular, the saturation of the first two inequalities guarantees that the minimum variances allowed by the quantum Cram\'er--Rao bound are achieved for the simultaneous estimation of $\Delta_0$ and $T$, thereby providing the highest attainable precision within the considered metrological framework.
\section{QFIM in the abscence of noisy channels}\label{sec:4}
In this section, we investigate the joint estimation of the parameters $\Delta_0$ and $T$ in the absence of noisy channels. Our analysis focuses on the ultimate precision limits quantified by the minimal variances $\mathrm{Var}(\Delta_0)_{\min}$ and $\mathrm{Var}(T)_{\min}$, as imposed by the quantum Cram\'er--Rao bound. By considering a noiseless scenario, we isolate the intrinsic quantum contributions to the estimation precision without the detrimental effects induced by environmental decoherence.

In particular, we analyze how the estimation performance depends on the temperature $T$, the energy level spacing $\omega$, and the initial state parameter $\Delta_0$. These quantities play a central role in determining the structure of the quantum Fisher information matrix (QFIM), and consequently govern the attainable lower bounds on the variances of the corresponding estimators.
Starting from Eqs.~(\ref{ineq1}) and (\ref{ineq2}), the minimal values of $\mathrm{Var}(\Delta_0)$ and $\mathrm{Var}(T)$ are given by
\begin{align}
	\mathrm{Var}(T)_{\text{min}}&=
	-\frac{2 T^{4}\, (\Delta_0-1)\,\left[ 1+2\cosh\!\left(\frac{\omega}{T}\right)\right] ^{2}}
	{\omega^{2}\,\left[ 2+\cosh\!\left(\frac{\omega}{T}\right)\right] } \left[ \frac{-1}{(\Delta_0+3)(\Delta_0-1)} \right] = \frac{2 T^{4}\, \,\left[ 1+2\cosh\!\left(\frac{\omega}{T}\right)\right] ^{2}}
	{\omega^{2}\,(\Delta_0+3)\left[ 2+\cosh\!\left(\frac{\omega}{T}\right)\right] }, \, \,\,\,\text{for }  \,\Delta_0 \neq\left( -3,1\right) \\[8pt]
\mathrm{Var}(\Delta_0)_{\text{min}}&=	-\frac{2 T^{4}\, (\Delta_0-1)\,\left[ 1+2\cosh\!\left(\frac{\omega}{T}\right)\right] ^{2}}
{\omega^{2}\,\left[ 2+\cosh\!\left(\frac{\omega}{T}\right)\right] } \left[ \frac{(3+\Delta_0)\,\omega^{2}\left(2+\cosh\!\left(\frac{\omega}{T}\right)\right)}
{2\,T^{4}\left(1+2\cosh\!\left(\frac{\omega}{T}\right)\right)^{2}}\right]  =3-2\Delta_0-\Delta_0^2, \, \,\,\,\text{for }  \, \Delta_0 \neq\left( -3,1\right)\label{varD}
\end{align}
We now consider the estimation of the parameters performed individually. 
In this scenario, we assume that the parameters are statistically independent, 
meaning that the optimal estimation of one parameter does not influence the 
precision achievable for the other. This condition is satisfied when the 
off-diagonal elements of the Fisher information matrix vanish, namely 
$\mathcal{F}_{ij}=0$ for $i \neq j$.
Under this assumption, the Cram\'er-Rao bounds reduce to
\begin{equation}
	\mathrm{Var}(\Delta_0)_{\text{min}}^{\text{Ind}} \geq (\mathcal{F}^{-1})_{\Delta_0\Delta_0}, 
	\qquad
	\mathrm{Var}(T)_{\text{min}}^{\text{Ind}} \geq (\mathcal{F}^{-1})_{TT}.
\end{equation}

The saturation of these inequalities provides the minimal achievable 
variances within the individual estimation strategy. Explicitly, we obtain
\begin{equation}
\mathrm{Var}(\Delta_0)^{\text{Ind}}_{\min}=\mathrm{Var}(\Delta_0)_{\min},\quad \text{and}\quad
	\mathrm{Var}(T)^{\text{Ind}}_{\min}=\mathrm{Var}(T)_{\min}
\end{equation}

These expressions characterize the ultimate precision limits attainable 
when the parameters $\gamma$ and $T$ are estimated independently, under 
the assumption of vanishing parameter correlations.
To enable a quantitative comparison between the two estimation strategies, we introduce a performance metric defined as the ratio of the total variance obtained in the simultaneous estimation scheme to that achieved in the individual estimation scheme. It is given by
\begin{equation}
	\Gamma = \frac{\Gamma_{\text{Sim}}}{\Gamma_{\text{Ind}}},\label{rap}
\end{equation}
with
\begin{equation}
	\Gamma_{\text{Ind}} 
	= \mathrm{Var}(\Delta_0)^{\text{Ind}}_{\min}
	+ \mathrm{Var}(T)^{\text{Ind}}_{\min},
\end{equation}
and
\begin{equation}
	\Gamma_{\text{Sim}} 
	= \frac{1}{2}
	\left(
	\mathrm{Var}(\Delta_0)_{\min}
	+ \mathrm{Var}(T)_{\min}
	\right).
\end{equation}
This ratio serves as an indicator of the relative efficiency of the simultaneous protocol with respect to the individual one.
\begin{figure}[H]
	\includegraphics[scale=0.58]{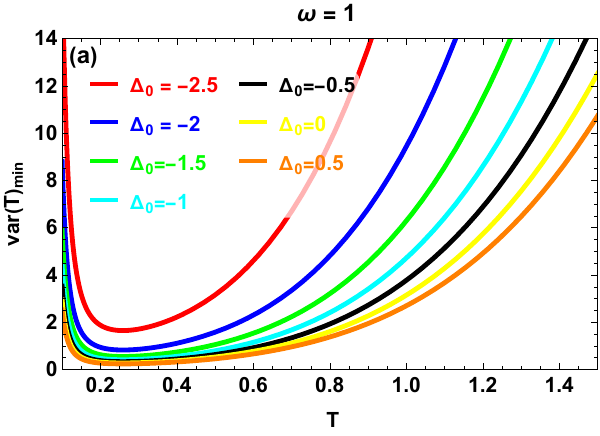}
	\includegraphics[scale=0.58]{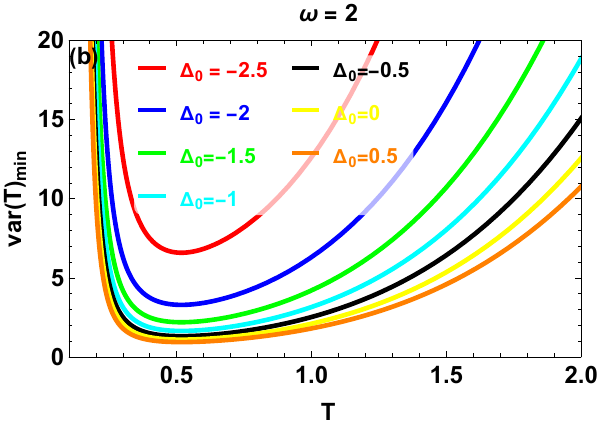}
	\includegraphics[scale=0.58]{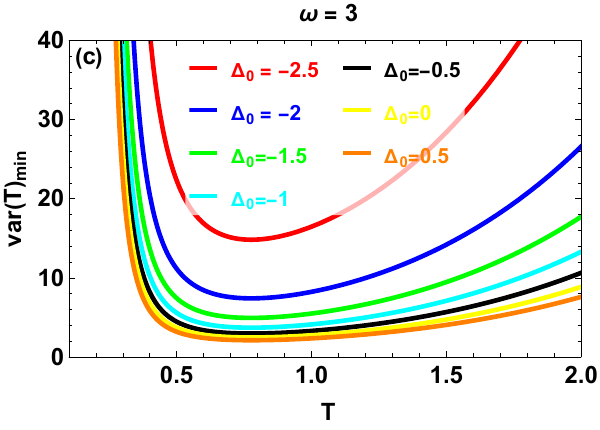}
\caption{
	Minimal variance for the simultaneous estimation of the temperature $T$ as a function of $\Delta_0$, shown for three values of the energy-level spacing: (a) $\omega=1$, (b) $\omega=2$, and (c) $\omega=3$.
}
	
	\label{fig:1}
\end{figure}
Figure~\ref{fig:1}(a)-(c) displays the minimal variance associated with the simultaneous estimation of the temperature $T$ for different values of the initial-state parameter $\Delta_0$ and energy level spacings $\omega$. For all cases, $\mathrm{Var}(T)_{\min}$ exhibits a non-monotonic dependence on $T$, reaching a minimum at an intermediate temperature, which identifies the optimal working regime for thermometric precision.
For fixed $\omega$, increasing $\Delta_0$ systematically reduces the minimal variance, highlighting the crucial role of the initial-state preparation in enhancing sensitivity. Furthermore, as $\omega$ increases, the overall magnitude of the variance becomes larger and the optimal temperature shifts toward higher values, reflecting the influence of the energy gap on the thermal response of the system.
\begin{figure}[H]
	\centering
	\includegraphics[scale=0.8]{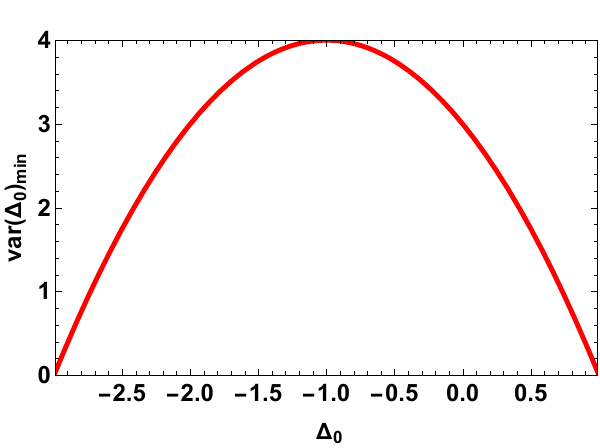}
	
	\caption{Minimal variances of the parameter $\Delta_0$}

	\label{fig:2}
\end{figure}
The behavior of the minimal variance for the estimation of $\Delta_0$ is illustrated in Figure~\ref{fig:2}. A purely quadratic dependence on $\Delta_0$ is observed, as shown in Eq.~(\ref{varD}), and remains entirely independent of the temperature $T$ and the energy level spacing $\omega$, indicating that the estimation precision is solely governed by the structure of the initial state.
The variance approaches its minimum value (approximately $0.04$) near the boundary values $\Delta_0 \to -3$ and $\Delta_0 \to 1$. Although these extremal values are not physically allowed, the strong suppression of the variance in their vicinity reflects an enhanced sensitivity to small variations of $\Delta_0$. In contrast, the maximal variance occurs around $\Delta_0=-1$, corresponding to the least sensitive configuration.

 In conclusion, the minimal variances for $T$ and $\Delta_0$ in the individual estimation scheme coincide with those obtained in the simultaneous scenario, owing to the vanishing off-diagonal elements of the quantum Fisher information matrix. This absence of correlations ensures that no precision loss arises in the multiparameter framework. Furthermore, the ratio defined in Eq.~(\ref{rap}) is equal to $0.5$, highlighting the metrological efficiency and full compatibility of the joint estimation strategy.
\section{Individual and Simultaneous Parameter Estimation in Markovian and Non-Markovian Regimes} \label{sec:5}
In this section, we present the formal framework describing correlated quantum channels acting on a bipartite two-qubit system initially prepared in the density operator $\varrho(0)$. Assuming that both qubits are subjected to identical local actions of a quantum map $\mathcal{E}$, the evolved state at time $t$ can be expressed in the Kraus representation as \cite{Daffer2004,Caruso2014}
\begin{equation}
	\varrho(t)=\mathcal{E}[\varrho(0)]=\sum_{m,n=0}^{3} M_{m,n}\varrho(0)M_{m,n}^{\dagger}.
\end{equation}
The Kraus operators are defined by
\begin{equation}
	M_{m,n}=\sqrt{p_{m,n}}\,\sigma_m\otimes\sigma_n,
\end{equation}
where $\sigma_0=\mathbb{I}_2$ denotes the identity matrix and $\sigma_{x,y,z}$ are the Pauli operators. The coefficients $p_m$ form a normalized probability distribution satisfying $p_m\ge0$ and $\sum_m p_m=1$, while complete positivity and trace preservation impose the condition $\sum_{m,n} M_{m,n}^{\dagger}M_{m,n}=\mathbb{I}_4$. To incorporate classical memory effects between successive channel applications, the joint probabilities are chosen as
\begin{equation}
	p_{m,n}=(1-\mu)p_m p_n+\mu p_m\delta_{m,n},
\end{equation}
where $\delta_{m,n}$ is the Kronecker delta and $\mu\in[0,1]$ quantifies the degree of classical correlation. The limiting cases $\mu=0$ and $\mu=1$ correspond to memoryless and fully correlated channels, respectively. We focus on a pure dephasing process characterized by $p_0=1-q$, $p_x=p_y=0$, and $p_z=q$, whose time dependence is described by the colored noise Hamiltonian \cite{Daffer2004,Caruso2014,HuZhou2019}
\begin{equation}
	H(t)=\hbar\Lambda(t)\sigma_z,
\end{equation}
where $\Lambda(t)=\Omega\xi(t)$ represents a random telegraph signal. The stochastic variable $\xi(t)$ follows a Poisson distribution with mean value $\langle\xi(t)\rangle=t/(2\tau)$, and $\Omega=\pm1$. For $\Omega=1$, the time-dependent dephasing parameter reads
\begin{equation}
	q=\frac{1-F(t)}{2}.
\end{equation}
In the non-Markovian regime, defined by $4\tau>1$, the memory kernel takes the form
\begin{equation}
	F(t)=e^{-\alpha t}\left[\cos(\beta t)+\frac{1}{\beta}\sin(\beta t)\right],
\end{equation}
whereas in the Markovian regime, characterized by $4\tau<1$, and with $\alpha=1/(2\tau)$ and $\beta=\sqrt{|\alpha^2-1|}$, it becomes
\begin{equation}
	F(t)=e^{-\alpha t}\left[\cosh(\beta t)+\frac{1}{\beta}\sinh(\beta t)\right].
\end{equation}
Applying this correlated dephasing map to an initial X-type two-qubit state yields the evolved density operator
\begin{equation}
	\varrho(t)=
	\begin{pmatrix}
		x&0&0&0\\
		0&z&\kappa(t)\,\delta&0\\
		0&\kappa(t)\,\delta&z&0\\
		0&0&0&y
	\end{pmatrix},
\end{equation}
where the effective decoherence factor is given by
\begin{equation}
	\kappa(t)=F^2(t)+\left[1-F^2(t)\right]\mu.
\end{equation}
This expression clearly shows that increasing the classical correlation parameter $\mu$ mitigates the decay of the off-diagonal elements, thereby delaying the loss of quantum coherence. Using the same methodology developed in Section \ref{sec:2}, the simultaneous and individual estimation of the parameters $T$ and $\Delta_0$ in the present scenario can be formulated as follows
\begin{equation}
\mathrm{Var}(T)_{\text{min}} =
	\frac{
		T^{4}\left(1 + 2\cosh\left(\frac{\omega}{T}\right)\right)
		\left[
		1 - \Delta_0
		+ (5+3\Delta_0)\kappa^{2}
		+ 4\left(1+(2+\Delta_0)\kappa^{2}\right)
		\cosh\left(\frac{\omega}{T}\right)
		- (\Delta_0-1)(1-\kappa^{2})
		\cosh\left(\frac{2\omega}{T}\right)
		\right]
	}{
		(3+\Delta_0)\,\omega^{2}
		\left[
		2 + 2(2+\Delta_0)\kappa^{2}
		+ \left(1-\Delta_0+2(1+\Delta_0)\kappa^{2}\right)
		\cosh\left(\frac{\omega}{T}\right)
		\right]
	},
\end{equation}
\begin{align}
	\mathrm{Var}(\Delta_0)_{\text{min}}
	&= \frac{e^{-\frac{2\omega}{T}} (3+\Delta_0)}
	{2\left(1 + 2\cosh\!\left(\frac{\omega}{T}\right)\right)
		\left[
		2 + 2(2+\Delta_0)\kappa^2
		+ \left(1-\Delta_0+2(1+\Delta_0)\kappa^2\right)
		\cosh\!\left(\frac{\omega}{T}\right)
		\right]}
	\nonumber\\[6pt]
	&\quad \times \Big[
	-2 e^{\frac{2\omega}{T}}
	\left(
	-12 + 5\Delta_0 - \Delta_0^2
	+ (3+\Delta_0+4\Delta_0^2)\kappa^2
	\right)
	\nonumber\\
	&\qquad
	-(\Delta_0-1)\left(
	2 + \kappa^2 + \Delta_0(-1+2\kappa^2)
	\right)\left(1+e^{\frac{4\omega}{T}}\right)
	\nonumber\\
	&\qquad
	+\left(e^{\frac{\omega}{T}}+e^{\frac{3\omega}{T}}\right)
	\left(
	13 + 2\kappa^2
	+ \Delta_0(-10+\Delta_0-6\Delta_0\kappa^2)
	\right)
	\Big],
\end{align}
\begin{equation}
\mathrm{Var}(T)_{\text{min}}^{\text{Ind}}=	\frac{
		4 e^{\frac{2\omega}{T}} T^{4}
		\left[1+2\cosh\!\left(\frac{\omega}{T}\right)\right]^{2}
		\left[
		2+\kappa+\Delta_0\kappa
		+(\Delta_0-1)(\kappa-1)\cosh\!\left(\frac{\omega}{T}\right)
		\right]
		\left[
		-2+\kappa+\Delta_0\kappa
		+(\Delta_0-1)(1+\kappa)\cosh\!\left(\frac{\omega}{T}\right)
		\right]
	}{
		A\,\omega^{2}
	}
\end{equation}
where \begin{equation}
	\begin{aligned}
		A = (3+\Delta_0)\Big[
		&2 e^{\frac{2\omega}{T}}
		\left[-12 + 5\Delta_0 - \Delta^{2}
		+ (3+\Delta_0+4\Delta_0^{2})\kappa^{2}\right]  \\
		&+(\Delta_0-1)\left[2+\kappa^{2}
		+\Delta_0(-1+2\kappa^{2})\right]
		\left(1+e^{\frac{4\omega}{T}}\right) \\
		&+\left(e^{\frac{\omega}{T}}+e^{\frac{3\omega}{T}}\right)
		\left[-13-2\kappa^{2}
		+\Delta_0\left(10+\Delta_0(-1+6\kappa^{2})\right)\right]
		\Big],
	\end{aligned}
\end{equation}
\begin{equation}
\mathrm{Var}(\Delta_0)_{\text{min}}^{\text{Ind}}=	-\frac{2 (3 + \Delta_0 )
		\left( 2 + \kappa + \Delta_0 \kappa 
		+ (-1 + \Delta_0 )(-1 + \kappa)\cosh\!\left(\frac{\omega}{T}\right) \right)
		\left( -2 + \kappa + \Delta_0  \kappa 
		+ (-1 + \Delta_0 )(1 + \kappa)\cosh\!\left(\frac{\omega}{T}\right) \right)}
	{1 - \Delta_0  
		+ (5 + 3\Delta_0 )\kappa^2
		+ 4\left(1 + (2 + \Delta_0 )\kappa^2\right)\cosh\!\left(\frac{\omega}{T}\right)
		+ (-1 + \Delta_0 )(-1 + \kappa^2)\cosh\!\left(\frac{2\omega}{T}\right)}.
\end{equation}
The explicit expression of the ratio $\Gamma$, defined in Eq.~(\ref{rap}), is written as $\Gamma=\frac{C}{D}$, with
\begin{align}
	C &= 
	\left[
	1 - \Delta + (5 + 3\Delta)\kappa^2
	+ 4\left(1 + (2+\Delta)\kappa^2\right)
	\cosh\!\left(\frac{\omega}{T}\right)
	+ (-1+\Delta)(-1+\kappa^2)
	\cosh\!\left(\frac{2\omega}{T}\right)
	\right] \notag\\
	&\quad \times
	\left[
	-12 + 5\Delta - \Delta^2
	+ (3+\Delta+4\Delta^2)\kappa^2
	+ \left(-13 -2\kappa^2
	+ \Delta\left(10+\Delta(-1+6\kappa^2)\right)\right)
	\cosh\!\left(\frac{\omega}{T}\right)
	\right. \notag\\
	&\qquad\left.
	+ (-1+\Delta)(2-\Delta+\kappa^2+2\Delta\kappa^2)
	\cosh\!\left(\frac{2\omega}{T}\right)
	\right],
\end{align}
and \begin{align}
	D &=
	\left[
	4\left(1 + 2\cosh\!\left(\frac{\omega}{T}\right)\right)
	\left(2+\kappa+\Delta\kappa
	+ (-1+\Delta)(-1+\kappa)
	\cosh\!\left(\frac{\omega}{T}\right)\right)
	\right. \notag\\
	&\qquad \times
	\left(-2+\kappa+\Delta\kappa
	+ (-1+\Delta)(1+\kappa)
	\cosh\!\left(\frac{\omega}{T}\right)\right)
	\notag\\
	&\qquad \left.
	\times
	\left(2+2(2+\Delta)\kappa^2
	+ (1-\Delta+2(1+\Delta)\kappa^2)
	\cosh\!\left(\frac{\omega}{T}\right)\right)
	\right].
\end{align}


Figure~\ref{fig:6} illustrates the time evolution of the minimal variance $\mathrm{Var}(T)_{\min}$ for the simultaneous estimation of the temperature $T$ in the Markovian regime. As observed in both panels, the precision strongly depends on the interplay between temperature and evolution time. In particular, for short interaction times, the variance remains relatively small within a limited temperature window, indicating enhanced sensitivity in the early dynamical stage. As time increases, $\mathrm{Var}(T)_{\min}$ generally grows, reflecting the progressive loss of information due to Markovian dissipation.
A comparison between panels \ref{fig:6}(a) and \ref{fig:6}(b) reveals the significant impact of the energy level spacing. For $\omega=0.5$, the variance remains moderately bounded over time, whereas for $\omega=0.9$ it increases dramatically, especially at longer times. This behavior shows that larger energy gaps amplify the detrimental effect of Markovian decoherence on thermometric precision. Consequently, optimal temperature estimation in the Markovian regime is achieved at relatively short times and smaller values of $\omega$, where information loss is less pronounced.
\begin{figure}[H]
	\includegraphics[scale=0.58]{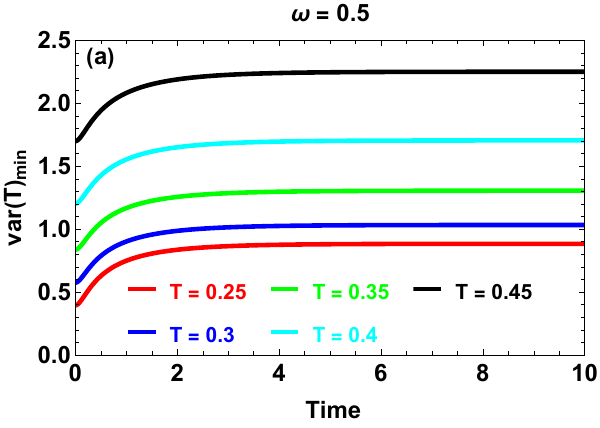}
	\includegraphics[scale=0.58]{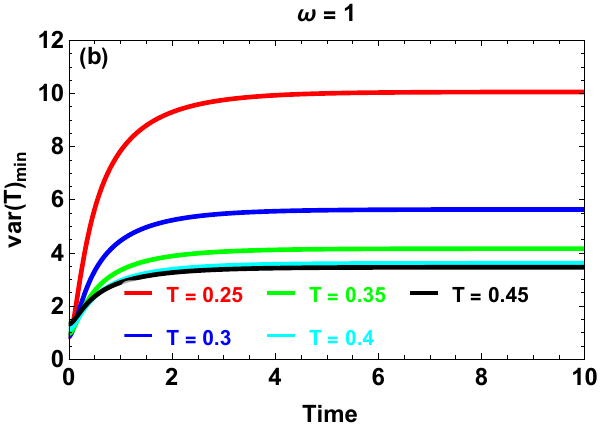}
	\includegraphics[scale=0.58]{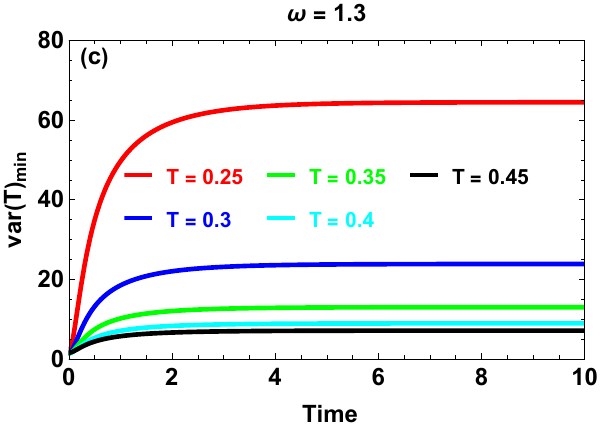}
\caption{Dynamical evolution of the minimal variance $\mathrm{Var}(T)_{\min}$ in the Markovian regime, for different values of the temperature $T$ and for three values of the parameter $\omega$: (a) $\omega = 0.5$, (b) $\omega = 1$, and (c) $\omega = 1.3$. The remaining parameters are fixed at $\tau = 0.1$, $\mu=0.6$, and $\Delta_0=-2$.}
	
	\label{fig:4}
\end{figure}
As depicted in Figure~\ref{fig:4}, the minimal variance $\mathrm{Var}(T)_{\min}$ increases monotonically with time in the Markovian regime for all considered temperatures and energy level spacings. After a rapid initial growth, the variance approaches a steady-state value, reflecting the irreversible loss of information due to Markovian dissipation. This behavior indicates that optimal thermometric precision is achieved at short interaction times.
For a fixed value of $\omega$, higher temperatures lead to larger asymptotic variances, implying reduced estimation accuracy in hotter regimes. Moreover, as shown in panels \ref{fig:4}(a)-(c), increasing the energy gap from $\omega=0.5$ to $\omega=1.3$ significantly amplifies the variance, demonstrating that larger spectral separations enhance the detrimental effect of Markovian dynamics on temperature estimation.
\begin{figure}[H]
	\includegraphics[scale=0.58]{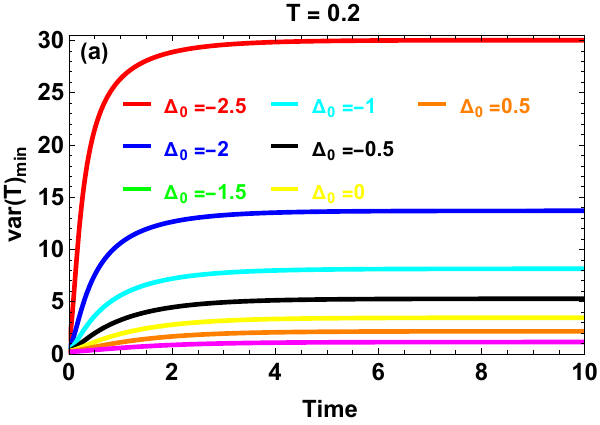}
	\includegraphics[scale=0.58]{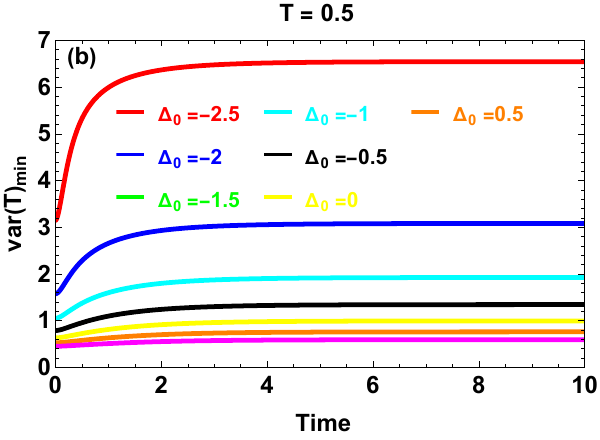}
	\includegraphics[scale=0.58]{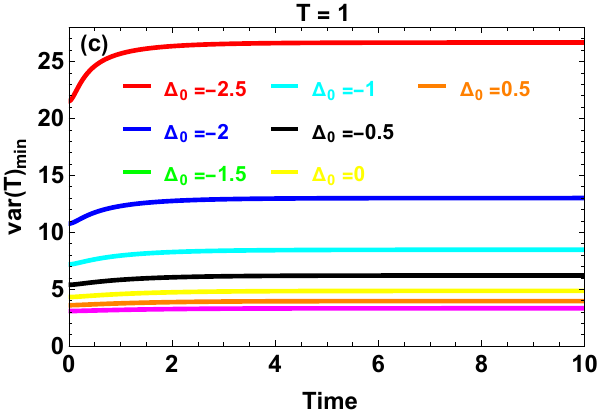}
\caption{Dynamical evolution of the minimal variance $\mathrm{Var}(T)_{\min}$ in the Markovian regime, for different values of the parameter $\Delta_0$ and for three values of the temperature $T$: (a) $T=0.2$, (b) $T=0.5$, and (c) $T=1$. The remaining parameters are fixed at $\tau = 0.1$, $\mu=0.6$, and $\omega=1$.}
	
	\label{fig:5}
\end{figure}
The results presented in Figure~\ref{fig:5} highlight the dynamical behavior of the minimal variance $\mathrm{Var}(\Delta_0)_{\min}$ in the Markovian regime for different initial-state parameters and temperatures. For all temperatures, the variance exhibits a rapid transient increase at short times followed by a clear saturation toward a stationary value. This pattern reflects the progressive stabilization of the information content as the system approaches its steady state under Markovian dissipation.
A strong dependence on the initial-state parameter $\Delta_0$ is observed: values of $\Delta_0$ farther from zero lead to significantly larger asymptotic variances, indicating reduced precision in those configurations. In contrast, states prepared closer to $\Delta_0=0$ maintain comparatively smaller variances over time, revealing more favorable estimation conditions.

Comparing panels \ref{fig:5}(a)-(c) shows that temperature modifies the overall scale of the variance but does not qualitatively alter the dynamical structure. While increasing $T$ changes the magnitude of the stationary values, the monotonic approach to equilibrium remains preserved, confirming that the dominant effect stems from the Markovian dissipative mechanism rather than thermal fluctuations alone.


The dynamical behavior displayed in Figure~\ref{fig:6} reveals a strikingly different pattern in the non-Markovian regime. In contrast to the monotonic behavior observed under Markovian dynamics, the minimal variance $\mathrm{Var}(T)_{\min}$ exhibits pronounced oscillations over time. These oscillatory features clearly signal the presence of memory effects, where information temporarily flows back from the environment to the system, partially restoring the precision of temperature estimation.
For $\omega=0.5$, the variance remains bounded and displays regular revivals, indicating that non-Markovian backflow mechanisms help preserve thermometric sensitivity over extended time intervals. When the energy gap increases to $\omega=0.9$, the overall scale of the variance becomes significantly larger; however, the oscillatory structure persists, confirming that the non-Markovian character of the dynamics continues to modulate the estimation precision.

These results demonstrate that memory effects can counteract dissipative degradation and generate temporal windows where the estimation precision is temporarily enhanced, making the timing of the measurement a crucial factor in optimizing thermometric protocols.
\begin{figure}[H]
	\includegraphics[scale=0.58]{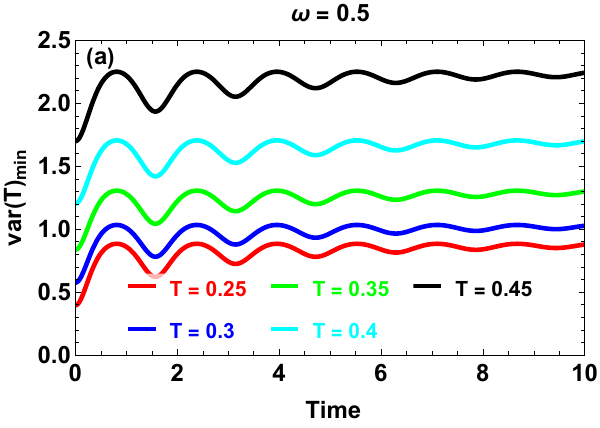}
	\includegraphics[scale=0.58]{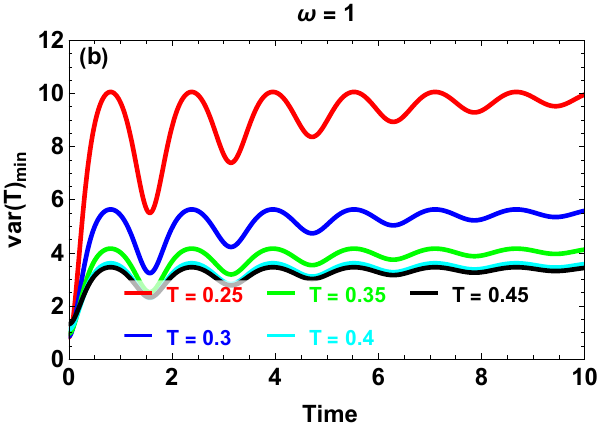}
	\includegraphics[scale=0.58]{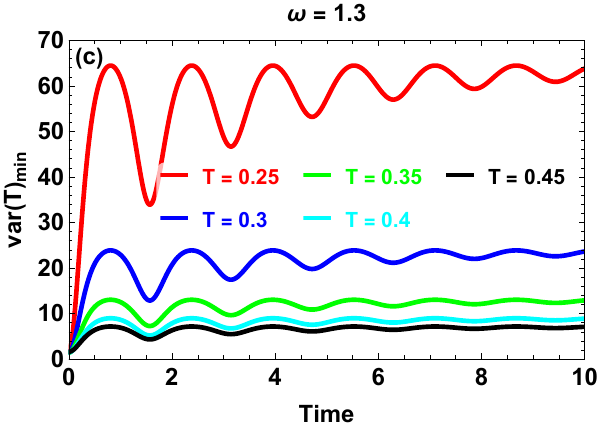}
	\caption{Dynamical evolution of the minimal variance $\mathrm{Var}(T)_{\min}$ in the non-Markovian regime, for different values of the temperature $T$ and for three values of the parameter $\omega$: (a) $\omega = 0.5$, (b) $\omega = 1$, and (c) $\omega = 1.3$. The remaining parameters are fixed at $\tau = 5$, $\mu=0.6$, and $\Delta_0=-2$.}
	
	\label{fig:7}
\end{figure}
Figure~\ref{fig:7} illustrates the time evolution of the minimal variance $\mathrm{Var}(T)_{\min}$ in the non-Markovian regime for different temperatures and energy level spacings. A clear oscillatory behavior is observed in all panels, revealing persistent memory effects that continuously modulate the estimation precision over time. Unlike the Markovian case, the variance does not converge monotonically toward a stationary value but instead exhibits damped revivals, indicating repeated exchanges of information between the system and its environment.

For each value of $\omega$, the curves remain ordered with respect to temperature: lower temperatures systematically yield smaller variances, while higher temperatures produce larger oscillation amplitudes and higher average values. Increasing the energy gap from $\omega=0.5$ to $\omega=1.3$ significantly amplifies both the magnitude of the variance and the amplitude of the oscillations, demonstrating that stronger spectral separation enhances the sensitivity of the dynamics to non-Markovian memory effects.

These results emphasize that, in the non-Markovian regime, the estimation precision is not only temperature-dependent but also strongly time-dependent, with specific temporal windows where the variance is locally reduced due to information backflow.
\begin{figure}[H]
	\includegraphics[scale=0.58]{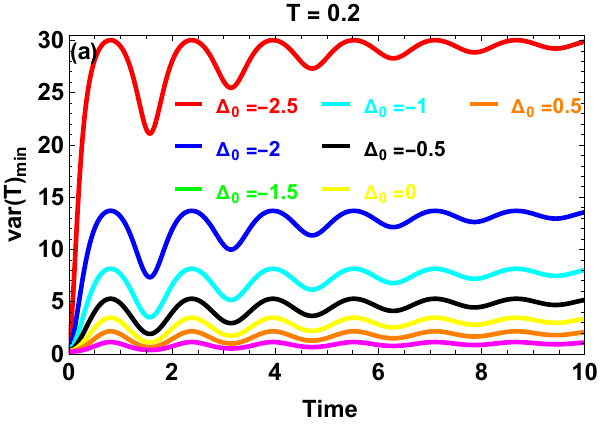}
	\includegraphics[scale=0.58]{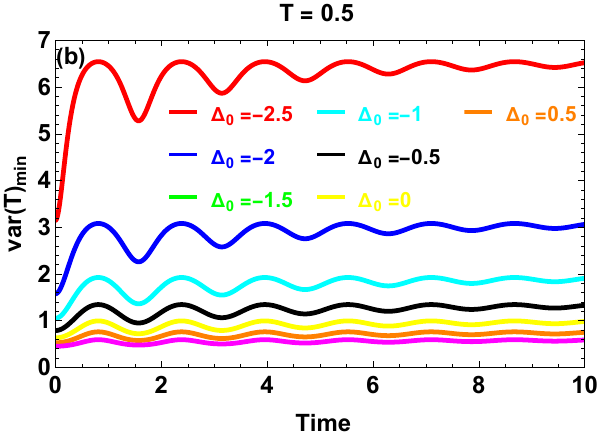}
	\includegraphics[scale=0.58]{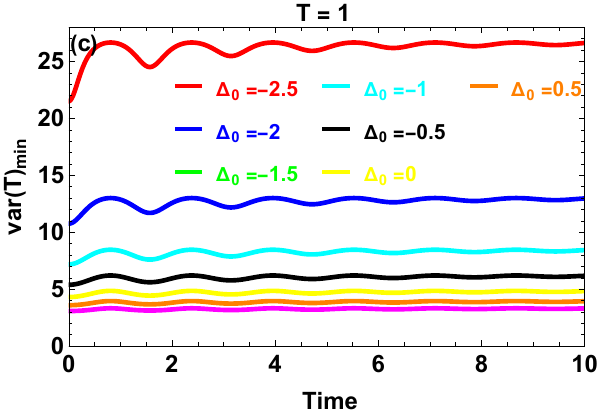}
	\caption{Dynamical evolution of the minimal variance $\mathrm{Var}(T)_{\min}$ in the non-Markovian regime, for different values of the parameter $\Delta_0$ and for three values of the temperature $T$: (a) $T=0.2$, (b) $T=0.5$, and (c) $T=1$. The remaining parameters are fixed at $\tau = 5$, $\mu=0.6$, and $\omega=1$.}
	
	\label{fig:8}
\end{figure}
The dynamics reported in Figure~\ref{fig:8} reveal that the minimal variance $\mathrm{Var}(T)_{\min}$ in the non-Markovian regime strongly depends on the initial-state parameter $\Delta_0$. For all considered temperatures, the variance exhibits damped oscillations over time, reflecting persistent memory effects and recurrent information backflow from the environment to the system. These oscillatory patterns indicate that the estimation precision is periodically modulated rather than irreversibly degraded.
A clear hierarchy with respect to $\Delta_0$ is observed: initial states with more negative values of $\Delta_0$ lead to significantly larger variances, whereas states prepared closer to $\Delta_0=0$ yield lower average values and reduced oscillation amplitudes. This demonstrates that the initial-state structure not only determines the baseline precision but also controls the sensitivity of the dynamics to non-Markovian memory effects.

Comparing panels \ref{fig:8}(a)-(c) further shows that increasing the temperature mainly rescales the magnitude of the variance while preserving the oscillatory character of the evolution. Hence, in the non-Markovian regime, both the preparation of the initial state and the thermal energy jointly shape the temporal windows where enhanced precision can be achieved.


The results displayed in Figure~\ref{fig:9} describe the joint dependence of the minimal variance $\mathrm{Var}(\Delta_0)_{\min}$ on the initial-state parameter $\Delta_0$ and the evolution time in the Markovian regime. A smooth parabolic profile is clearly observed along the $\Delta_0$ direction for all times, confirming that the estimation precision is primarily governed by the intrinsic quadratic structure of the parameter space. The variance attains its maximum around the central region of $\Delta_0$, while it decreases toward the boundaries of the allowed domain.

As time progresses, the surface gradually elevates and approaches a stationary configuration, reflecting the cumulative effect of Markovian dissipation on the information content. The comparison between panels \ref{fig:9}(a) and \ref{fig:9}(b) shows that increasing the temperature mainly rescales the overall magnitude of the variance without altering its qualitative parabolic structure. Hence, while thermal effects influence the scale of the precision, the dominant feature of the estimation landscape remains dictated by the geometric dependence on $\Delta_0$.
\begin{figure}[H]
	\includegraphics[scale=0.58]{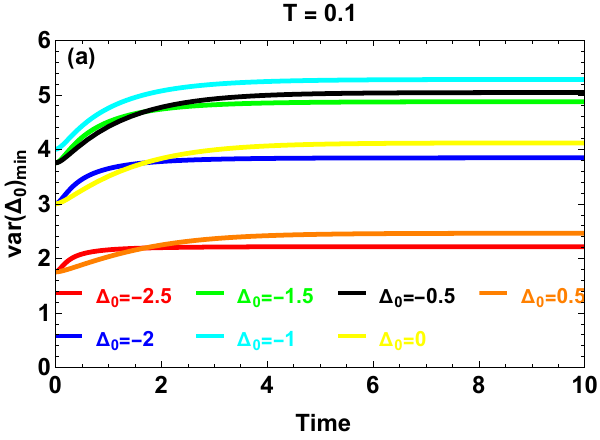}
	\includegraphics[scale=0.58]{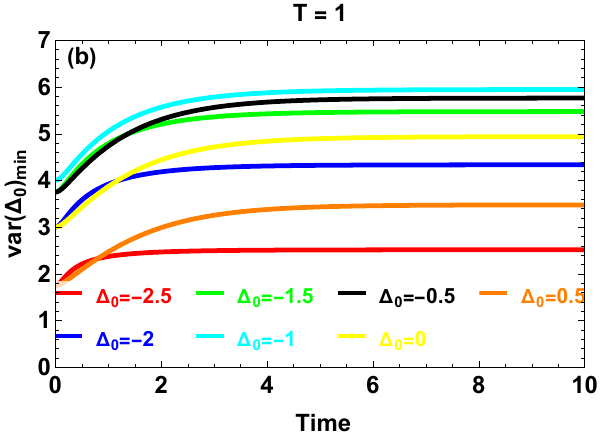}
	\includegraphics[scale=0.58]{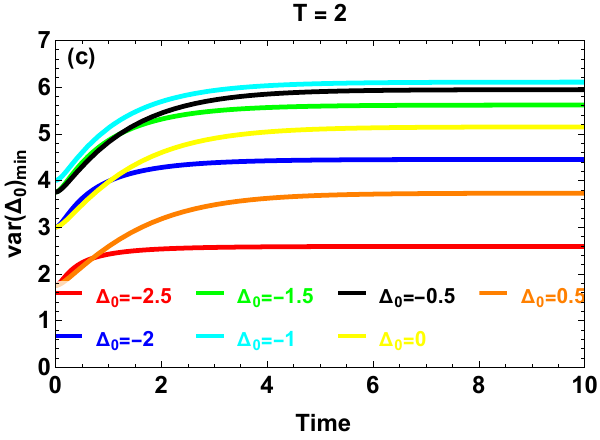}
	\caption{Dynamical evolution of the minimal variance $\mathrm{Var}(\Delta_0)_{\min}$ in the Markovian regime, for different values of  $\Delta_0$ and for three values of the parameter $T$: (a) $T=0.1$, (b) $T=1$, and (c) $T=2$. The remaining parameters are fixed at $\tau = 0.1$, $\mu=0.6$, and $\omega=1$.}
	
	\label{fig:10}
\end{figure}
The temporal profiles presented in Figure~\ref{fig:10} show that, in the Markovian regime, the minimal variance $\mathrm{Var}(\Delta_0)_{\min}$ increases smoothly with time before reaching a well-defined steady-state value for all temperatures. The absence of oscillatory behavior confirms that memoryless dissipation drives the system toward a stationary configuration where the available information becomes time-independent.
For each temperature, the curves remain clearly ordered with respect to $\Delta_0$, indicating that the initial-state parameter determines the hierarchy of achievable precision throughout the evolution. States characterized by more negative values of $\Delta_0$ systematically yield larger asymptotic variances, while configurations closer to $\Delta_0=0$ lead to lower stationary values and therefore improved estimation performance.

A comparison between panels \ref{fig:10}(a)-(c) reveals that increasing the temperature progressively elevates the stationary variance without modifying the qualitative dynamical structure. This demonstrates that, under Markovian dynamics, temperature primarily affects the overall precision scale, whereas the temporal behavior itself remains governed by irreversible dissipative relaxation.


A markedly different dynamical structure emerges in the non-Markovian regime, as depicted in Figure~\ref{fig:11}. The minimal variance $\mathrm{Var}(\Delta_0)_{\min}$ exhibits clear oscillatory modulations in time, superimposed on the underlying parabolic dependence with respect to $\Delta_0$. These temporal oscillations reflect memory effects that periodically enhance and reduce the attainable estimation precision.

Instead of approaching a stationary surface monotonically, the variance displays successive revivals, signaling recurrent information backflow from the environment to the system. Although the quadratic profile in $\Delta_0$ remains preserved throughout the evolution, increasing the temperature mainly amplifies both the overall magnitude of the variance and the amplitude of the oscillations. This confirms that, in the non-Markovian regime, the estimation precision is shaped by the combined influence of parameter geometry and environmental memory effects.
\begin{figure}[H]
	\includegraphics[scale=0.58]{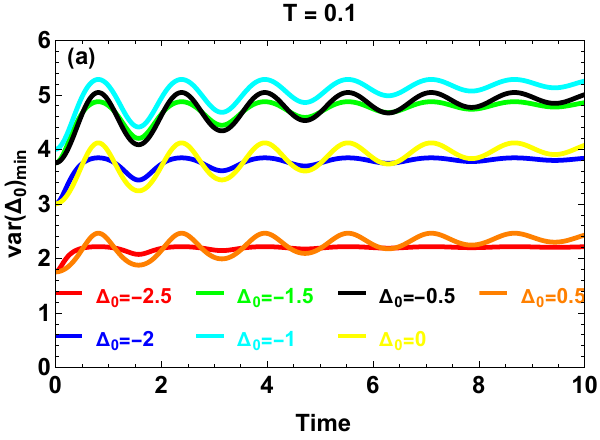}
	\includegraphics[scale=0.58]{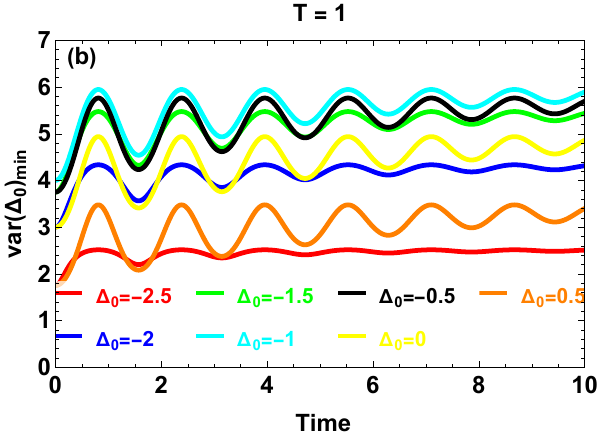}
	\includegraphics[scale=0.58]{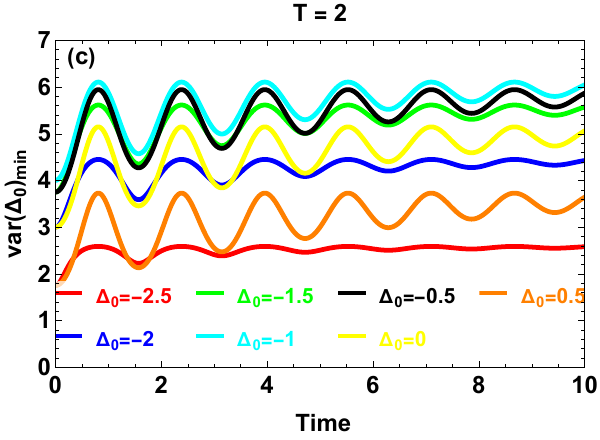}
	\caption{Dynamical evolution of the minimal variance $\mathrm{Var}(\Delta_0)_{\min}$ in the non-Markovian regime, for different values of  $\Delta_0$ and for three values of the parameter $T$: (a) $T=0.1$, (b) $T=1$, and (c) $T=2$. The remaining parameters are fixed at $\tau = 5$, $\mu=0.6$, and $\omega=1$.}
	
	\label{fig:12}
\end{figure}
Non-Markovian dynamics imprint a clearly oscillatory signature on the estimation of $\Delta_0$, as evidenced in Figure~\ref{fig:12}. The minimal variance $\mathrm{Var}(\Delta_0)_{\min}$ undergoes persistent damped oscillations over time for all considered temperatures, indicating recurrent information exchange between the system and its environment. These oscillations prevent the variance from settling monotonically into a stationary value, in contrast to the Markovian case.
For each temperature, the curves preserve a distinct ordering with respect to $\Delta_0$, showing that the initial-state preparation determines the overall hierarchy of achievable precision throughout the evolution. More negative values of $\Delta_0$ systematically lead to higher average variances, whereas states closer to $\Delta_0=0$ exhibit improved robustness against environmental disturbances.

Increasing the temperature from $T=0.1$ to $T=2$ mainly enhances the amplitude of the oscillations and slightly raises the mean variance, while leaving the qualitative temporal structure unchanged. This demonstrates that memory effects dominate the dynamical behavior, whereas thermal contributions primarily rescale the precision landscape.
\begin{figure}[H]
	\centering
	\includegraphics[scale=0.73]{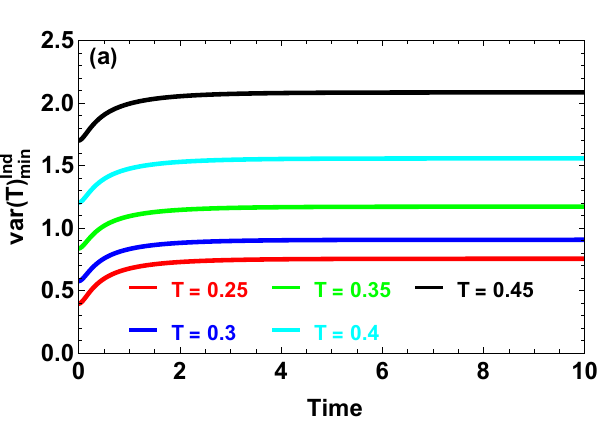}
	\hspace*{0.7cm}
	\includegraphics[scale=0.73]{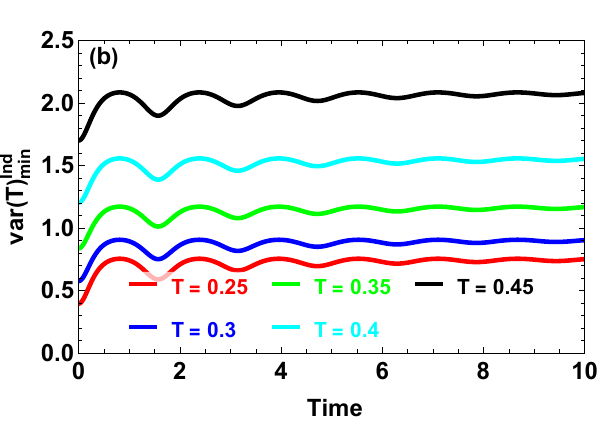}

\caption{The minimal variance associated with the individual estimation of temperature, $\mathrm{Var}(T)_{\min}^{\text{Ind}}$, as a function of time $t$ for different temperatures. Panel (a) corresponds to the Markovian regime with $\tau = 0.1$, while panel (b) represents the non-Markovian regime with $\tau = 5$. The remaining parameters are fixed at $\mu = 0.6$, $\Delta_0 = -2$, and $\omega = 0.5$.}

	\label{fig:13}
\end{figure}
The behavior of the minimal variance $\mathrm{Var}(T)_{\min}^{\mathrm{Ind}}$ reveals distinct dynamical signatures in the Markovian and non-Markovian regimes, as illustrated in Figure~\ref{fig:13}. In the Markovian case (panel \ref{fig:13}(a)), the variance increases smoothly with time and rapidly approaches a stationary value for all considered temperatures. This monotonic convergence reflects irreversible information loss induced by memoryless dissipation, leading to a time-independent precision bound at long times.
By contrast, the non-Markovian regime (panel \ref{fig:13}(b)) exhibits clear damped oscillations in the variance, indicating recurrent information exchange between the system and its environment. These temporal revivals periodically enhance and reduce the attainable precision, preventing a purely monotonic relaxation toward equilibrium.

For both regimes, the curves remain ordered with respect to temperature: higher temperatures systematically correspond to larger variances, implying reduced thermometric precision. The comparison between panels therefore highlights that, while Markovian dynamics impose a steady degradation of precision, non-Markovian memory effects can temporarily mitigate this loss and create optimal time windows for temperature estimation.

\begin{figure}[H]
	\centering
	\includegraphics[scale=0.73]{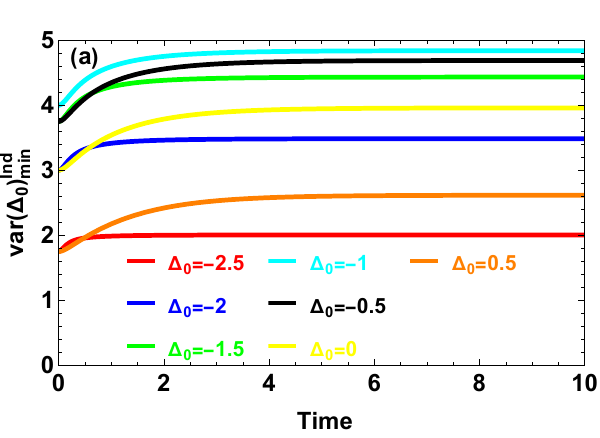}
		\hspace*{0.7cm}
	\includegraphics[scale=0.73]{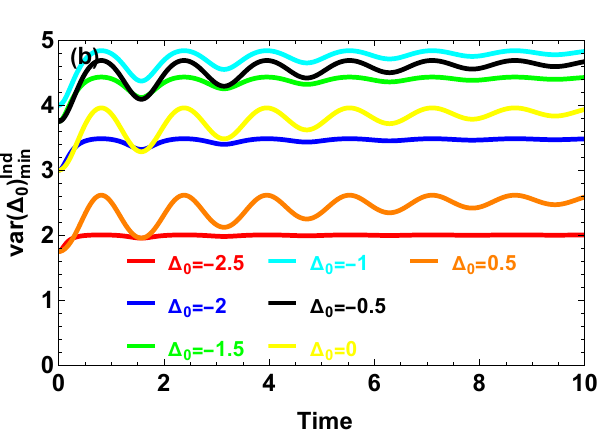}
	
\caption{The minimal variance for the individual estimation of the initial state, $\mathrm{Var}(\Delta_0)_{\min}^{\mathrm{Ind}}$, as a function of time $t$ for different values of $\Delta_0$. Panel (a) corresponds to the Markovian regime ($\tau = 0.1$), while panel (b) represents the non-Markovian regime ($\tau = 5$). The remaining parameters are fixed at $\mu = 0.6$, $T=0.1$, and $\omega = 0.2$.}
	\label{fig:14}
\end{figure}
A comparison between the two dynamical regimes highlights distinct behaviors in the individual estimation of the initial-state parameter $\Delta_0$, as shown in Figure~\ref{fig:14}. In the Markovian case (panel \ref{fig:14}(a)), the minimal variance $\mathrm{Var}(\Delta_0)_{\min}^{\mathrm{Ind}}$ increases smoothly and approaches a steady value for all initial configurations. The absence of oscillations reflects the irreversible character of memoryless dissipation, which drives the system toward a stationary precision bound.
In contrast, the non-Markovian regime (panel \ref{fig:14}(b)) displays damped oscillatory patterns over time, revealing the influence of environmental memory effects. These temporal modulations indicate that the estimation precision is periodically enhanced and reduced due to recurrent information backflow.

For both regimes, a clear ordering with respect to $\Delta_0$ is preserved throughout the evolution. States prepared with more negative values of $\Delta_0$ exhibit lower variances and thus better estimation performance, whereas configurations closer to zero lead to higher precision bounds. This demonstrates that, even in the individual estimation scenario, the achievable accuracy is strongly conditioned by the initial-state structure and its interaction with the surrounding environment.
\begin{figure}[H]
	\includegraphics[scale=0.58]{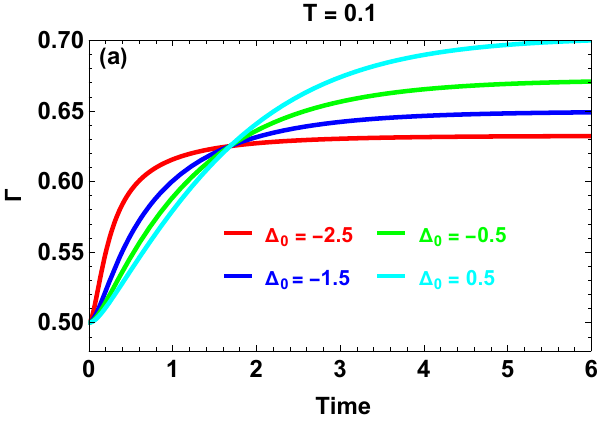}
	\includegraphics[scale=0.58]{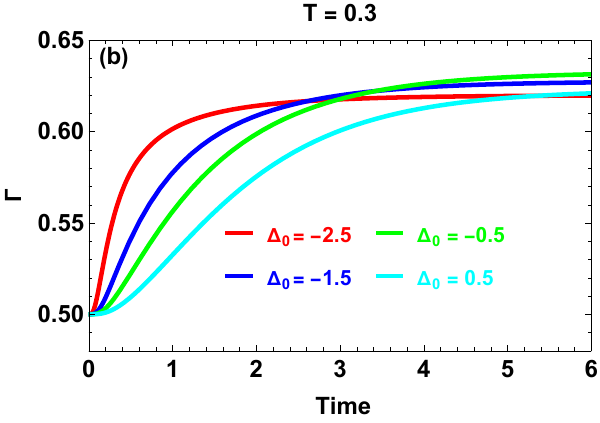}
	\includegraphics[scale=0.58]{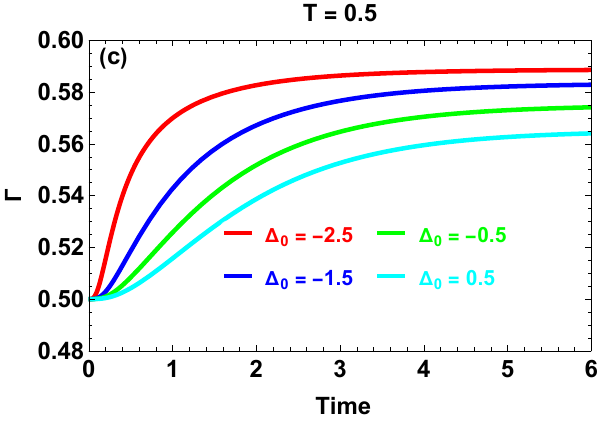}
\caption{The ratio between the minimal total variances in the estimation of the parameters $T$ and $\Delta_0$ in the Markovian regime as a function of time, for different values of $\Delta_0$, and for three values of the temperature: (a) $T = 0.1$, (b) $T = 0.3$, and (c) $T = 0.5$. The other parameters are fixed as $\mu = 0.6$, $\tau = 0.1$, and $\omega = 1$.}
	
	\label{fig:15}
\end{figure}
The temporal behavior of the performance ratio $\Gamma$, defined in Eq.~(\ref{rap}), provides direct insight into the relative efficiency of the simultaneous estimation strategy in the Markovian regime, as illustrated in Figure~\ref{fig:15}. For all considered temperatures and initial-state parameters, the ratio starts from a value close to $0.5$ at short times and gradually increases before reaching a steady-state plateau.
Since $\Gamma < 1$ throughout the evolution, the simultaneous estimation protocol consistently outperforms the individual one in terms of total variance. The initial value $\Gamma \approx 0.5$ reflects the absence of information trade-off between the parameters, confirming their statistical independence. As time progresses, the increase of $\Gamma$ indicates that Markovian dissipation progressively reduces the relative advantage of the joint strategy, although it never eliminates it.

The influence of $\Delta_0$ appears mainly in the rate at which the stationary value is approached, while increasing the temperature slightly modifies the asymptotic magnitude of $\Gamma$ without altering its qualitative monotonic behavior. Overall, the results demonstrate that, even under memoryless dynamics, the simultaneous scheme maintains a clear metrological advantage over the individual estimation protocol.

\begin{figure}[H]
	\includegraphics[scale=0.58]{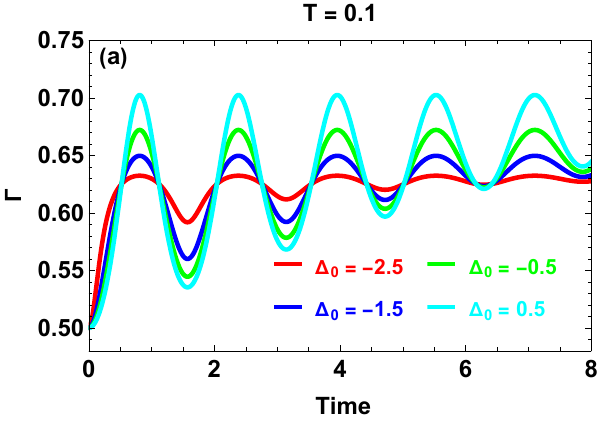}
	\includegraphics[scale=0.58]{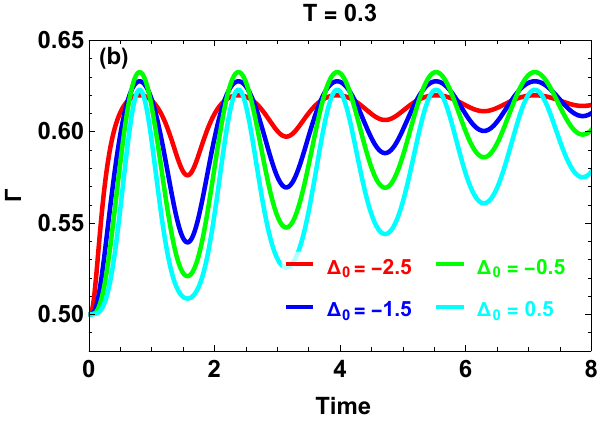}
	\includegraphics[scale=0.58]{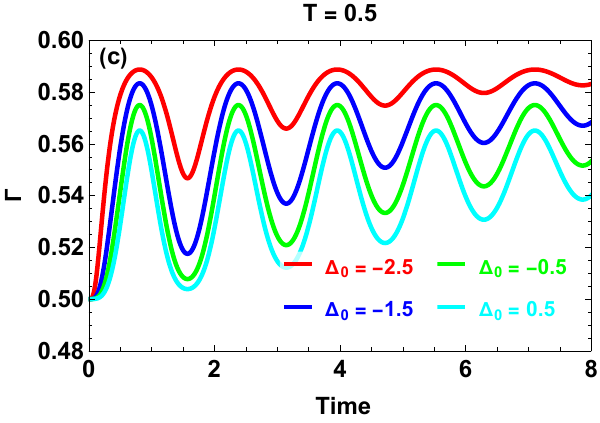}
\caption{The ratio between the minimal total variances in the estimation of the parameters $T$ and $\Delta_0$ in the non-Markovian regime as a function of time, for different values of $\Delta_0$, and for three values of the temperature: (a) $T = 0.1$, (b) $T = 0.3$, and (c) $T = 0.5$. The other parameters are fixed as $\mu = 0.6$, $\tau = 5$, and $\omega = 1$.}

	\label{fig:16}
\end{figure}
The evolution of the performance ratio $\Gamma$ in the non-Markovian regime exhibits a qualitatively distinct behavior, as shown in Figure~\ref{fig:16}. Instead of a monotonic increase toward a stationary value, the ratio displays pronounced oscillations over time, reflecting the impact of environmental memory effects on the relative efficiency of the simultaneous estimation strategy. These oscillations indicate that the metrological advantage of the joint protocol is dynamically modulated by information backflow processes.
Although $\Gamma$ remains below unity throughout the evolution—confirming the persistent superiority of the simultaneous scheme—the amplitude of its oscillations depends on both the initial-state parameter $\Delta_0$ and the temperature. For lower temperatures, the oscillatory features are more pronounced, whereas increasing $T$ tends to reduce their amplitude and slightly shift the average value of $\Gamma$.

Overall, the results demonstrate that, in the presence of memory effects, the efficiency of the simultaneous protocol is not only preserved but also periodically enhanced within specific temporal windows, emphasizing the crucial role of measurement timing in optimizing multiparameter estimation.

In summary, the analysis of parameter estimation in the Markovian and non-Markovian regimes reveals two fundamentally distinct dynamical behaviors. Under Markovian dynamics, the minimal variances exhibit a monotonic evolution toward stationary values, reflecting the irreversible loss of information induced by memoryless dissipation. In this regime, the attainable precision is primarily determined by the competition between decoherence and the intrinsic sensitivity of the quantum probe, leading to well-defined asymptotic bounds.

In contrast, the non-Markovian regime is characterized by pronounced oscillatory patterns in the variances, signaling the presence of memory effects and information backflow from the environment to the system. These temporal revivals generate specific time windows in which the estimation precision can be temporarily enhanced, highlighting the crucial role of measurement timing.

Overall, while Markovian dynamics impose a steady degradation of precision, non-Markovian effects can partially mitigate this loss and dynamically modulate the metrological performance. The comparison demonstrates that environmental memory constitutes a valuable resource for optimizing multiparameter quantum estimation protocols.
\section{Individual and Simultaneous Parameter Estimation in Correlated Quantum Channels}\label{sec:6}
The interaction of quantum systems with their surrounding environment inevitably induces decoherence, leading to the progressive degradation of quantum coherence and the suppression of fragile quantum correlations. This irreversible process transforms initially pure states into mixed states and significantly affects nonclassical resources such as entanglement.

To quantitatively describe environmental effects, several standard quantum noise models are commonly employed, including the amplitude damping (AD), phase flip (PF), and phase damping (PD) channels~\cite{NielsenChuang2010}. These decohering channels effectively simulate different types of system–environment interactions responsible for dissipation and dephasing mechanisms.

For an initial bipartite quantum state $\varrho_{12}$, the evolved state under the action of a generic decohering channel can be rigorously formulated within the Kraus operator representation as
\begin{equation}
	\hat{\varepsilon}\!\left(\varrho^{X}_{Y\bar{Y}}\right)
	=
	\sum_{k,l}
	\mathcal{K}_{kl}\,
	\varrho_{12}\,
	\mathcal{K}_{kl}^{\dagger},
	\label{kraus}
\end{equation}
where the composite Kraus operators are defined as
\begin{equation}
	\mathcal{K}_{kl} =\mathcal{K}_k \otimes \mathcal{K}_l,
\end{equation}
with $\mathcal{K}_k$ and $\mathcal{K}_l$ denoting the single-qubit Kraus operators acting locally on each subsystem.
For the map $\hat{\varepsilon}$ to represent a physically admissible quantum operation, the Kraus operators must satisfy the completeness (trace-preserving) condition
\begin{equation}
	\sum_{k,l}
	\mathcal{K}_{kl}^{\dagger}
	\mathcal{K}_{kl}
	=
	\mathbb{I},
	\label{closure_condition}
\end{equation}
which guarantees the preservation of the trace of the density operator and ensures the complete positivity of the quantum channel.
\subsection{Amplitude Damping Channel}
The amplitude damping (AD) channel constitutes a paradigmatic model for energy dissipation in open quantum systems, particularly describing spontaneous emission processes. It captures the irreversible decay of an excited state into the ground state due to the interaction with the surrounding environment.

For a single qubit, the AD channel is described within the Kraus representation through the operators
\begin{equation}
	\mathcal{K}_1 =
	\begin{pmatrix}
		1 & 0 \\
		0 & \sqrt{1-s}
	\end{pmatrix},
	\qquad
	\mathcal{K}_2 =
	\begin{pmatrix}
		0 & \sqrt{s} \\
		0 & 0
	\end{pmatrix},
	\label{AD_kraus}
\end{equation}
where the decoherence parameter is defined as
\begin{equation}
	s = 1 - e^{-v t}, \qquad 0 \le s \le 1,
\end{equation}
with $v$ denoting the decay rate and $t$ the interaction time. The parameter $s$ therefore quantifies the strength of environmental dissipation.
By substituting Eqs.~(\ref{densit M}) and (\ref{AD_kraus}) into Eq.~(\ref{kraus}), the evolved bipartite density matrix under the influence of the AD channel reads
\begin{equation}
	{\varrho}^{AD}_{12} =
	\begin{pmatrix}
		\lambda_{11} & 0 & 0 & 0\\
		0 & \lambda_{22} &\lambda_{23} & 0 \\
		0 & \lambda_{23} & \lambda_{22} & 0 \\
		0 & 0 & 0 & \lambda_{44}
	\end{pmatrix}.
	\label{rho_AD}
\end{equation}

The nonvanishing matrix elements are given by
\begin{equation}
	\begin{aligned}
		\lambda_{11} &= x + s \left( 2z + s\, \delta \right), \\
		\lambda_{22} &=(1 - s)\left( z + s \,\delta \right), \\
		\lambda_{23} &= (1 - s)\,\delta, \\
		\lambda_{44} &= (1 - s)^2 y.
	\end{aligned}
\end{equation}

Extending the analysis of Section~\ref{sec:2}, we now derive the closed-form expressions of 
$\mathrm{Var}(T)_{\min}$, $\mathrm{Var}(\Delta_0)_{\min}$, 
$\mathrm{Var}(T)_{\min}^{\mathrm{Ind}}$, and 
$\mathrm{Var}(\Delta_0)_{\min}^{\mathrm{Ind}}$, 
associated with the joint and individual estimation of the parameters $T$ and $\Delta_0$. 
They take the form
\begin{equation}
	\mathrm{Var}(T)_{\min}=\dfrac{E}{F},
\end{equation}
with
\begin{align}
	E &= 
	4 e^{-\frac{\omega}{2T}}
	\left(1 + e^{\frac{\omega}{T}} s\right)
	T^4
	\left(1 + 2\cosh\!\left(\frac{\omega}{T}\right)\right)
	\Bigg[
	-3 - 2s - \Delta_0 + 2s\Delta \nonumber
	\\
	&\quad
	+ \left(-3 + s(-9+\Delta) - \Delta_0\right)
	\cosh\!\left(\frac{\omega}{T}\right)
	- \left(3 + \Delta_0 + s\left(1 + 3s + (-1+s)\Delta_0\right)\right)
	\cosh\!\left(\frac{2\omega}{T}\right)
	\\
	&\quad
	+ \left(3 + s(-1+\Delta_0) + \Delta_0\right)
	\sinh\!\left(\frac{\omega}{T}\right)
	+ \left(3 + \Delta_0 + s\left(-1 + \Delta - s(3+\Delta_0)\right)\right)
	\sinh\!\left(\frac{2\omega}{T}\right)
	\Bigg],\nonumber
\end{align}
and 
\begin{align}
	F &= 
	(-1+s)(3+\Delta_0)\omega^2
	\Bigg[
	\left(3 + \Delta_0 + s(23 + 5\Delta_0 + 4s(1+\Delta_0))\right)
	\cosh\!\left(\frac{\omega}{2T}\right)\nonumber
	\\
	&\quad
	+ \left(4(3+\Delta_0) + s(1-\Delta_0 + 2s(1+\Delta_0))\right)
	\cosh\!\left(\frac{3\omega}{2T}\right)
	+ (3+\Delta)\cosh\!\left(\frac{5\omega}{2T}\right)\nonumber
	\\
	&\quad
	+ \left(-1 + s(-5+4s)\right)(3+\Delta_0)
	\sinh\!\left(\frac{\omega}{2T}\right)
	+ \left(-4(3+\Delta_0) + s(1-\Delta_0 + 2s(1+\Delta_0))\right)
	\sinh\!\left(\frac{3\omega}{2T}\right)
	\\
	&\quad
	- (3+\Delta_0)
	\sinh\!\left(\frac{5\omega}{2T}\right)
	\Bigg].\nonumber
\end{align}
\begin{equation}
	\mathrm{Var}(\Delta_0)_{\min}=\dfrac{G}{H},
\end{equation}
with 
\begin{align}
	G &= (3+\Delta_0)\Bigg[
	\Big(6(-1+\Delta_0)(3+\Delta_0)
	-2 s^3(3+\Delta_0)(3+5\Delta_0)
	+s^2(-139+(-38+\Delta_0)\Delta_0)
	+s(-113+3\Delta_0(10+\Delta_0))\Big)
	\cosh\!\left(\frac{\omega}{2T}\right)\nonumber
	\\
	&\quad
	-2\Big(9+3s(7+s(3+5s))
	+2(-3-7s+7s^3)\Delta_0
	+(-1+s)(3+s(2+3s))\Delta_0^2\Big)
	\cosh\!\left(\frac{3\omega}{2T}\right)\nonumber
	\\
	&\quad
	-\Big(6s(-1+\Delta_0)^2
	-5(-1+\Delta_0)(3+\Delta_0)
	+2s^3(1+\Delta_0)(3+\Delta_0)
	-s^2(-1+\Delta_0)(5+3\Delta_0)\Big)
	\cosh\!\left(\frac{5\omega}{2T}\right)\nonumber
	\\
	&\quad
	-\Big(-3+s(-1+\Delta_0)-\Delta_0\Big)(-1+\Delta_0)
	\cosh\!\left(\frac{7\omega}{2T}\right)\nonumber
	\\
	&\quad
	-\Big(4(-1+\Delta_0)(3+\Delta_0)
	+2s^3(3+\Delta_0)(7+\Delta_0)
	+s^2(41+(10-3\Delta_0)\Delta_0)
	-3s(13+\Delta_0(2+\Delta_0))\Big)
	\sinh\!\left(\frac{\omega}{2T}\right)
	\\
	&\quad
	-2\Big(8s(-2+\Delta_0)
	+3(-1+\Delta_0)(3+\Delta_0)
	+s^3(3+\Delta_0)(5+3\Delta_0)
	-2s^2(-8+\Delta_0+3\Delta_0^2)\Big)
	\sinh\!\left(\frac{3\omega}{2T}\right)\nonumber
	\\
	&\quad
	-\Big(-4s(-1+\Delta_0)^2
	+5(-1+\Delta_0)(3+\Delta_0)
	+2s^3(1+\Delta_0)(3+\Delta_0)
	-s^2(-1+\Delta_0)(5+3\Delta_0)\Big)
	\sinh\!\left(\frac{5\omega}{2T}\right)\nonumber
	\\
	&\quad
	+\Big(-3+s(-1+\Delta_0)-\Delta_0\Big)(-1+\Delta)
	\sinh\!\left(\frac{7\omega}{2T}\right)
	\Bigg],\nonumber
\end{align}
and 
\begin{align}
H &= (-1+s)\left(1+2\cosh\!\left(\frac{\omega}{T}\right)\right)
	\Bigg[
	\left(3+\Delta_0+s(23+5\Delta_0+4s(1+\Delta_0))\right)
	\cosh\!\left(\frac{\omega}{2T}\right)\nonumber
	\\
	&\quad
	+\left(4(3+\Delta_0)+s(1-\Delta_0+2s(1+\Delta_0))\right)
	\cosh\!\left(\frac{3\omega}{2T}\right)
	+(3+\Delta_0)\cosh\!\left(\frac{5\omega}{2T}\right)
	\\
	&\quad
	+\left(-1+s(-5+4s)\right)(3+\Delta)
	\sinh\!\left(\frac{\omega}{2T}\right)
	+\left(-4(3+\Delta)+s(1-\Delta_0+2s(1+\Delta_0))\right)
	\sinh\!\left(\frac{3\omega}{2T}\right)
	-(3+\Delta_0)\sinh\!\left(\frac{5\omega}{2T}\right)
	\Bigg].\nonumber
\end{align}
\begin{equation}
	\mathrm{Var}(T)_{\min}^{\mathrm{Ind}}=\dfrac{I}{J}
\end{equation}
with
\begin{align}
	I&=
	-4 e^{-\frac{\omega}{2T}}
	\left(1 + e^{\frac{\omega}{T}} s\right)
	T^4
	\left(1 + 2\cosh\!\left(\frac{\omega}{T}\right)\right)^2\nonumber
	\\
	&\quad \times
	\Big(
	1 - \Delta80
	+ (2 - 2\Delta_0 + s(3+\Delta_0))
	\cosh\!\left(\frac{\omega}{T}\right)
	+ s(3+\Delta_0)
	\sinh\!\left(\frac{\omega}{T}\right)
	\Big)
	\\
	&\quad \times
	\Big(
	4s
	+ (3+\Delta_0 + s(2-2\Delta_0 + s(3+\Delta_0)))
	\cosh\!\left(\frac{\omega}{T}\right)
	+ (-1+s^2)(3+\Delta_0)
	\sinh\!\left(\frac{\omega}{T}\right)
	\Big),\nonumber
\end{align} 
and
\begin{align}
	J&= (-1+s)(3+\Delta_0)\omega^2
	\Bigg[
	\Big(
	-6(-1+\Delta_0)(3+\Delta_0)
	+2s^3(3+\Delta_0)(3+5\Delta_0)
	+s^2(139-(-38+\Delta_0)\Delta_0)\nonumber
	\\
	&\quad+s(113-3\Delta_0(10+\Delta_0))
	\Big)\cosh\!\left(\frac{\omega}{2T}\right)
	+2\Big(
	9+3s(7+s(3+5s))
	+2(-3-7s+7s^3)\Delta_0
	+(-1+s)(3+s(2+3s))\Delta_0^2
	\Big)\cosh\!\left(\frac{3\omega}{2T}\right)\nonumber
	\\
	&\quad
	+\Big(
	6s(-1+\Delta_0)^2
	-5(-1+\Delta_0)(3+\Delta_0)
	+2s^3(1+\Delta_0)(3+\Delta_0)
	-s^2(-1+\Delta_0)(5+3\Delta_0)
	\Big)\cosh\!\left(\frac{5\omega}{2T}\right)\nonumber
	\\
	&\quad
	+\Big(
	-3+s(-1+\Delta_0)-\Delta_0
	\Big)(-1+\Delta_0)\cosh\!\left(\frac{7\omega}{2T}\right)
	\\
	&\quad
	+\Big(
	4(-1+\Delta_0)(3+\Delta_0)
	+2s^3(3+\Delta_0)(7+\Delta_0)
	+s^2(41+(10-3\Delta_0)\Delta_0)
	-3s(13+\Delta_0(2+\Delta_0))
	\Big)\sinh\!\left(\frac{\omega}{2T}\right)\nonumber
	\\
	&\quad
	+2\Big(
	8s(-2+\Delta_0)
	+3(-1+\Delta_0)(3+\Delta_0)
	+s^3(3+\Delta_0)(5+3\Delta_0)
	-2s^2(-8+\Delta_0+3\Delta_0^2)
	\Big)\sinh\!\left(\frac{3\omega}{2T}\right)\nonumber
	\\
	&\quad
	+\Big(
	-4s(-1+\Delta_0)^2
	+5(-1+\Delta_0)(3+\Delta_0)
	+2s^3(1+\Delta)(3+\Delta_0)
	-s^2(-1+\Delta_0)(5+3\Delta_0)
	\Big)\sinh\!\left(\frac{5\omega}{2T}\right)\nonumber
	\\
	&\quad
	-\Big(
	-3+s(-1+\Delta_0)-\Delta_0
	\Big)(-1+\Delta_0)\sinh\!\left(\frac{7\omega}{2T}\right)\nonumber
	\Bigg].
\end{align}
\begin{equation}
	\mathrm{Var}(\Delta_0)_{\min}^{\mathrm{Ind}}=\dfrac{K}{L},
\end{equation}
with
\begin{align}
	K &= (3+\Delta_0)
	\Bigg[1-\Delta_0
	+ (2-2\Delta_0 + s(3+\Delta_0))
	\cosh\!\left(\frac{\omega}{T}\right)
	+ s(3+\Delta_0)
	\sinh\!\left(\frac{\omega}{T}\right) \Bigg]
	\\
	&\quad \times
	\Bigg[ 4s
	+ (3+\Delta_0 + s(2-2\Delta_0 + s(3+\Delta_0)))
	\cosh\!\left(\frac{\omega}{T}\right)
	+ (-1+s^2)(3+\Delta_0)
	\sinh\!\left(\frac{\omega}{T}\right) \Bigg],\nonumber
\end{align}
and
\begin{align}
L &= (-1+s)\Bigg[
	-3-2s-\Delta_0+2s\Delta_0
	+(-3+s(-9+\Delta_0)-\Delta_0)\cosh\!\left(\frac{\omega}{T}\right)\nonumber
	\\
	&\quad
	-(3+\Delta_0+s(1+3s+(-1+s)\Delta_0))\cosh\!\left(\frac{2\omega}{T}\right)
	+(3+s(-1+\Delta_0)+\Delta_0)\sinh\!\left(\frac{\omega}{T}\right)
	\\
	&\quad
	+(3+\Delta_0+s(-1+\Delta_0-s(3+\Delta_0)))\sinh\!\left(\frac{2\omega}{T}\right)
	\Bigg].\nonumber
\end{align}
The explicit analytical expression of the ratio $\Gamma$, as defined in Eq.~(\ref{rap}), can be written in the compact form $\Gamma = \frac{M}{N}$, where 
\begin{align}
	M &= 
	\Big[
	s(76-12\Delta_0)
	-5(-1+\Delta_0)(3+\Delta_0)
	+4s^3(-1+\Delta_0)(3+\Delta_0)
	+s^2(7+\Delta_0)^2\nonumber
	\\
	&\quad
	+e^{-\frac{3\omega}{T}}(-3+s(-1+\Delta_0)-\Delta_0)(-1+\Delta_0)
	+5e^{-\frac{2\omega}{T}}(-3+s(-1+\Delta_0)-\Delta_0)(-1+\Delta_0)\nonumber
	\\
	&\quad
	+e^{-\frac{\omega}{T}}\!\left(-6(-1+\Delta)(3+\Delta_0)
	+s^2(-1+\Delta)(7+5\Delta_0)
	+s(37+(-22+\Delta_0)\Delta_0)\right)\nonumber
	\\
	&\quad
	+e^{\frac{\omega}{T}}\!\left(-2s^2(-15+\Delta_0)(3+\Delta_0)
	-(-1+\Delta_0)(3+\Delta_0)
	+2s^3(3+\Delta_0)(5+3\Delta_0)
	+s(37-3\Delta_0(6+\Delta_0))\right)
	\\
	&\quad
	+e^{\frac{2\omega}{T}} s(1-\Delta_0+s(3+\Delta_0))(5-\Delta_0+2s(5+3\Delta_0))
	+e^{\frac{3\omega}{T}} s(1-\Delta_0+2s(1+\Delta_0))(1-\Delta_0+s(3+\Delta_0))
	\Big]\nonumber
	\\
	&\quad\times
	\Big[
	3+2s+\Delta_0-2s\Delta_0
	+(3-s(-9+\Delta_0)+\Delta_0)\cosh\!\left(\frac{\omega}{T}\right)
	+(3+\Delta_0+s(1+3s+(-1+s)\Delta_0))\cosh\!\left(\frac{2\omega}{T}\right)\nonumber
		\\
	&\quad
	-(3+s(-1+\Delta_0)+\Delta_0)\sinh\!\left(\frac{\omega}{T}\right)
	+(-3+s+3s^2+(-1+(-1+s)s)\Delta_0)\sinh\!\left(\frac{2\omega}{T}\right)
	\Big],\nonumber
\end{align}
and
\begin{align}
	N &= 
	2\big(1+2\cosh(\tfrac{\omega}{T})\big)
	\Big(
	1-\Delta_0+(2-2\Delta_0+s(3+\Delta_0))\cosh(\tfrac{\omega}{T})
	+s(3+\Delta_0)\sinh(\tfrac{\omega}{T})
	\Big)\nonumber
	\\
	&\quad\times
	\Big(
	4s+(3+\Delta_0+s(2-2\Delta_0+s(3+\Delta_0)))\cosh(\tfrac{\omega}{T})
	+(-1+s^2)(3+\Delta_0)\sinh(\tfrac{\omega}{T})
	\Big)\nonumber
	\\
	&\quad\times
	\Big(
	3+19s-4s^2+\Delta_0+5s\Delta_0
	+4(3+s+\Delta_0+s^2(2+\Delta_0))\cosh(\tfrac{\omega}{T})
	\\
	&\qquad
	+(3+\Delta_0+s(1-\Delta_0+2s(1+\Delta_0)))\cosh(\tfrac{2\omega}{T})
	+4(-1+s)(3+\Delta_0+s(2+\Delta_0))\sinh(\tfrac{\omega}{T})\nonumber
	\\
	&\qquad
	+(-1+s)(3+\Delta_0+2s(1+\Delta_0))\sinh(\tfrac{2\omega}{T})
	\Big).\nonumber
\end{align}


The influence of the amplitude damping (AD) channel on the precision of temperature estimation is illustrated in Figure~\ref{fig:17}, where the minimal variance $\mathrm{Var}(T)_{\min}$ is plotted as a function of the temperature $T$ and the decoherence parameter $s$. The results show a strong dependence of the estimation precision on the damping strength.
In the weak-noise regime ($s \approx 0$), the variance remains relatively small over a broad temperature range, indicating that the thermal sensitivity of the probe is well preserved. As the decoherence parameter $s$ increases, $\mathrm{Var}(T)_{\min}$ grows significantly, reflecting a gradual reduction of the temperature-dependent information encoded in the quantum state.

A comparison between panels \ref{fig:17}(a) and \ref{fig:17}(b) reveals the role of the energy-level spacing $\omega$. Increasing $\omega$ slightly modifies the curvature of the variance surface and shifts the region of optimal precision, while the overall qualitative behavior remains unchanged: stronger amplitude damping systematically enlarges the estimation uncertainty. These findings emphasize that both the spectral structure of the system (through $\omega$) and the strength of decoherence jointly determine the achievable thermometric performance under the AD channel.

\begin{figure}[H]
	\includegraphics[scale=0.58]{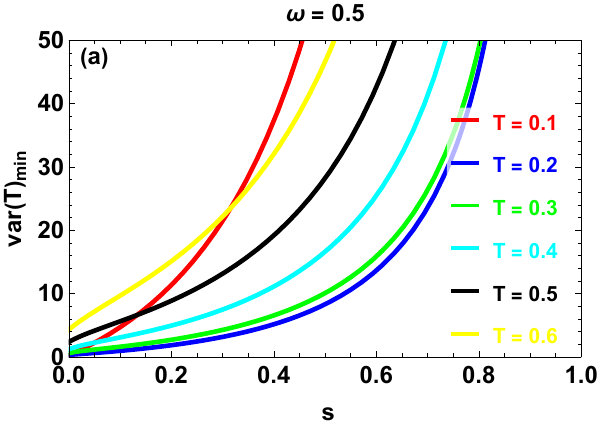}
	\includegraphics[scale=0.58]{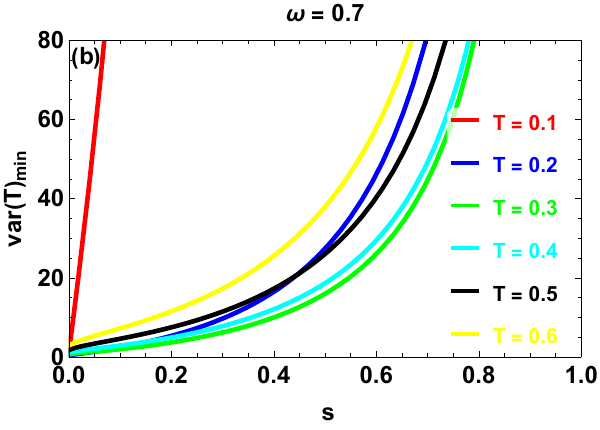}
	\includegraphics[scale=0.58]{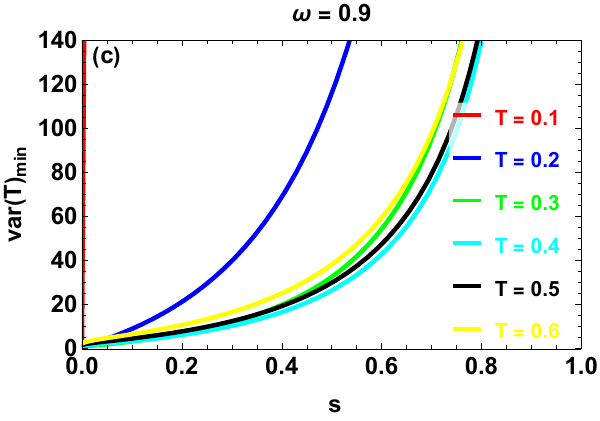}
\caption{Minimal variance $\mathrm{Var}(T)_{\min}$ plotted as a function of the decoherence parameter $s$ for different values of the temperature $T$ under the amplitude damping (AD) channel. Panel (a) corresponds to $\omega = 0.5$, panel (b) to $\omega = 0.6$, and panel (c) to $\omega = 0.9$. The initial state parameter is fixed at $\Delta_0 = -2$.}

	\label{fig:18}
\end{figure}
Figure~\ref{fig:18} illustrates the dependence of the minimal variance $\mathrm{Var}(T)_{\min}$ on the decoherence parameter $s$ for different temperatures under the amplitude damping (AD) channel. For all values of the energy-level spacing $\omega$, the variance increases monotonically with $s$, indicating that stronger damping progressively diminishes the temperature sensitivity of the quantum probe.
A clear ordering with respect to temperature is observed in each panel. In the weak-damping regime, lower temperatures yield comparatively smaller variances, while higher temperatures lead to a more pronounced growth of $\mathrm{Var}(T)_{\min}$ as $s$ increases. As the decoherence parameter approaches unity, the variance rises sharply, revealing a substantial loss of estimation precision in the strong-noise regime.

The comparison between panels \ref{fig:18}(a)-(c) shows that increasing the energy-level spacing $\omega$ modifies the overall magnitude and steepness of the variance curves, although the qualitative monotonic behavior with respect to $s$ remains unchanged. This confirms that decoherence plays a dominant role in determining the attainable thermometric precision under the AD channel.

\begin{figure}[H]
	\includegraphics[scale=0.58]{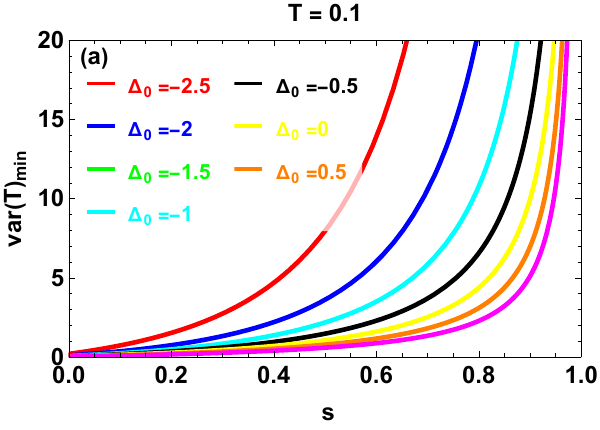}
	\includegraphics[scale=0.58]{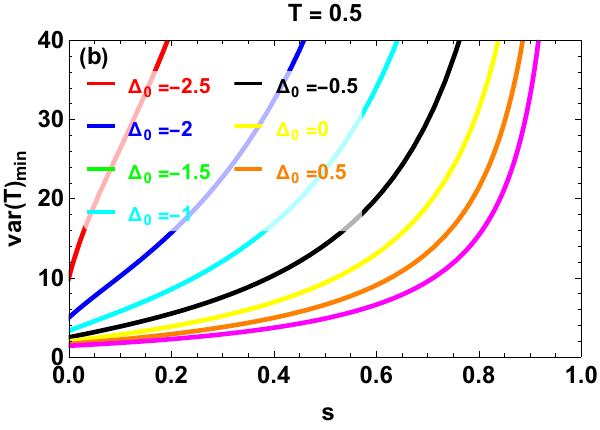}
	\includegraphics[scale=0.58]{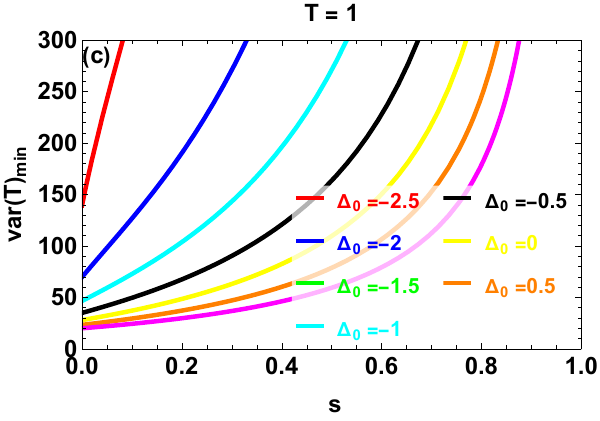}
\caption{Minimal variance $\mathrm{Var}(T)_{\min}$ plotted as a function of the decoherence parameter $s$ for different values of  $\Delta_0$ under the amplitude damping (AD) channel. Panel (a) corresponds to $T=0.1$, panel (b) to $T=0.5$, and panel (c) to $T =1$. The  energy level spacing is fixed at $\omega = 0.3$.}
	\label{fig:19}
\end{figure}
The dependence of the minimal variance $\mathrm{Var}(T)_{\min}$ on the decoherence parameter $s$ for different initial-state parameters $\Delta_0$ under the amplitude damping channel is presented in Figure~\ref{fig:19}. For all temperatures, the variance increases monotonically with $s$, confirming that stronger damping progressively reduces the precision of temperature estimation.
A pronounced ordering with respect to $\Delta_0$ is observed throughout the evolution. Initial states characterized by more negative values of $\Delta_0$ exhibit significantly larger variances, while configurations closer to $\Delta_0=0$ yield comparatively smaller values over the entire range of $s$. This indicates that the preparation of the initial state strongly influences the robustness of thermometric precision against decoherence.

Comparing panels \ref{fig:19}(a)-(c) shows that increasing the temperature substantially amplifies the overall magnitude of the variance and accelerates its growth as $s$ approaches unity. In particular, at higher temperatures the divergence of $\mathrm{Var}(T)_{\min}$ becomes much more pronounced, demonstrating that thermal effects magnify the sensitivity of the estimation process to amplitude damping noise.


Figure~\ref{fig:20} presents the minimal variance $\mathrm{Var}(\Delta_0)_{\min}$ as a function of the initial-state parameter $\Delta_0$ and the decoherence parameter $s$ under the amplitude damping channel. For both values of the energy-level spacing, the variance surface exhibits a smooth increase along the $s$ direction, indicating that stronger damping systematically deteriorates the precision in estimating $\Delta_0$.
Along the $\Delta_0$ axis, the variance preserves its characteristic quadratic-like profile, with smaller values obtained near the boundary of the allowed parameter domain and larger values toward the central region. This confirms that the intrinsic geometrical dependence on $\Delta_0$ remains intact even in the presence of decoherence, although its overall magnitude becomes increasingly sensitive to the damping strength.

Comparing panels \ref{fig:20}(a) and \ref{fig:20}(b) shows that increasing the energy-level spacing $\omega$ modifies the overall scale of the variance without altering the qualitative structure of the surface. In both cases, the combined dependence on $\Delta_0$ and $s$ highlights that decoherence primarily controls the growth rate of the estimation uncertainty, while the parameter geometry dictates its distribution along the $\Delta_0$ direction.
\begin{figure}[H]
	\includegraphics[scale=0.58]{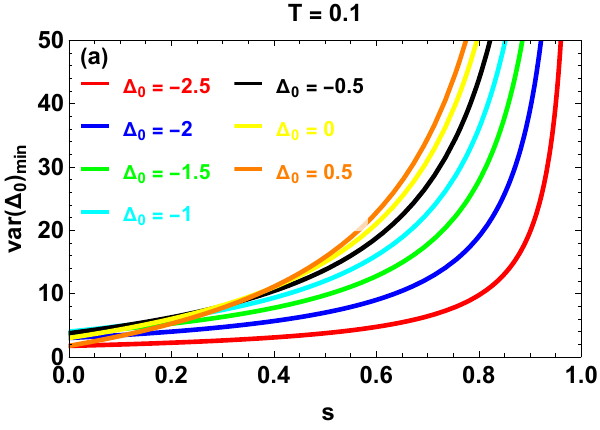}
	\includegraphics[scale=0.58]{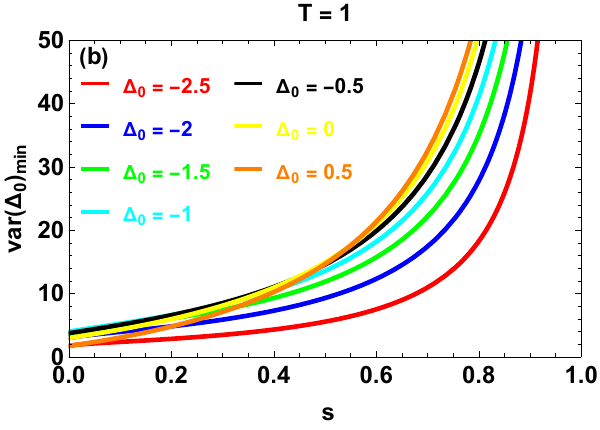}
	\includegraphics[scale=0.58]{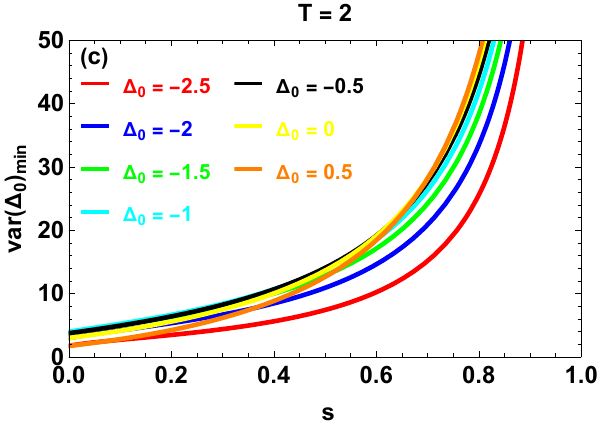}
		\caption{Minimal variance $\mathrm{Var}(\Delta_0)_{\min}$ plotted as a function of the decoherence parameter $s$  under the amplitude damping (AD) channel.  Panel (a) corresponds to $T=0.1$, panel (b) to $T=1$, and panel (c) to $T=2$. The  energy level spacing is fixed at $\omega = 1$.}
	
	\label{fig:21}
\end{figure}
The behavior of the minimal variance $\mathrm{Var}(\Delta_0)_{\min}$ under the amplitude damping channel is depicted in Figure~\ref{fig:21} for different temperatures. In all panels, the variance increases monotonically with the decoherence parameter $s$, showing that stronger damping progressively deteriorates the precision in estimating the initial-state parameter.
For small values of $s$, the curves remain relatively close to each other, indicating limited sensitivity to noise in the weak-damping regime. However, as $s$ approaches unity, the variance grows rapidly and the separation between different $\Delta_0$ configurations becomes more pronounced. This divergence reflects the increasing susceptibility of the estimation process to decoherence effects.

Comparing panels \ref{fig:21}(a)-(c) reveals that raising the temperature does not qualitatively modify the monotonic structure of the curves, but it alters their relative ordering and slightly enhances the overall magnitude of the variance. Hence, while the intrinsic dependence on $\Delta_0$ governs the hierarchy of the curves, the temperature mainly acts as a scaling factor that influences the strength of the decoherence-induced degradation.

\begin{figure}[H]
	\centering
	\includegraphics[scale=0.73]{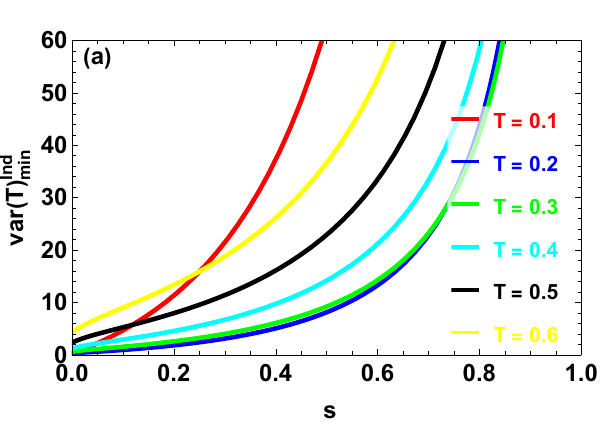}
	\hspace*{0.7cm}
	\includegraphics[scale=0.73]{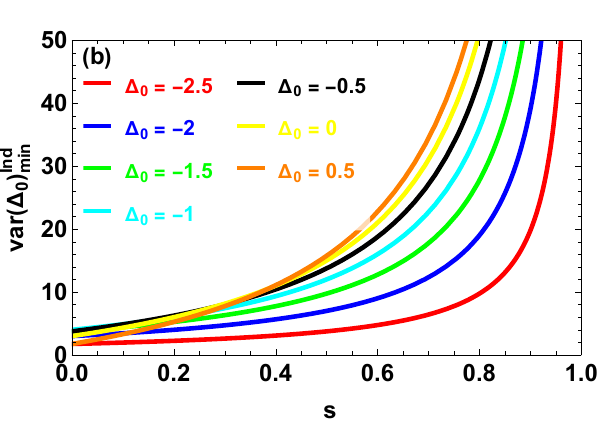}
	\caption{The minimal variance for the individual estimation as a function of the decoherence parameter $s$ under the amplitude damping (AD) channel. 
		(a) Minimal variance $\mathrm{Var}(T)_{\min}^{\mathrm{Ind}}$ plotted versus $s$, with fixed parameters $\omega = 0.5$ and $\Delta_0 = -2$. 
		(b) Minimal variance $\mathrm{Var}(\Delta_0)_{\min}^{\mathrm{Ind}}$ plotted versus $s$, with $\omega = 0.5$ and $T = 0.1$.}
	\label{fig:22}
\end{figure}
The behavior of the minimal variances in the individual estimation scenario under the amplitude damping channel is presented in Figure~\ref{fig:22}. Panel (a) shows $\mathrm{Var}(T)_{\min}^{\mathrm{Ind}}$ as a function of the decoherence parameter $s$, while panel (b) depicts $\mathrm{Var}(\Delta_0)_{\min}^{\mathrm{Ind}}$ under the same noise conditions.
For temperature estimation (panel \ref{fig:22}(a)), the variance increases rapidly with $s$ for all considered temperatures, indicating a strong sensitivity of thermometric precision to amplitude damping noise. The separation between the curves becomes more pronounced as $s$ grows, revealing that higher temperatures lead to a steeper degradation of precision in the strong-noise regime.
A similar monotonic trend is observed for the estimation of the initial-state parameter (panel \ref{fig:22}(b)). The variance $\mathrm{Var}(\Delta_0)_{\min}^{\mathrm{Ind}}$ rises significantly as $s$ approaches unity, and a clear hierarchy with respect to $\Delta_0$ is maintained throughout the evolution. Initial configurations with more negative values of $\Delta_0$ tend to exhibit larger variances, whereas states closer to zero display comparatively improved robustness.

Overall, these results confirm that, in the individual estimation framework, amplitude damping noise systematically enlarges the uncertainty for both parameters, with the degradation becoming particularly severe in the strong decoherence regime.

\begin{figure}[H]
	\includegraphics[scale=0.58]{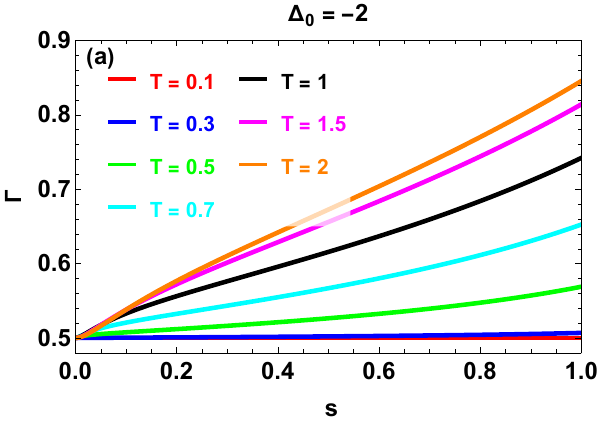}
	\includegraphics[scale=0.58]{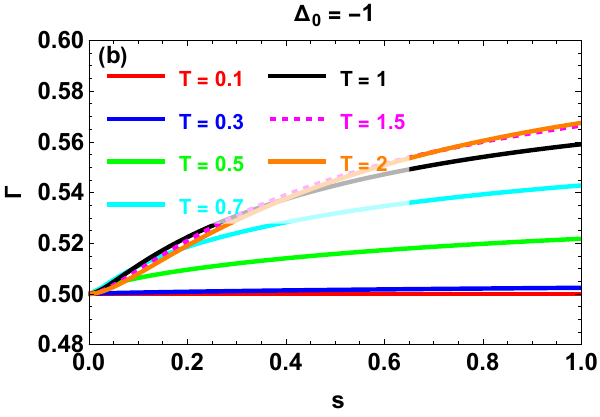}
	\includegraphics[scale=0.58]{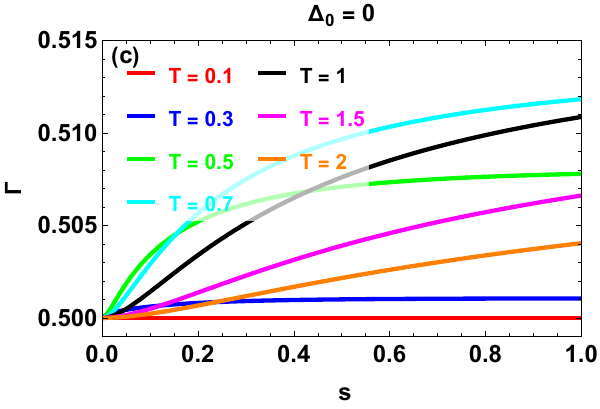}
\caption{The ratio between the minimal total variances in the estimation of the parameters $T$ and $\Delta_0$ in the amplitude damping (AD) channel as a function of the decoherence parameter, for different values of the temperature $T$, and for three values of $\Delta_0$: (a) $\Delta_0 = -2$, (b) $\Delta_0 = -1$, and (c) $\Delta_0 = 0$. The energy level spacing is fixed at $\omega = 1$.}
	\label{fig:23}
\end{figure}
The behavior of the performance ratio $\Gamma$ under the amplitude damping channel is displayed in Figure~\ref{fig:23} for different temperatures and initial-state parameters. In all panels, $\Gamma$ starts from a value close to $0.5$ in the weak-noise limit ($s \approx 0$), confirming that the simultaneous estimation strategy initially retains its optimal advantage over the individual one.

As the decoherence parameter increases, $\Gamma$ grows monotonically, indicating a gradual reduction of the relative efficiency of the simultaneous protocol. This increase becomes more pronounced for larger temperatures, as clearly observed in panel (a) for $\Delta_0=-2$, where the ratio approaches significantly higher values as $s$ tends to unity. 

The dependence on the initial-state parameter is also evident: when $\Delta_0$ moves from $-2$ to $0$ (figures \ref{fig:23}(a)–(c)), the growth of $\Gamma$ with $s$ becomes progressively weaker. In particular, for $\Delta_0=0$, the ratio remains very close to $0.5$ over the entire range of $s$, indicating that the simultaneous scheme preserves nearly the same relative efficiency even in the presence of strong damping.

Overall, these results show that, although amplitude damping reduces the metrological advantage of the joint estimation strategy, this degradation strongly depends on both the temperature and the initial-state configuration.
\subsection{Phase flip channel}
This channel describes a stochastic dynamical process occurring with probability $s$. 
Its action on the initial state corresponds to a unitary transformation generated by 
the Pauli operator $\tau_z$, which induces a phase-flip (PF) error. 
The single-qubit PF decoherence channel can be represented in the Kraus operator formalism as
\begin{equation}
	\mathcal{K}_1 =
	\begin{pmatrix}
		\sqrt{s} & 0 \\
		0 & \sqrt{s}
	\end{pmatrix},
	\qquad
	\mathcal{K}_2 =
	\begin{pmatrix}
		\sqrt{1-s} & 0 \\
		0 & -\sqrt{1-s}
	\end{pmatrix}.
\end{equation}

Under the action of this channel, the evolved two-qubit density matrix takes the form
\begin{equation}
	{\varrho}^{\,\mathrm{PF}}_{12} =
	\begin{pmatrix}
		x & 0 & 0 & 0 \\
		0 & z & \pi_{23} & 0 \\
		0 & \pi_{23} & z & 0 \\
		0 & 0 & 0 & y
	\end{pmatrix},
\end{equation}
where the modified coherence terms are given by
\begin{equation}
	\pi_{23} = \delta(1 - 2s)^2.
\end{equation}

It is evident that the phase-flip channel preserves the diagonal populations, 
while the off-diagonal coherences are attenuated by the factor $(1 - 2s)^2$, 
which characterizes the decohering nature of the process.

 Building upon the framework established in Section~\ref{sec:2}, we now present the explicit analytical expressions of $\mathrm{Var}(T)_{\min}$, $\mathrm{Var}(\Delta_0)_{\min}$, $\mathrm{Var}(T)_{\min}^{\mathrm{Ind}}$, and $\mathrm{Var}(\Delta_0)_{\min}^{\mathrm{Ind}}$, corresponding to the joint and separate estimation strategies of the parameters $T$ and $\Delta_0$. These quantities can be written as follows
\begin{equation}
\mathrm{Var}(T)_{\min}=\dfrac{O}{P},
\end{equation}
with
\begin{align}
	O &= 2
	T^4\bigg(1+2\cosh(\tfrac{\omega}{T})\bigg)
	\Bigg[
	3+\Delta_0
	+4(-1+s)s(1+2(-1+s)s)(5+3\Delta_0)
	\\
	&\quad
	+2\bigg(3+\Delta_0+8(-1+s)s(1+2(-1+s)s)(2+\Delta_0)\bigg)\cosh(\tfrac{\omega}{T})
	+4(-1+s)s(1+2(-1+s)s)(-1+\Delta_0)\cosh(\tfrac{2\omega}{T})
	\Bigg],\nonumber
\end{align}
and
\begin{equation}
	P =
	(3+\Delta_0)\omega^2
	\Bigg[
	2\big(3+\Delta_0+8(-1+s)s(1+2(-1+s)s)(2+\Delta_0)\big)
	+\big(3+\Delta_0+16(-1+s)s(1+2(-1+s)s)(1+\Delta_0)\big)\cosh(\tfrac{\omega}{T})
	\Bigg].
\end{equation}
\begin{equation}
	\mathrm{Var}(\Delta_0)_{\min}=\dfrac{Q}{R},
\end{equation}
\begin{align}
	Q &=
	-\,e^{\frac{2\omega}{T}}(3+\Delta_0)
	\Bigg[
	2e^{\frac{2\omega}{T}}\Bigg(-9+24(-1+s)s(1+2(-1+s)s)+6\Delta_0
	+8(-1+s)s(1+2(-1+s)s)\Delta_0\nonumber
	\\
	&\quad
	+(1+8(-1+s)s)(3+8(-1+s)s)\Delta_0^2\Bigg)
	+(-1+\Delta_0)\big(3+\Delta_0+8(-1+s)s(1+2(-1+s)s)(1+2\Delta_0)\big)\nonumber
	\\
	&\quad
	+e^{\frac{4\omega}{T}}(-1+\Delta_0)\bigg(3+\Delta_0+8(-1+s)s(1+2(-1+s)s)(1+2\Delta_0)\bigg)
	\\
	&\quad
	+\big(e^{\frac{\omega}{T}}+e^{\frac{3\omega}{T}}\big)
	\Bigg[5(-1+\Delta_0)(3+\Delta_0)
	+16(-1+s)s(1+2(-1+s)s)(-1+3\Delta_0^2)\Bigg]\nonumber
	\Bigg]
	\\
	&\quad\times
	\bigg(1+2\cosh(\tfrac{\omega}{T})\bigg)
	\bigg(3+\Delta_0+4(-1+s)s(1+\Delta_0)
	+4(-1+s)s(-1+\Delta_0)\cosh(\tfrac{\omega}{T})\bigg)\nonumber
	\\
	&\quad\times
	\bigg(-1+\Delta_0+4(-1+s)s(1+\Delta_0)
	+2(1+2(-1+s)s)(-1+\Delta_0)\cosh(\tfrac{\omega}{T})\bigg),\nonumber
\end{align} and
\begin{equation}
	\begin{aligned}
		R &= 
		2\bigg(1+e^{\frac{\omega}{T}}+e^{\frac{2\omega}{T}}\bigg)^2
		\\
		&\qquad\times
		\Bigg[
		(1+2(-1+s)s)(-1+\Delta_0)(1+e^{\frac{2\omega}{T}})
		+e^{\frac{\omega}{T}}\big(-1+\Delta_0+4(-1+s)s(1+\Delta_0)\big)
		\Bigg]
		\\
		&\qquad\times
			\Bigg[
		2(-1+s)s(-1+\Delta_0)(1+e^{\frac{2\omega}{T}})
		+e^{\frac{\omega}{T}}\big(3+\Delta_0+4(-1+s)s(1+\Delta_0)\big)
		\Bigg]
		\\
		&\qquad\times
		\Bigg[
		2(3+\Delta_0+8(-1+s)s(1+2(-1+s)s)(2+\Delta_0))
		+(3+\Delta_0+16(-1+s)s(1+2(-1+s)s)(1+\Delta_0))\cosh(\tfrac{\omega}{T})
		\Bigg].
	\end{aligned}
\end{equation}
\begin{equation}
\mathrm{Var}(T)_{\min}^{\mathrm{Ind}}=\frac{S}{T},
\end{equation}
with
\begin{equation}
	\begin{aligned}
		S &= 
		4e^{-\frac{2\omega}{T}}
		\bigg(1+e^{\frac{\omega}{T}}+e^{\frac{2\omega}{T}}\bigg)^2
		T^4
		\\
		&\quad\times
		\Bigg[
		(1+2(-1+s)s)(-1+\Delta_0)(1+e^{\frac{2\omega}{T}})
		+e^{\frac{\omega}{T}}\big(-1+\Delta_0+4(-1+s)s(1+\Delta_0)\big)
		\Bigg]
		\\
		&\quad\times
		\Bigg[
		2(-1+s)s(-1+\Delta_0)(1+e^{\frac{2\omega}{T}})
		+e^{\frac{\omega}{T}}\big(3+\Delta_0+4(-1+s)s(1+\Delta_0)\big)
			\Bigg],
	\end{aligned}
\end{equation}
and 
\begin{align}
	T &= (3+\Delta_0)\omega^2
	\Bigg[
	2e^{\frac{2\omega}{T}}
	\Bigg(-9+24(-1+s)s(1+2(-1+s)s)\nonumber
	\\
	&\quad
	+6\Delta_0
	+8(-1+s)s(1+2(-1+s)s)\Delta_0
	+(1+8(-1+s)s)(3+8(-1+s)s)\Delta_0^2
	\Bigg)\nonumber
	\\
	&\quad
	+(1+e^{\frac{4\omega}{T}})
	(-1+\Delta_0)
	\Bigg(3+\Delta_0
	+8(-1+s)s(1+2(-1+s)s)(1+2\Delta_0)\Bigg)
	\\
	&\quad
	+(e^{\frac{\omega}{T}}+e^{\frac{3\omega}{T}})
	\Bigg(
	5(-1+\Delta_0)(3+\Delta_0)\nonumber
	\\
	&\qquad
	+16(-1+s)s(1+2(-1+s)s)(-1+3\Delta_0^2)
	\Bigg)
	\Bigg].\nonumber
\end{align}
\begin{equation}
	\mathrm{Var}(\Delta_0)_{\min}^{\mathrm{Ind}}=\frac{U}{V},
\end{equation}
with
\begin{align}
	U &= 
	-(3+\Delta_0)
	\Bigg[
	3+\Delta_0
	+4(-1+s)s(1+\Delta_0)
	+4(-1+s)s(-1+\Delta_0)\cosh\!\left(\frac{\omega}{T}\right)
	\Bigg]\nonumber
	\\
	&\quad\times
	\Bigg[
	-1+\Delta_0
	+4(-1+s)s(1+\Delta_0)
	+2(1+2(-1+s)s)(-1+\Delta_0)
	\cosh\!\left(\frac{\omega}{T}\right)
	\Bigg],
\end{align}
and 
\begin{align}
	V	 &=
	3+\Delta_0
	+4(-1+s)s(1+2(-1+s)s)(5+3\Delta_0)
	+2\Bigg[
	3+\Delta_0
	+8(-1+s)s(1+2(-1+s)s)(2+\Delta_0)
	\Bigg]
	\cosh\!\left(\frac{\omega}{T}\right)\nonumber
	\\
	&\quad
	+4(-1+s)s(1+2(-1+s)s)(-1+\Delta_0)
	\cosh\!\left(\frac{2\omega}{T}\right).
\end{align}
The ratio $\Gamma$ defined in Eq.~(\ref{rap}) admits the following explicit expression, which can be conveniently expressed as $\Gamma = \frac{W}{X}$, with
\begin{align}
	W &= 
	\Bigg[
	2(-1+s)s(1+2(-1+s)s)(-1+\Delta_0)\,(1+e^{\frac{4\omega}{T}})
	+(e^{\frac{\omega}{T}}+e^{\frac{3\omega}{T}})
	\big(3+\Delta_0+8(-1+s)s(1+2(-1+s)s)(2+\Delta_0)\big)\nonumber
	\\
	&\quad
	+e^{\frac{2\omega}{T}}
	\big(3+\Delta_0+4(-1+s)s(1+2(-1+s)s)(5+3\Delta_0)\big)
	\Bigg]\nonumber
	\\
	&\quad\times
	\Bigg[
	2e^{\frac{2\omega}{T}}\!\left[-9+24(-1+s)s(1+2(-1+s)s)
	+6\Delta_0
	+8(-1+s)s(1+2(-1+s)s)\Delta_0
	\right.
	\\
	&\qquad\left.
	+(1+8(-1+s)s)(3+8(-1+s)s)\Delta_0^2\right]\nonumber
	\\
	&\quad
	+(1+e^{\frac{4\omega}{T}})(-1+\Delta_0)
	\big(3+\Delta_0+8(-1+s)s(1+2(-1+s)s)(1+2\Delta_0)\big)\nonumber
	\\
	&\quad
	+(e^{\frac{\omega}{T}}+e^{\frac{3\omega}{T}})
	\big(5(-1+\Delta_0)(3+\Delta_0)
	+16(-1+s)s(1+2(-1+s)s)(-1+3\Delta_0^2)\big)
	\Bigg].\nonumber
\end{align}
and
\begin{align}
X &= 
	2\bigg(1+e^{\frac{\omega}{T}}+e^{\frac{2\omega}{T}}\bigg)
	\Bigg[
	(1+2(-1+s)s)(-1+\Delta_0)(1+e^{\frac{2\omega}{T}})
	+e^{\frac{\omega}{T}}
	\big(-1+\Delta_0+4(-1+s)s(1+\Delta_0)\big)
	\Bigg]\nonumber
	\\
	&\quad\times
	\Bigg[
	2(-1+s)s(-1+\Delta_0)(1+e^{\frac{2\omega}{T}})
	+e^{\frac{\omega}{T}}
	\big(3+\Delta_0+4(-1+s)s(1+\Delta_0)\big)
\Bigg]
	\\
	&\quad\times
	\Bigg[
	(3+\Delta_0+16(-1+s)s(1+2(-1+s)s)(1+\Delta_0))(1+e^{\frac{2\omega}{T}})
	+4e^{\frac{\omega}{T}}
	\big(3+\Delta_0+8(-1+s)s(1+2(-1+s)s)(2+\Delta_0)\big)
	\Bigg].\nonumber
\end{align}


The influence of the phase flip (PF) channel on the temperature estimation precision is illustrated in Figure~\ref{fig:24}. In Figure~\ref{fig:24}(a), corresponding to $\omega=0.6$, the minimal variance $\mathrm{Var}(T)_{\min}$ exhibits a smooth dependence on both the temperature $T$ and the decoherence parameter $s$. The growth of the variance with increasing $s$ remains moderate, indicating a gradual degradation of thermometric precision under phase noise.
In Figure~\ref{fig:24}(b), associated with $\omega=1$, the overall magnitude of the variance becomes significantly larger, particularly in regions of stronger dephasing. Although the qualitative dependence on $s$ is preserved, the estimation uncertainty is clearly amplified for larger energy-level spacing, demonstrating that the spectral structure affects the robustness of temperature estimation under the PF channel.
\begin{figure}[H]
	\includegraphics[scale=0.58]{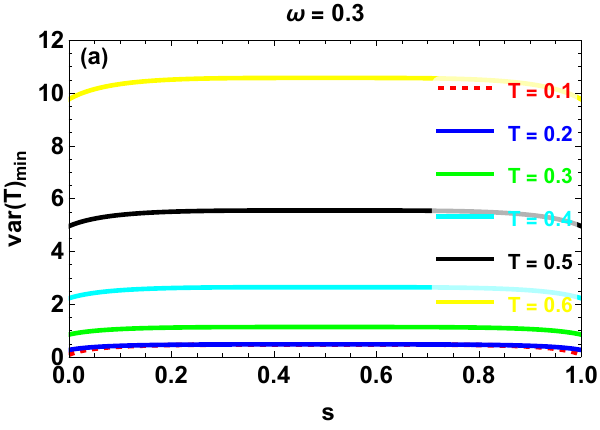}
	\includegraphics[scale=0.58]{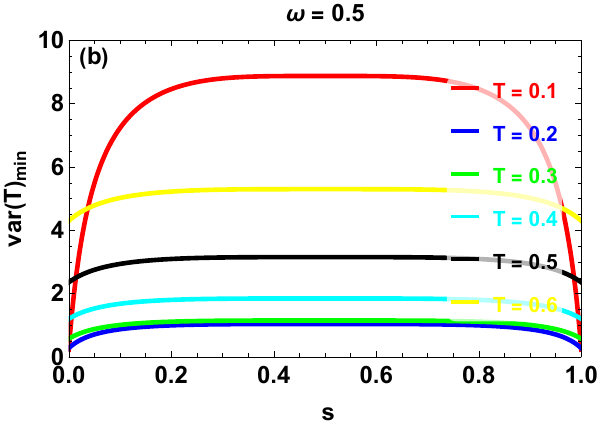}
	\includegraphics[scale=0.58]{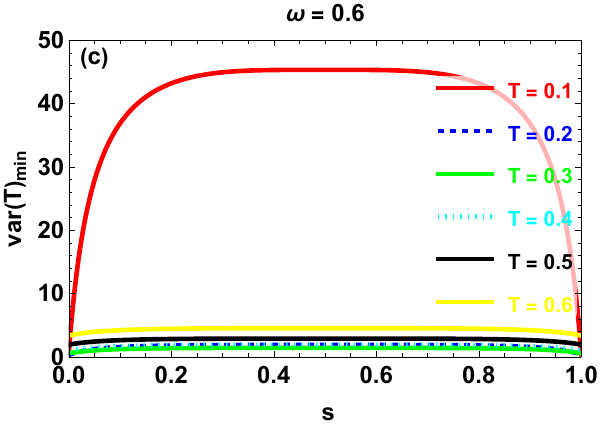}
	\caption{Minimal variance $\mathrm{Var}(T)_{\min}$ plotted as a function of the decoherence parameter $s$ for different values of the temperature $T$ under the phase flip (PF) channel. Panel (a) corresponds to $\omega = 0.3$, panel (b) to $\omega = 0.5$, and panel (c) to $\omega = 0.6$. The initial state parameter is fixed at $\Delta_0 = -2$.}
	
	\label{fig:25}
\end{figure}
The behavior of the minimal variance $\mathrm{Var}(T)_{\min}$ under the phase flip (PF) channel is presented in Figure~\ref{fig:25} as a function of the decoherence parameter $s$ for different temperatures. In contrast to the amplitude damping case, the variance does not exhibit a simple monotonic growth with $s$. Instead, for each value of $\omega$, the curves display a nontrivial dependence characterized by an initial increase followed by a saturation and, in some cases, a slight decrease as $s$ approaches unity.
A clear temperature ordering is maintained throughout the evolution: higher temperatures generally correspond to larger variances, indicating reduced thermometric precision. However, the sensitivity to phase noise strongly depends on the energy-level spacing. For $\omega=0.3$, the overall magnitude of the variance remains relatively moderate and the curves are smooth across the entire range of $s$. As $\omega$ increases to $0.5$ and $0.6$, the variance becomes significantly amplified, and the dependence on $s$ becomes more pronounced, particularly at low temperatures.

These results indicate that, under the PF channel, the degradation of temperature estimation precision is primarily governed by the interplay between dephasing strength and spectral structure, with phase noise affecting the estimation process in a qualitatively different manner compared to dissipative channels.

\begin{figure}[H]
	\includegraphics[scale=0.58]{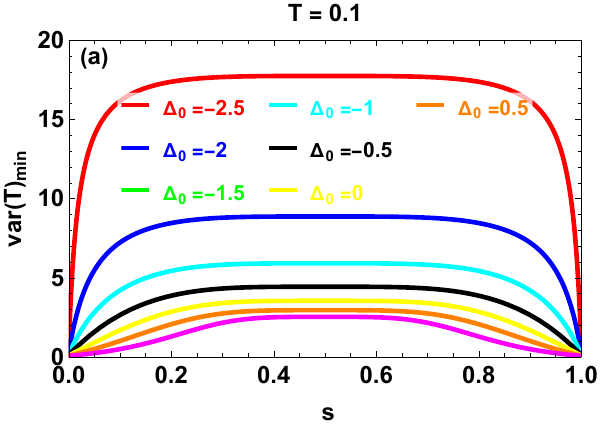}
	\includegraphics[scale=0.58]{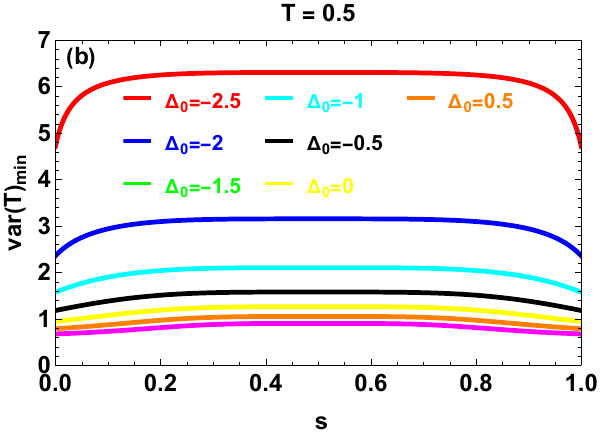}
	\includegraphics[scale=0.58]{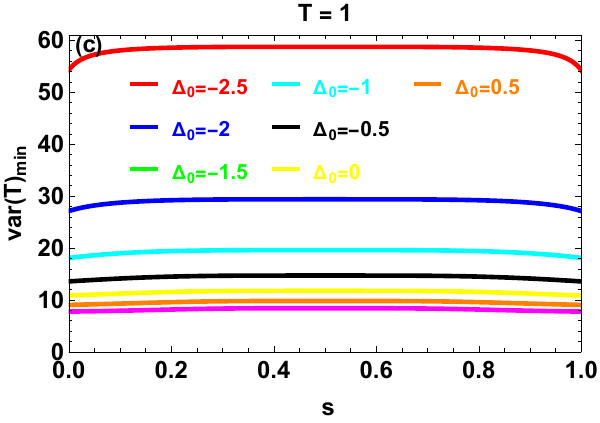}
	\caption{Minimal variance $\mathrm{Var}(T)_{\min}$ plotted as a function of the decoherence parameter $s$ for different values of the  $\Delta_0$ under the phase flip (PF) channel. Panel (a) corresponds to $T = 0.1$, panel (b) to $T = 0.5$, and panel (c) to $T = 1$.  The  energy level spacing is fixed at $\omega = 0.5$.}
	
	\label{fig:25a}
\end{figure}
The dependence of the minimal variance $\mathrm{Var}(T)_{\min}$ on the decoherence parameter $s$ for different values of $\Delta_0$ under the phase flip (PF) channel is illustrated in Figure~\ref{fig:25a}. In contrast to the monotonic behavior observed under dissipative noise, the curves here exhibit a symmetric, non-monotonic structure with respect to $s$. In particular, the variance vanishes at $s=0$ and $s=1$, while reaching a maximum around intermediate values of $s$.
This characteristic profile reflects the specific action of phase flip noise, which preserves population probabilities while altering phase coherence. As a result, the estimation precision is maximally degraded at intermediate dephasing strengths, whereas it is partially restored in the limiting cases.

A clear hierarchy with respect to $\Delta_0$ is maintained in all three subfigures. More negative values of $\Delta_0$ (e.g., $\Delta_0=-2.5$) systematically correspond to larger peak variances, while configurations closer to $\Delta_0=0$ yield lower maxima. 

Comparing the three temperatures shows that increasing $T$ significantly amplifies the overall magnitude of the variance. At $T=1$, the peak values become substantially larger, indicating that higher temperatures enhance the sensitivity of the estimation process to phase noise. Nevertheless, the qualitative bell-shaped dependence on $s$ remains preserved across all temperatures.


The minimal variance $\mathrm{Var}(\Delta_0)_{\min}$ under the phase flip (PF) channel is depicted in Figure~\ref{fig:26a} as a function of both the initial-state parameter $\Delta_0$ and the decoherence parameter $s$. In both cases, the surface preserves a pronounced quadratic profile along the $\Delta_0$ direction, with smaller values near the boundaries of the allowed domain and larger values around the central region. 

Along the $s$ direction, a non-monotonic dependence is observed. The variance remains small in the weak-dephasing regime ($s \approx 0$), increases toward intermediate values of $s$, and then decreases again as $s$ approaches unity. This bell-shaped structure reflects the specific action of phase flip noise, which modifies phase coherence without altering population distributions, leading to maximal estimation uncertainty at intermediate dephasing strengths.

Comparing the two subfigures reveals a significant influence of the energy-level spacing $\omega$. For $\omega=0.1$, the variance surface reaches considerably larger values, with a broad region of high uncertainty. In contrast, for $\omega=2$, the overall magnitude of $\mathrm{Var}(\Delta_0)_{\min}$ is noticeably reduced and the surface appears more compressed along the vertical direction. Although the qualitative structure remains unchanged, increasing $\omega$ clearly suppresses the maximal estimation uncertainty under the PF channel.

These results indicate that, in the presence of phase noise, the precision of $\Delta_0$ estimation is governed by the intrinsic quadratic dependence on the parameter and is modulated by the strength of dephasing and the spectral spacing of the system.
\begin{figure}[H]
	\includegraphics[scale=0.58]{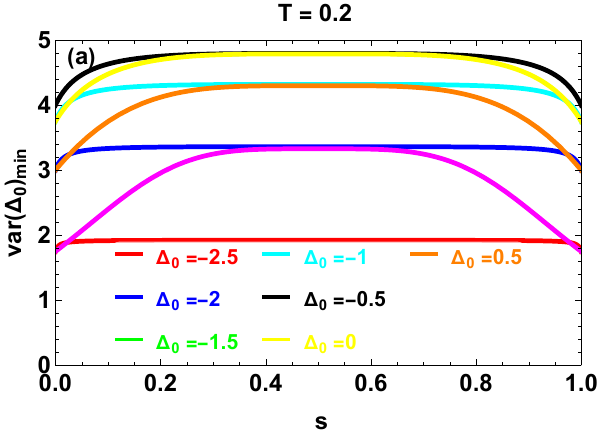}
	\includegraphics[scale=0.58]{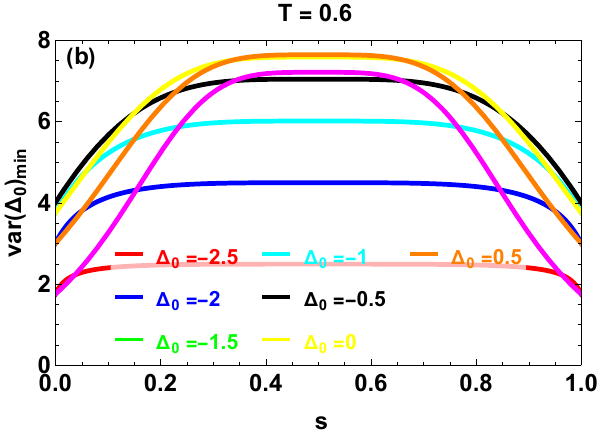}
	\includegraphics[scale=0.58]{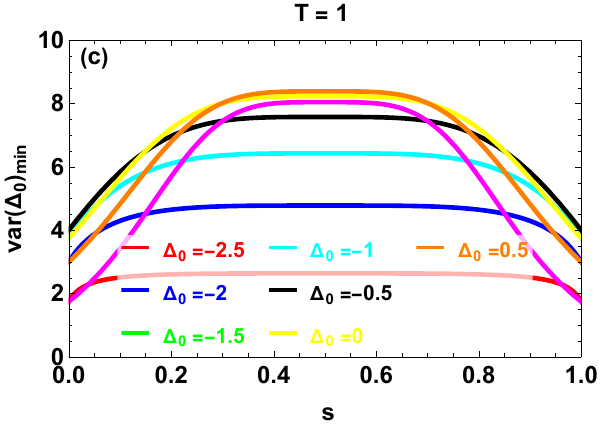}
	\caption{Minimal variance $\mathrm{Var}(\Delta_0)_{\min}$ plotted as a function of the decoherence parameter $s$  under the phase flip (PF) channel.  Panel (a) corresponds to $T=0.2$, panel (b) to $T=0.6$, and panel (c) to $T=1$. The  energy level spacing is fixed at $\omega =0.5$.}
	
	\label{fig:26}
\end{figure}
The behavior of the minimal variance $\mathrm{Var}(\Delta_0)_{\min}$ under the phase flip (PF) channel is illustrated in Figure~\ref{fig:26} as a function of the decoherence parameter $s$ for different temperatures. In all three subfigures, the variance exhibits a characteristic bell-shaped dependence on $s$: it increases from small values in the weak-dephasing regime, reaches a maximum at intermediate values of $s$, and then decreases again as $s$ approaches unity.

This symmetric, non-monotonic structure reflects the nature of phase flip noise, which alters phase coherence without modifying population distributions. Consequently, the estimation uncertainty is maximal at intermediate dephasing strengths, whereas it is reduced in the limiting cases.
A clear ordering with respect to $\Delta_0$ is maintained throughout the evolution. More negative values of $\Delta_0$ generally correspond to lower variances, while configurations closer to $\Delta_0=0$ yield higher peaks. 

Comparing panels \ref{fig:26}(a)-(c) reveals that increasing the temperature significantly amplifies the height of the bell-shaped curves. At $T=1$, the maximal values of $\mathrm{Var}(\Delta_0)_{\min}$ become substantially larger than those observed at $T=0.2$, indicating that higher temperatures enhance the sensitivity of the estimation process to phase noise. Nevertheless, the qualitative dependence on $s$ remains unchanged across all temperatures.
\begin{figure}[H]
	\centering
	\includegraphics[scale=0.73]{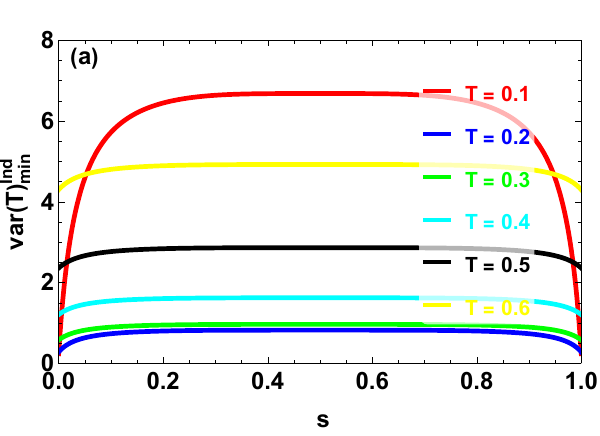}
	\hspace*{0.7cm}
	\includegraphics[scale=0.73]{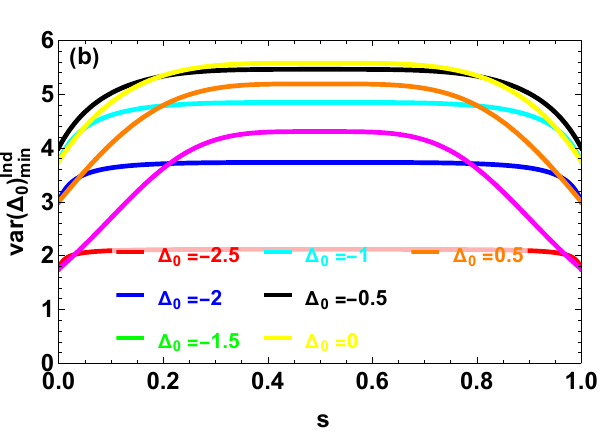}
	\caption{The minimal variance for the individual estimation as a function of the decoherence parameter $s$ under  the phase flip (PF) channel. 
		(a) Minimal variance $\mathrm{Var}(T)_{\min}^{\mathrm{Ind}}$ plotted versus $s$, with fixed parameters $\omega = 0.5$ and $\Delta_0 = -2$. 
		(b) Minimal variance $\mathrm{Var}(\Delta_0)_{\min}^{\mathrm{Ind}}$ plotted versus $s$, with $\omega = 0.5$ and $T = 0.3$.}
	\label{fig:27}
\end{figure}
The individual estimation scenario under the phase flip (PF) channel is presented in Figure~\ref{fig:27}, where the minimal variances are plotted as functions of the decoherence parameter $s$. In both cases, the curves exhibit a pronounced non-monotonic dependence on $s$, characterized by a rise from $s=0$, a maximum at intermediate dephasing strengths, and a subsequent decrease as $s$ approaches unity.
For temperature estimation (figure \ref{fig:27}(a)), the variance $\mathrm{Var}(T)_{\min}^{\mathrm{Ind}}$ increases significantly at intermediate values of $s$, with higher temperatures leading to larger peak values. This indicates that phase noise affects thermometric precision most strongly in the partial dephasing regime, while the limiting cases remain comparatively less detrimental.
A similar bell-shaped structure is observed for the estimation of the initial-state parameter (figure \ref{fig:27}(b)). The variance $\mathrm{Var}(\Delta_0)_{\min}^{\mathrm{Ind}}$ preserves a clear hierarchy with respect to $\Delta_0$, where values closer to zero generally correspond to higher peaks, whereas more negative configurations yield comparatively lower maximal uncertainties.

Overall, these results show that, under pure dephasing noise, the degradation of estimation precision in the individual strategy is not monotonic but instead exhibits a symmetric dependence on the decoherence strength, with maximal uncertainty occurring at intermediate phase flip probabilities.
\begin{figure}[H]
	\includegraphics[scale=0.58]{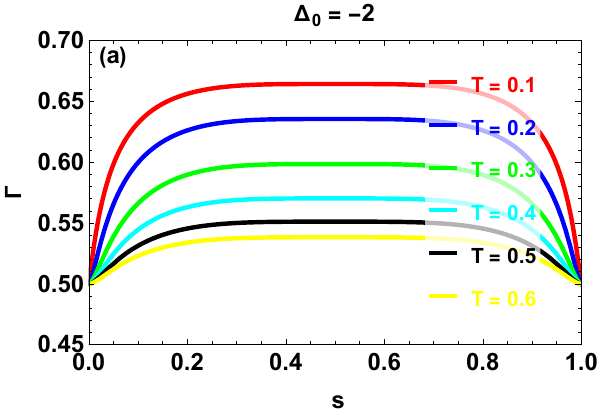}
	\includegraphics[scale=0.58]{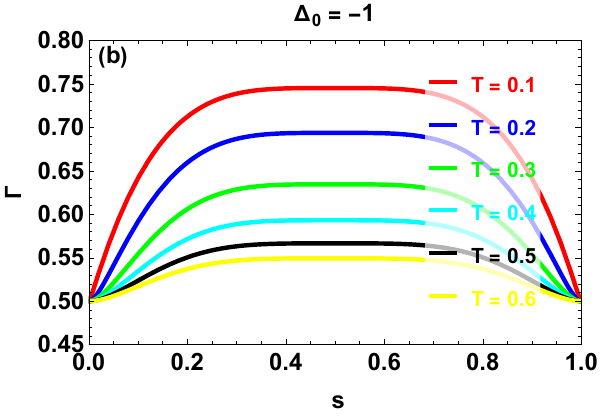}
	\includegraphics[scale=0.58]{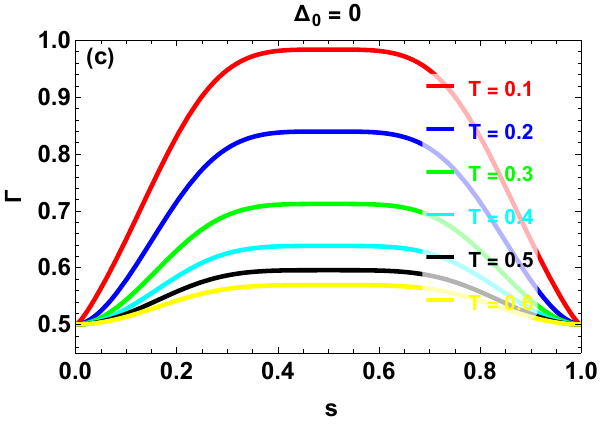}
\caption{The ratio between the minimal total variances in the estimation of the parameters $T$ and $\Delta_0$ in the phase plip (PF) channel as a function of the decoherence parameter, for different values of the temperature $T$, and for three values of $\Delta_0$: (a) $\Delta_0 = -2$, (b) $\Delta_0 = -1$, and (c) $\Delta_0 = 0$. The energy level spacing is fixed at $\omega = 0.5$.}

	\label{fig:28}
\end{figure}
The behavior of the performance ratio $\Gamma$ under the phase flip (PF) channel is illustrated in Figure~\ref{fig:28} for different temperatures and initial-state parameters. In all three subfigures, $\Gamma$ exhibits a non-monotonic dependence on the decoherence parameter $s$. Starting from a value close to $0.5$ in the weak-noise limit ($s \approx 0$), the ratio increases toward a maximum at intermediate dephasing strengths and then decreases again as $s$ approaches unity.
This symmetric profile reflects the characteristic action of phase flip noise, which induces maximal estimation imbalance between the simultaneous and individual strategies in the partial dephasing regime. The fact that $\Gamma$ remains below unity for all values of $s$ confirms that the simultaneous estimation protocol consistently outperforms the individual one, even in the presence of phase noise.

A clear hierarchy with respect to temperature is observed in each subfigure: lower temperatures correspond to larger peak values of $\Gamma$, whereas increasing $T$ reduces both the maximal value and the sensitivity of the ratio to $s$. Furthermore, the influence of the initial-state parameter is evident when comparing $\Delta_0=-2$, $\Delta_0=-1$, and $\Delta_0=0$. As $\Delta_0$ approaches zero, the peak of $\Gamma$ becomes significantly higher, indicating that the relative advantage of the simultaneous strategy is more strongly affected by phase noise for these initial configurations.

Overall, these results demonstrate that, under the PF channel, the efficiency of the simultaneous estimation scheme is dynamically modulated by dephasing strength, with the largest deviation from the weak-noise limit occurring at intermediate values of the decoherence parameter.
\subsection{Phase Damping channel}
Phase-damping (PD) decoherence channels describe a type of quantum noise that 
selectively suppresses quantum phase coherence while preserving the system's 
energy populations. The Kraus operators corresponding to a single-qubit PD channel are given by
\begin{equation}
	\mathcal{K}_1 =
	\begin{pmatrix}
		1 & 0 \\
		0 & \sqrt{1-s}
	\end{pmatrix},
	\qquad
	\mathcal{K}_2 =
	\begin{pmatrix}
		0 & 0 \\
		0 & \sqrt{s}
	\end{pmatrix}.
\label{PD}
\end{equation} 

By substituting Eqs.~(\ref{densit M}) and (\ref{PD}) into Eq.~(\ref{kraus}), the density matrix of the system 
under the influence of the PD channel can be written as
\begin{equation}
	{\varrho}^{\,\mathrm{PD}}_{12} =
	\begin{pmatrix}
		x & 0 & 0 & 0 \\
		0 & z & \upsilon_{23} & 0 \\
		0 & \upsilon_{23} & z & 0 \\
		0 & 0 & 0 & y
	\end{pmatrix},
\end{equation}
where the coherence terms are modified according to
\begin{equation}
	\upsilon_{23} = \delta\,(1 - s).
\end{equation}

This result clearly shows that the phase-damping channel leaves the diagonal 
elements of the density matrix invariant, while the off-diagonal terms are 
attenuated by a factor $(1 - s)$, reflecting the loss of phase coherence induced by the environment.

Using the approach outlined in Section~\ref{sec:2}, we now compute the analytical expressions of 
$\mathrm{Var}(T)_{\min}$, $\mathrm{Var}(\Delta_0)_{\min}$, 
$\mathrm{Var}(T)_{\min}^{\mathrm{Ind}}$, and 
$\mathrm{Var}(\Delta_0)_{\min}^{\mathrm{Ind}}$ 
for the simultaneous and individual estimation scenarios. 
These quantities read as follows
\begin{equation}
	\mathrm{Var}(T)_{\min}=\dfrac{Y}{Z},
\end{equation}
with
\begin{align}
	Y &= 
	T^4\bigg(1+2\cosh(\tfrac{\omega}{T})\bigg)
	\Bigg[
	2(3+\Delta_0)+(-2+s)s(5+3\Delta_0)
	+4(3+\Delta_0+(-2+s)s(2+\Delta_0))\cosh(\tfrac{\omega}{T})
	\\
	&\qquad
	+(-2+s)s(-1+\Delta_0)\cosh(\tfrac{2\omega}{T})
	\Bigg],\nonumber
\end{align}
and 
\begin{align}
	Z =
	(3+\Delta_0)\omega^2
	\Bigg[
	2(3+\Delta_0+(-2+s)s(2+\Delta_0))
	+(3+\Delta_0+2(-2+s)s(1+\Delta_0))
	\cosh(\tfrac{\omega}{T})
	\Bigg].
\end{align}
\begin{equation}
	\mathrm{Var}(\Delta_0)_{\min}=\dfrac{A}{B},
\end{equation}
with
\begin{align}
	A &=
	-(3+\Delta_0)
	\Bigg[
	2e^{\frac{2\omega}{T}}
	\bigg[3(-3+s)(1+s)
	+(6+(-2+s)s)\Delta_0
	+(3+4(-2+s)s)\Delta_0^2\bigg]\nonumber
	\\
	&\quad
	+(1+e^{\frac{4\omega}{T}})
	(-1+\Delta_0)\bigg[3+\Delta_0+(-2+s)s(1+2\Delta_0)\bigg]
	\\
	&\quad
	+(e^{\frac{\omega}{T}}+e^{\frac{3\omega}{T}})
	\bigg[5(-1+\Delta_0)(3+\Delta_0)
	+2(-2+s)s(-1+3\Delta_0^2)\bigg]
	\Bigg],\nonumber
\end{align}
and
\begin{align}
	B &=
	\bigg(1+e^{\frac{\omega}{T}}+e^{\frac{2\omega}{T}}\bigg)
	\Bigg[
	(3+\Delta_0+2(-2+s)s(1+\Delta_0))(1+e^{\frac{2\omega}{T}})
	+4e^{\frac{\omega}{T}}
	\big(3+\Delta_0+(-2+s)s(2+\Delta_0)\big)
	\Bigg].
\end{align}
\begin{equation}
	\mathrm{Var}(T)_{\text{min}}^{\text{Ind}}=\dfrac{A'}{B'},
\end{equation}
with
\begin{align}
	A' &= 
	e^{-\frac{2\omega}{T}}
	\bigg(1+e^{\frac{\omega}{T}}+e^{\frac{2\omega}{T}}\bigg)^2
	T^4
	\Bigg[
	s(-1+\Delta_0)(1+e^{\frac{2\omega}{T}})
	+2e^{\frac{\omega}{T}}\big(-3+s+(-1+s)\Delta_0\big)
	\Bigg]
	\\
	&\quad\times
\Bigg[
	(-2+s)(-1+\Delta_0)(1+e^{\frac{2\omega}{T}})
	+2e^{\frac{\omega}{T}}\big(1+s+(-1+s)\Delta_0\big)
	\Bigg],\nonumber
\end{align}
and
\begin{align}
	B' &=
	(3+\Delta_0)\omega^2
	\Bigg[
	2e^{\frac{2\omega}{T}}
	\big[3(-3+s)(1+s)
	+(6+(-2+s)s)\Delta_0
	+(3+4(-2+s)s)\Delta_0^2\big]\nonumber
	\\
	&\quad
	+(1+e^{\frac{4\omega}{T}})
	(-1+\Delta_0)\bigg[3+\Delta_0+(-2+s)s(1+2\Delta_0)\bigg]
	+(e^{\frac{\omega}{T}}+e^{\frac{3\omega}{T}})
	\bigg[5(-1+\Delta_0)(3+\Delta_0)
	+2(-2+s)s(-1+3\Delta_0^2)\bigg]
	\Bigg].
\end{align}
\begin{align}
\mathrm{Var}(\Delta_0)_{\text{min}}^{\text{Ind}}=	&-\frac{
		2(3+\Delta_0)
		\Bigg[
		1+s+(-1+s)\Delta_0
		+(-2+s)(-1+\Delta_0)\cosh\!\left(\frac{\omega}{T}\right)
		\Bigg]
	\Bigg[
		-3+s+(-1+s)\Delta_0
		+s(-1+\Delta_0)\cosh\!\left(\frac{\omega}{T}\right)
		\Bigg]
	}{
		2(3+\Delta_0)
		+(-2+s)s(5+3\Delta_0)
		+4\left(3+\Delta_0+(-2+s)s(2+\Delta_0)\right)
		\cosh\!\left(\frac{\omega}{T}\right)
		+(-2+s)s(-1+\Delta_0)
		\cosh\!\left(\frac{2\omega}{T}\right)
	}.
\end{align}
In the same manner, the ratio $\Gamma$, defined in Eq.~(\ref{rap}), is explicitly expressed as $\Gamma =\dfrac{C^\prime}{D^\prime}$, where $C^\prime$ and $D^\prime$ are given by
\begin{align}
		C^\prime &=
	\Bigg[
	(-2+s)s(-1+\Delta_0)(1+e^{\frac{4\omega}{T}})
	+4(e^{\frac{\omega}{T}}+e^{\frac{3\omega}{T}})\bigg(3+\Delta_0+(-2+s)s(2+\Delta_0)\bigg)\nonumber
	\\
	&\quad
	+2e^{\frac{2\omega}{T}}\bigg(2(3+\Delta_0)+(-2+s)s(5+3\Delta_0)\bigg)
	\Bigg]\nonumber
	\\
	&\quad\times
\Bigg[
	2e^{\frac{2\omega}{T}}\bigg(3(-3+s)(1+s)+(6+(-2+s)s)\Delta_0+(3+4(-2+s)s)\Delta_0^2\bigg)
	\\
	&\quad
	+(1+e^{\frac{4\omega}{T}})(-1+\Delta_0)\bigg(3+\Delta_0+(-2+s)s(1+2\Delta_0)\bigg)\nonumber
	\\
	&\quad
	+(e^{\frac{\omega}{T}}+e^{\frac{3\omega}{T}})\bigg(5(-1+\Delta_0)(3+\Delta_0)+2(-2+s)s(-1+3\Delta_0^2)\bigg)\nonumber
	\Bigg],
\end{align}
and 
\begin{align}
	D^\prime &=
	2\big(1+e^{\frac{\omega}{T}}+e^{\frac{2\omega}{T}}\big)\nonumber
	\\
	&\quad\times
	\Bigg[
	s(-1+\Delta_0)(1+e^{\frac{2\omega}{T}})
	+2e^{\frac{\omega}{T}}(-3+s+(-1+s)\Delta_0)
	\Bigg]\nonumber
	\\
	&\quad\times
		\Bigg[
	(-2+s)(-1+\Delta_0)(1+e^{\frac{2\omega}{T}})
	+2e^{\frac{\omega}{T}}(1+s+(-1+s)\Delta_0)
	\Bigg]
	\\
	&\quad\times
	\Bigg[
	(3+\Delta_0+2(-2+s)s(1+\Delta_0))(1+e^{\frac{2\omega}{T}})
	+4e^{\frac{\omega}{T}}(3+\Delta_0+(-2+s)s(2+\Delta_0))
	\Bigg].\nonumber
\end{align}


	The influence of the phase damping (PD) channel on the temperature estimation precision is illustrated in Figure~\ref{fig:29}, where the minimal variance $\mathrm{Var}(T)_{\min}$ is plotted as a function of the temperature $T$ and the decoherence parameter $s$. In both subfigures, the variance increases smoothly along the $s$ direction, indicating that stronger dephasing progressively degrades thermometric precision.
	For $\omega=0.6$, the surface remains relatively moderate in magnitude, with $\mathrm{Var}(T)_{\min}$ exhibiting a gradual variation over the explored temperature range. The dependence on $T$ is smooth and no abrupt amplification is observed, suggesting that moderate spectral spacing limits the impact of phase damping on estimation accuracy.
	In contrast, for $\omega=1$, the variance surface becomes significantly steeper and its overall magnitude increases dramatically. In particular, large values of $\mathrm{Var}(T)_{\min}$ emerge in regions of strong dephasing, reflecting a substantial loss of temperature-sensitive information. Although the qualitative monotonic dependence on $s$ is preserved, the estimation uncertainty is strongly amplified for larger energy-level spacing.
	
	These results indicate that, under the PD channel, dephasing noise systematically deteriorates temperature estimation precision, with the severity of this degradation being highly sensitive to the spectral structure of the system.
	\begin{figure}[H]
		\includegraphics[scale=0.58]{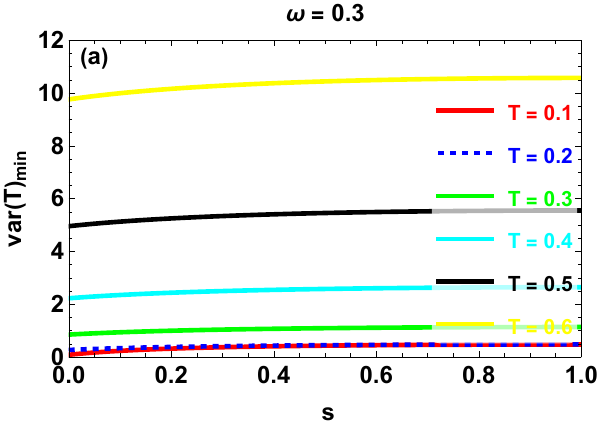}
		\includegraphics[scale=0.58]{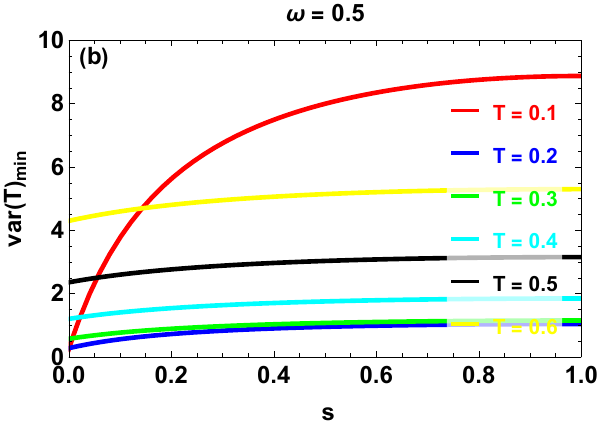}
		\includegraphics[scale=0.58]{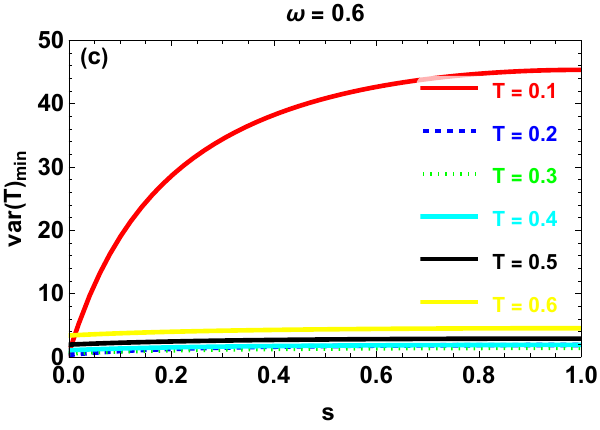}
		\caption{Minimal variance $\mathrm{Var}(T)_{\min}$ plotted as a function of the decoherence parameter $s$ for different values of the temperature $T$ under the phase damping (PD) channel. Panel (a) corresponds to $\omega = 0.3$, panel (b) to $\omega = 0.5$, and panel (c) to $\omega = 0.6$. The initial state parameter is fixed at $\Delta_0 = -2$.}
		
		\label{fig:30}

\end{figure}
The dependence of the minimal variance $\mathrm{Var}(T)_{\min}$ on the decoherence parameter $s$ under the phase damping (PD) channel is presented in Figure~\ref{fig:30} for different temperatures and energy-level spacings. In contrast to the symmetric behavior observed under the PF channel, the variance here exhibits a monotonic increase with $s$, followed by a gradual saturation for large dephasing strengths.
For $\omega=0.3$ (figure \ref{fig:30}(a)), the curves display a moderate growth with $s$, and the separation between different temperatures remains relatively smooth. As the energy-level spacing increases to $\omega=0.5$ (figure \ref{fig:30}(b)), the sensitivity to dephasing becomes more pronounced, particularly for low temperatures such as $T=0.1$, where the variance rises sharply before stabilizing.
A dramatic amplification is observed for $\omega=0.6$ (figure \ref{fig:30}(c)). In this case, the variance corresponding to low temperatures increases rapidly and reaches significantly higher values compared to the other temperatures. This indicates that, under stronger spectral separation, phase damping can severely deteriorate thermometric precision, especially in the low-temperature regime.

Overall, these results show that, in the PD channel, decoherence induces a steady degradation of temperature estimation accuracy, with the magnitude of this effect strongly influenced by both the energy-level spacing and the thermal configuration.

		\begin{figure}[H]
	\includegraphics[scale=0.58]{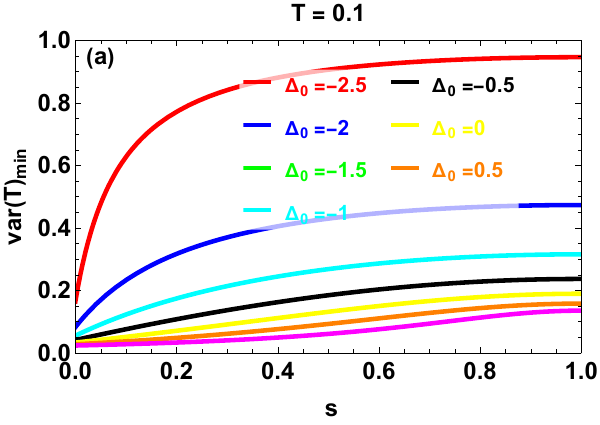}
	\includegraphics[scale=0.58]{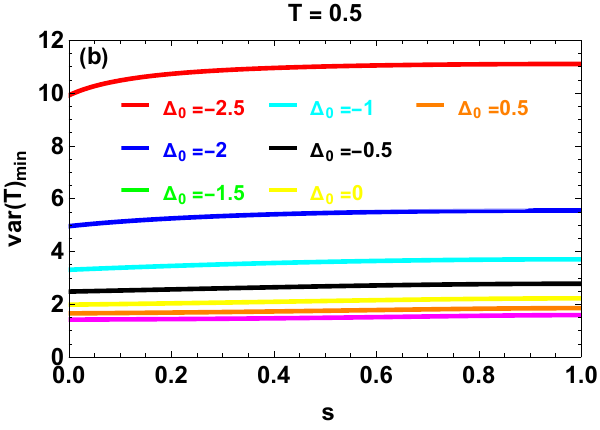}
	\includegraphics[scale=0.58]{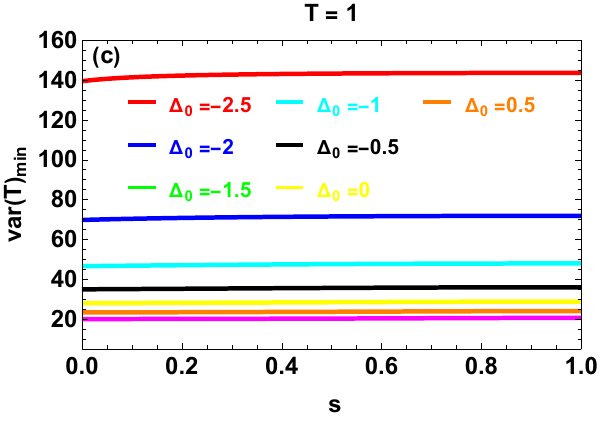}
	\caption{Minimal variance $\mathrm{Var}(T)_{\min}$ plotted as a function of the decoherence parameter $s$ for different values of  $\Delta_0$ under the phase damping (PD) channel. Panel (a) corresponds to $T = 0.1$, panel (b) to $T = 0.5$, and panel (c) to $T=1$. The energy level spacing is fixed at $\omega = 0.3$.}
	
	\label{fig31}
	
\end{figure}
The influence of the initial-state parameter $\Delta_0$ on the temperature estimation precision under the phase damping (PD) channel is illustrated in Figure~\ref{fig31}. In all three subfigures, the minimal variance $\mathrm{Var}(T)_{\min}$ increases monotonically with the decoherence parameter $s$ and gradually approaches a saturation value for strong dephasing.
A clear ordering with respect to $\Delta_0$ is preserved throughout the evolution. More negative values of $\Delta_0$ (e.g., $\Delta_0=-2.5$) systematically yield larger variances, while configurations closer to zero exhibit comparatively lower uncertainty. This hierarchy remains stable across the entire range of $s$, indicating that the initial-state preparation plays a crucial role in determining the robustness of thermometric precision under phase damping noise.

Comparing the three temperatures reveals a substantial amplification of the variance as $T$ increases. At $T=1$, the magnitude of $\mathrm{Var}(T)_{\min}$ becomes dramatically larger than at $T=0.1$, with the curves reaching significantly higher saturation levels. This demonstrates that higher thermal energy enhances the sensitivity of the estimation process to dephasing effects, even though the qualitative monotonic dependence on $s$ remains unchanged.
Overall, these results show that, under the PD channel, decoherence induces a steady and temperature-dependent degradation of thermometric precision, while preserving a consistent hierarchy determined by the initial-state parameter.


The minimal variance $\mathrm{Var}(\Delta_0)_{\min}$ under the phase damping (PD) channel is depicted in Figure~\ref{fig32} as a function of the initial-state parameter $\Delta_0$ and the decoherence parameter $s$. In both subfigures, the surface preserves a smooth quadratic profile along the $\Delta_0$ direction, with lower values near the boundaries of the allowed domain and higher values in the central region.
Along the $s$ direction, the variance increases monotonically and gradually approaches a saturation plateau for strong dephasing. This steady growth reflects the cumulative loss of phase coherence induced by the PD channel, which progressively limits the precision achievable in estimating $\Delta_0$.

Comparing the two subfigures shows that increasing the energy-level spacing from $\omega=0.3$ to $\omega=2$ slightly modifies the curvature and overall scale of the variance surface, while leaving its qualitative structure unchanged. In both cases, the dependence on $\Delta_0$ remains dominant in shaping the geometry of the surface, whereas the decoherence parameter mainly controls the amplitude of the estimation uncertainty.

These results confirm that, under phase damping noise, the estimation of the initial-state parameter retains its intrinsic geometric dependence, but the attainable precision is steadily degraded as dephasing becomes stronger.

\begin{figure}[H]
	\includegraphics[scale=0.58]{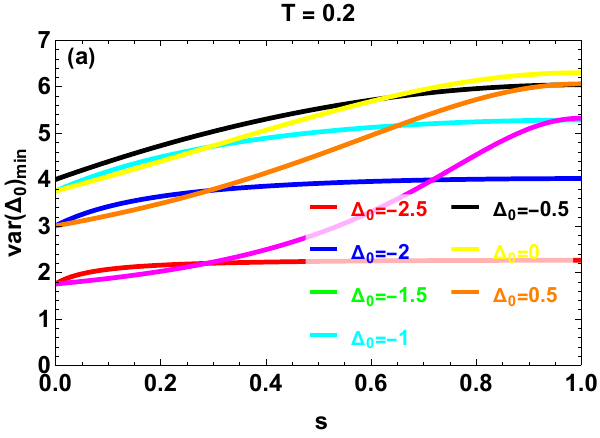}
	\includegraphics[scale=0.58]{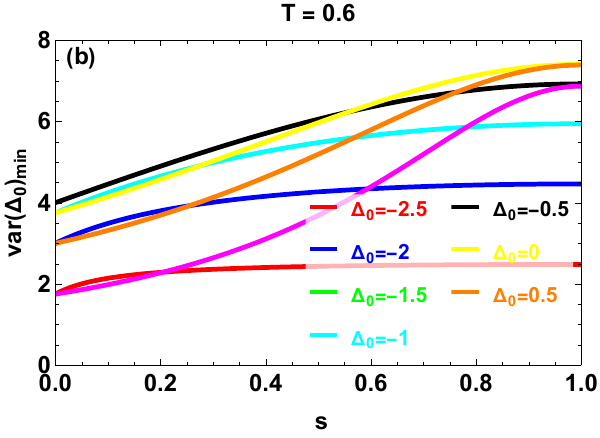}
	\includegraphics[scale=0.58]{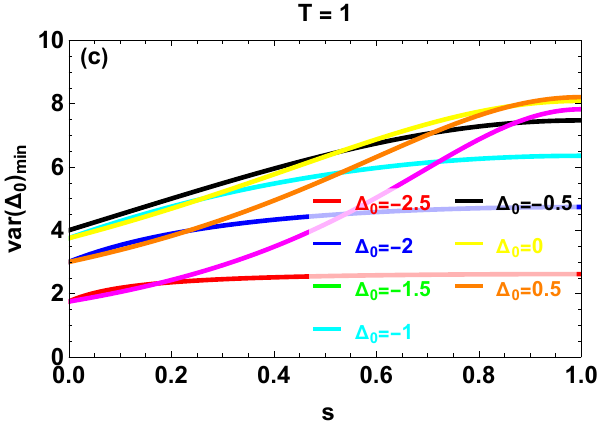}
	\caption{Minimal variance $\mathrm{Var}(\Delta_0)_{\min}$ plotted as a function of the decoherence parameter $s$  under the phase damping (PD) channel.  Panel (a) corresponds to $T=0.2$, panel (b) to $T=0.6$, and panel (c) to $T=1$. The  energy level spacing is fixed at $\omega =1$.}
	
	\label{fig33}
\end{figure}
The behavior of the minimal variance $\mathrm{Var}(\Delta_0)_{\min}$ under the phase damping (PD) channel is illustrated in Figure~\ref{fig33} for three different temperatures. In all subfigures, the variance increases monotonically with the decoherence parameter $s$ and tends to approach a saturation regime as $s$ approaches unity. No oscillatory or symmetric structure is observed, in contrast to the PF channel, indicating that PD induces a steady and cumulative degradation of estimation precision.
A well-defined hierarchy with respect to $\Delta_0$ is clearly preserved throughout the evolution. More negative values of $\Delta_0$ (e.g., $\Delta_0=-2.5$) correspond to lower variances across the entire range of $s$, whereas configurations closer to zero (e.g., $\Delta_0=0$ or $0.5$) lead to significantly larger uncertainties. This ordering remains stable even in the strong-dephasing regime.

Comparing the three temperatures reveals a substantial amplification of the overall variance as $T$ increases. While the qualitative monotonic behavior with respect to $s$ remains unchanged, the magnitude of $\mathrm{Var}(\Delta_0)_{\min}$ grows markedly from $T=0.2$ to $T=1$, with the highest temperature producing the steepest increase and largest saturation values.

These results demonstrate that, under phase damping noise, the estimation of the initial-state parameter is governed by a robust geometric dependence on $\Delta_0$, while dephasing strength and temperature primarily control the scale of the attainable precision.
\begin{figure}[H]
	\centering
	\includegraphics[scale=0.73]{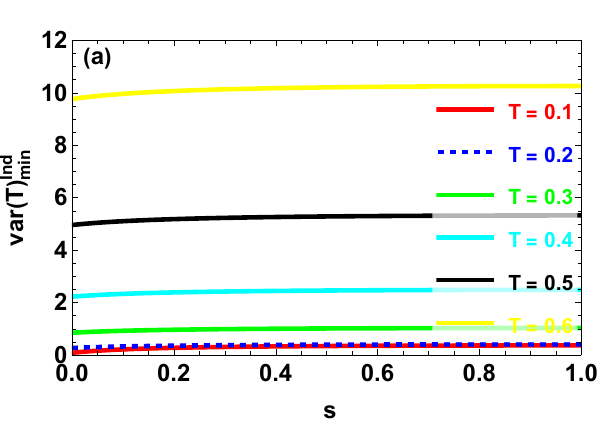}
	\hspace*{0.7cm}
	\includegraphics[scale=0.73]{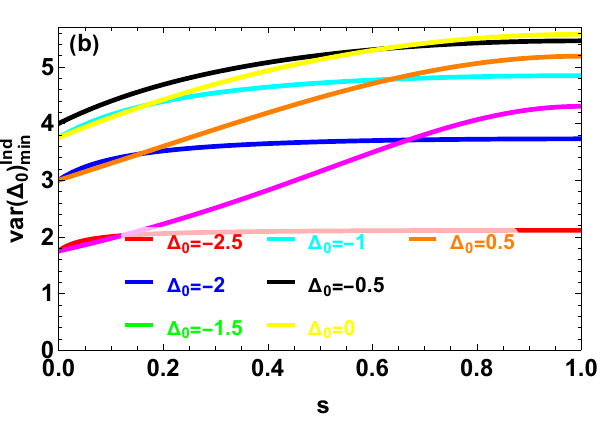}
	\caption{The minimal variance for the individual estimation as a function of the decoherence parameter $s$ under the phase damping (PD) channel. 
		(a) Minimal variance $\mathrm{Var}(T)_{\min}^{\mathrm{Ind}}$ plotted versus $s$, with fixed parameters $\omega = 0.3$ and $\Delta_0 = -2$. 
		(b) Minimal variance $\mathrm{Var}(\Delta_0)_{\min}^{\mathrm{Ind}}$ plotted versus $s$, with $\omega = 0.5$ and $T = 0.3$.}

	\label{fig:33}
\end{figure}
The individual estimation scenario under the phase damping (PD) channel is presented in Figure~\ref{fig:33}. In subfigure (a), the minimal variance $\mathrm{Var}(T)_{\min}^{\mathrm{Ind}}$ exhibits a monotonic increase with the decoherence parameter $s$ for all temperatures. The growth is relatively smooth and tends to approach a saturation value for large $s$, indicating a gradual loss of thermometric precision under increasing dephasing. A clear ordering with respect to temperature is preserved: higher temperatures systematically correspond to larger variances over the entire range of $s$.
Subfigure (b) displays the behavior of $\mathrm{Var}(\Delta_0)_{\min}^{\mathrm{Ind}}$ as a function of $s$ for different initial-state parameters. Here again, a monotonic increase followed by saturation is observed, without any oscillatory features. The hierarchy with respect to $\Delta_0$ remains stable throughout the evolution: configurations closer to $\Delta_0=0$ produce larger variances, while more negative values yield comparatively lower uncertainties. 

Overall, these results confirm that, under the PD channel, dephasing induces a steady and cumulative degradation of estimation precision in the individual strategy. The qualitative structure of the curves remains robust, with temperature and initial-state preparation primarily controlling the scale of the variance, while the decoherence parameter governs its progressive amplification.

\begin{figure}[H]
	\includegraphics[scale=0.58]{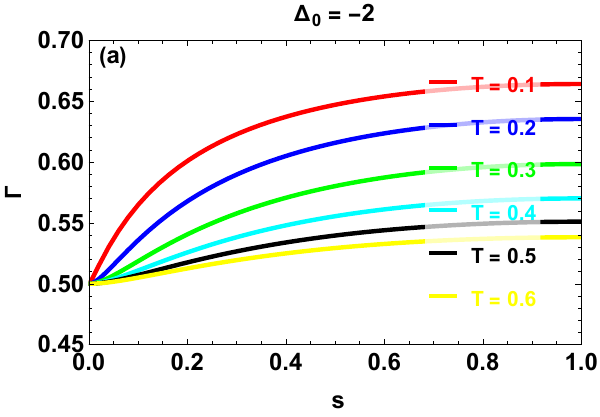}
	\includegraphics[scale=0.58]{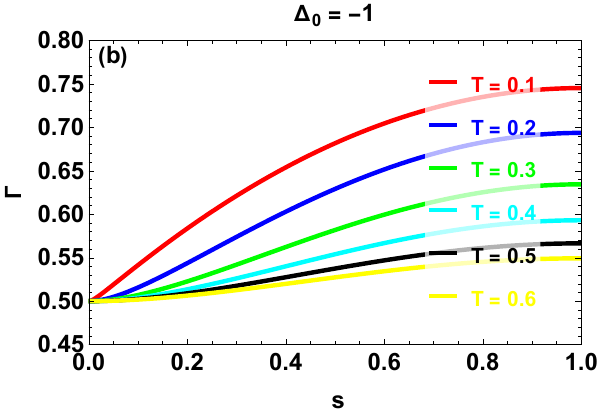}
	\includegraphics[scale=0.58]{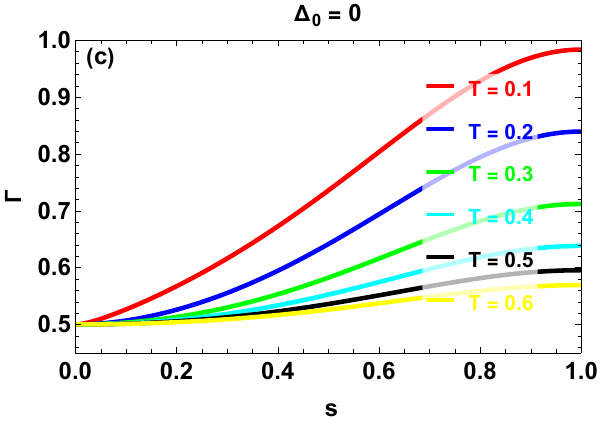}
\caption{The ratio between the minimal total variances in the estimation of the parameters $T$ and $\Delta_0$ in the phase damping (PD) channel as a function of the decoherence parameter, for different values of the temperature $T$, and for three values of $\Delta_0$: (a) $\Delta_0 = -2$, (b) $\Delta_0 = -1$, and (c) $\Delta_0 = 0$. The energy level spacing is fixed at $\omega = 0.5$.}
	\label{fig:34}
\end{figure}
The evolution of the performance ratio $\Gamma$ under the phase damping (PD) channel is presented in Figure~\ref{fig:34} for different temperatures and initial-state parameters. In all three subfigures, $\Gamma$ starts from a value close to $0.5$ in the weak-dephasing limit ($s \approx 0$) and increases monotonically with the decoherence parameter, eventually approaching a saturation value for strong phase damping.
For $\Delta_0=-2$ (subfigure \ref{fig:34}(a)), the growth of $\Gamma$ remains moderate, and the ratio stabilizes well below unity even at $s=1$. As the initial-state parameter moves toward $\Delta_0=-1$ (subfigure \ref{fig:34}(b)), the increase becomes more pronounced, leading to higher asymptotic values. The most significant amplification is observed for $\Delta_0=0$ (subfigure \ref{fig:34}(c)), where $\Gamma$ approaches values close to unity for low temperatures, indicating a substantial reduction of the relative advantage of the simultaneous estimation strategy in this configuration.

A clear temperature ordering is maintained in all cases: lower temperatures produce larger values of $\Gamma$ throughout the evolution, whereas increasing $T$ reduces both the slope and the asymptotic magnitude of the ratio. Despite this growth, $\Gamma$ remains below one for all parameters, confirming that the simultaneous estimation protocol consistently outperforms the individual one, even in the presence of phase damping noise.

These results show that, under the PD channel, dephasing progressively diminishes the metrological advantage of the joint estimation scheme, with the extent of this reduction being strongly dependent on the initial-state preparation and moderately influenced by temperature.

In summary, the analysis of the amplitude damping (AD), phase flip (PF), and phase damping (PD) channels reveals qualitatively distinct impacts of decoherence on multiparameter quantum estimation. 

The AD channel induces a monotonic and often rapid degradation of precision as the decoherence strength increases, reflecting the strong sensitivity of both temperature and initial-state estimation to dissipative noise. The relative advantage of the simultaneous estimation strategy, quantified by the ratio $\Gamma$, decreases progressively with increasing damping, although it remains below unity in all explored regimes.

In contrast, the PF channel exhibits a non-monotonic dependence on the decoherence parameter, characterized by symmetric, bell-shaped behaviors in the minimal variances and in the performance ratio. The estimation uncertainty is maximal at intermediate dephasing strengths, while it is partially reduced in the limiting cases. This distinctive structure highlights the fundamentally different role of phase-flip noise compared to dissipative mechanisms.

The PD channel, while also associated with pure dephasing, produces a monotonic increase and subsequent saturation of the variances and of $\Gamma$, without the symmetric behavior observed in the PF case. The degradation of precision is steady and cumulative, with its magnitude strongly influenced by the initial-state preparation and moderately by temperature.

Overall, these results demonstrate that the nature of the noise channel critically determines not only the scale of estimation uncertainty but also its dynamical structure. While decoherence generally reduces metrological performance, the extent and functional form of this degradation depend sensitively on whether the noise is dissipative or purely dephasing. Importantly, the simultaneous estimation protocol consistently maintains a metrological advantage over the individual strategy across all considered channels.
\section{CONCLUSION}   \label{sec:8}

In this work, we have investigated the ultimate precision limits for the estimation of the temperature $T$ and the initial-state parameter $\Delta_0$ in a bipartite system of uniformly accelerated Unruh--DeWitt detectors. By employing the quantum Fisher information matrix (QFIM) together with a vectorization-based analytical framework, we derived exact expressions for the corresponding quantum Cram\'er--Rao bounds and characterized both individual and simultaneous estimation strategies.
In the absence of noise, we have shown that the parameters $T$ and $\Delta_0$ are quantum compatible, as their associated symmetric logarithmic derivatives share a common eigenbasis. This guarantees the saturability of the multiparameter quantum Cram\'er--Rao bound and explains why the simultaneous estimation strategy does not lead to any loss of precision compared to the individual one. While the estimation precision of $\Delta_0$ is solely determined by the structure of the initial state, the precision associated with $T$ is strongly influenced by the effective Unruh temperature and by the detector energy-level spacing.
The analysis of the Markovian and non-Markovian regimes reveals two qualitatively distinct dynamical behaviors. In the Markovian regime, dissipation induces a monotonic and irreversible degradation of quantum information, leading to a progressive increase of the minimal variances toward stationary values. By contrast, the non-Markovian regime exhibits pronounced temporal oscillations in the estimation precision, reflecting information backflow from the environment to the system. These memory effects generate temporal windows in which the precision is temporarily enhanced, highlighting the metrological relevance of non-Markovian dynamics.
We have further examined the robustness of the estimation protocols under correlated noisy channels, namely amplitude damping (AD), phase flip (PF), and phase damping (PD). Our results demonstrate that the structure of the noise plays a crucial role in determining the achievable precision. The AD channel, affecting both populations and coherences, induces a stronger degradation of precision. In contrast, PF and PD channels, being predominantly dephasing in nature, lead to comparatively milder modifications of the estimation performance. In all cases, classical correlations in the noise mitigate the loss of precision and improve the overall metrological stability.
\begin{figure}[H]
	\centering
	\includegraphics[scale=0.78]{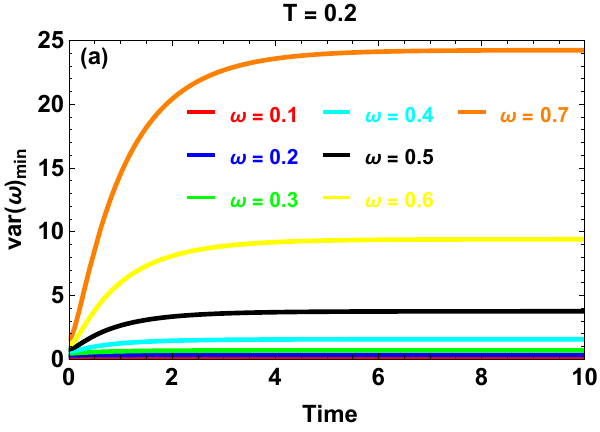}
	\includegraphics[scale=0.78]{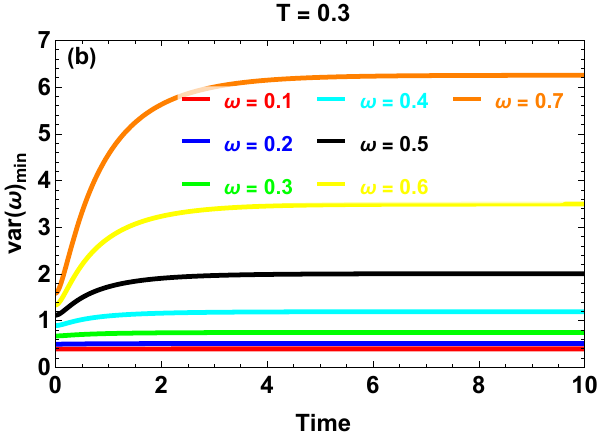}
	\caption{Dynamical evolution of the minimal variance $\mathrm{Var}(\omega)_{\min}$ in the Markovian regime, for different values of energy $\omega$ and for two values of  temperature $T$: (a) $T = 0.2$ and (b). The remaining parameters are fixed at $\tau = 0.1$, $\mu=0.4$, and $\Delta_0=-1.5$.}
	
	\label{fig:clc1}
\end{figure}
\begin{figure}[H]
	\centering
	\includegraphics[scale=0.78]{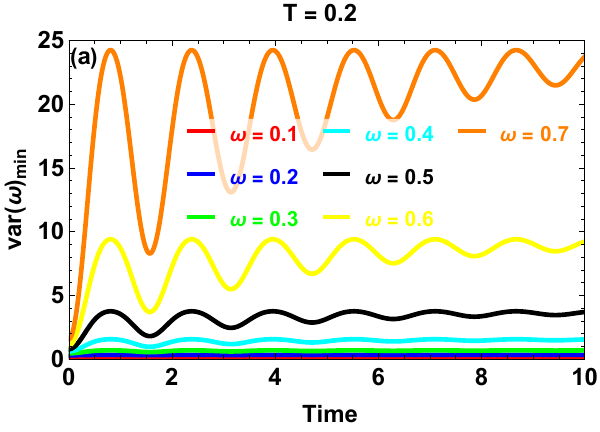}
	\includegraphics[scale=0.78]{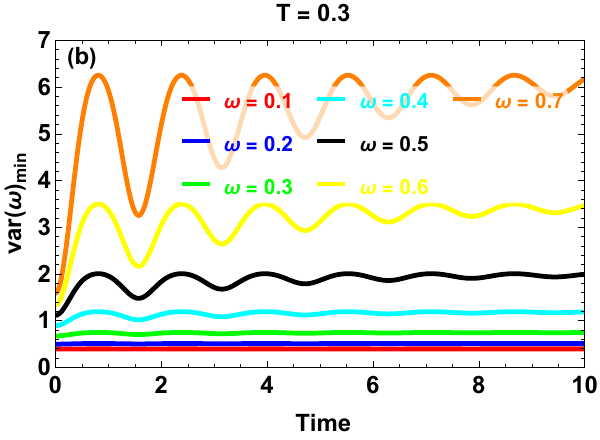}
	\caption{Dynamical evolution of the minimal variance $\mathrm{Var}(\omega)_{\min}$ in the non-Markovian regime, for different values of energy $\omega$ and for two values of  temperature $T$: (a) $T = 0.2$ and (b). The remaining parameters are fixed at $\tau = 5$, $\mu=0.4$, and $\Delta_0=-1.5$.}
	
	\label{fig:clc2}
\end{figure}
Although the core of this work focuses on the estimation of $T$ and $\Delta_0$, we have also explored, as a perspective, the estimation of the  parameter $\omega$. The behavior of the minimal variance associated with $\omega$, shown in Figs.~\ref{fig:clc1} and \ref{fig:clc2}, reveals a strong sensitivity to the dynamical nature of the environment. In the Markovian regime, the estimation precision degrades monotonically with time, whereas in the non-Markovian regime it exhibits oscillatory features induced by memory effects. These results indicate that $\omega$ may serve as an additional probe of environmental structure and relativistic thermal response.

Overall, our findings clarify the interplay between acceleration-induced thermality, initial-state structure, environmental memory, and noise mechanisms in relativistic quantum metrology. They provide a unified framework for assessing precision limits in accelerated open quantum systems and open promising directions for extended multiparameter estimation schemes including spectral and dynamical quantities.

\section*{Acknowledgments}

The research work was supported by Princess Nourah bint Abdulrahman University Researchers Supporting Project number (PNURSP2026R399), Princess Nourah bint Abdulrahman University, Riyadh, Saudi Arabia. The authors are thankful to the Deanship of Graduate Studies and Scientific Research at University of Bisha for supporting this work through the Fast-Track Research Support Program.


\end{document}